%% file: 000-paper.tex
\documentclass[manuscript]{acmart} % arxiv

\usepackage{soul}
\usepackage{enumitem}
\usepackage{rotating}
\usepackage{multirow}
\usepackage{pdflscape}
\usepackage{lineno}
\usepackage{longtable, longfbox}
\usepackage{xcolor, colortbl}
\usepackage{tikz}
\usepackage{tcolorbox}
\usepackage{diagbox}
\usepackage{wrapfig}
\usepackage{siunitx}
\usepackage{graphicx,marvosym}
\usepackage{caption, subcaption}
\usepackage{array}
\usepackage{textcomp}
\usepackage{mathtools}
\usepackage{adjustbox}
\usepackage{ifthen}

\AtBeginDocument{%
  \providecommand\BibTeX{{%
    \normalfont B\kern-0.5em{\scshape i\kern-0.25em b}\kern-0.8em\TeX}}}

\setcopyright{acmcopyright}
\copyrightyear{9999}
\acmYear{9999}
\acmDOI{99.9999/9999999.9999999}

\begin{document}
%\title{How to Find and Fix AI Inclusivity Bugs:\\A Field Study}
%\title{How to Find and Fix AI Inclusivity Bugs:\\What Three AI Product Teams Did}
%\title{AI Inclusivity Bugs:\\How 3 AI Product Teams Found Them and Fixed Them}
\title{``Over-the-Hood'' AI Inclusivity Bugs\\and How 3 AI Product Teams Found  and Fixed Them}
%\title{6 AI Inclusivity Bugs: \\What in-the-trenches AI product developer teams found}

\include{z-commands}
\include{z-defined-colors}
\include{z-GM-Adapt-Wordings/01-Original-GM-Wording}
\include{z-GM-Adapt-Wordings/02-Post-Action-Fork-Wording}
\include{z-GM-Adapt-Wordings/03-Pre-Action-Fork-Wording}

\include{z-GM-Adapt-Wordings/04-Unimplemented-Adaptation-Wordings}
\include{z-shorthand-commands}

\include{z-quotation-commands}
\include{z-statistical-commands}

\author{Andrew Anderson}
\email{anderan2@oregonstate.edu}
\orcid{0000-0003-4964-6059}
\affiliation{\institution{Oregon State University}}

\author{Fatima A. Moussaoui}
\email{moussaof@oregonstate.edu}
\affiliation{\institution{Oregon State University}}

\author{Jimena Noa Guevara}
\email{noaguevg@oregonstate.edu}
\affiliation{\institution{Oregon State University}}

\author{Md Montaser Hamid}
\email{hamidmd@oregonstate.edu}
\affiliation{\institution{Oregon State University}}
\orcid{0000-0002-5701-621X}

\author{Margaret Burnett}
\email{burnett@eecs.oregonstate.edu}
\affiliation{\institution{Oregon State University}}

% The default list of authors is too long for headers.
\renewcommand{\shortauthors}{Anderson et al.}
\renewcommand{\shorttitle}{AI Inclusivity Bugs}

\begin{abstract} % **MMB on 10/4/25 says D3** :-)

%1. what's the problem (motivations)
%2. why is the problem a problem (motivations)
%3. what did we do about it (objectives & methods)
%4. what does this do for the world OR what are some punchy results (results &/or discussion)
%\boldify{%1. Here is a design-time question: Evaluating AI is hard because it doesn't respond the same all the time}

% 1. are there AI inclusivity bugs 
While much research has shown the presence of AI's ``under-the-hood'' biases (e.g., algorithmic, training data, etc.), what about ``over-the-hood'' inclusivity biases: barriers in user-facing AI products that disproportionately exclude users with certain problem-solving approaches?
%2. Recent research says yes, but what do they look like, how common, and what can devls do about them?
Recent research has begun to report the existence of such biases---but what do they look like, how prevalent are they, and how can developers find and fix them?
% 3. To find out...
To find out, we conducted a field study with 3 AI product teams, to investigate what kinds of AI inclusivity bugs exist \textit{uniquely} in user-facing AI products, and whether/how AI product teams might harness an existing (non-AI-oriented) inclusive design method to find and fix them.
% 4. The results were...
The teams' work revealed 83 instances of 6 AI inclusivity bug types unique to user-facing AI products, their fixes covering 47 bug instances, and a new GenderMag inclusive design method variant, GenderMag-for-AI, that is especially effective at detecting certain kinds of AI inclusivity bugs. 

\end{abstract}

%FIXME
%%
%% The code below is generated by the tool at http://dl.acm.org/ccs.cfm.
%% Please copy and paste the code instead of the example below.
%%
\begin{CCSXML}
<ccs2012>
 <concept>
  <concept_id>10010520.10010553.10010562</concept_id>
  <concept_desc>Computer systems organization~Embedded systems</concept_desc>
  <concept_significance>500</concept_significance>
 </concept>
 <concept>
  <concept_id>10010520.10010575.10010755</concept_id>
  <concept_desc>Computer systems organization~Redundancy</concept_desc>
  <concept_significance>300</concept_significance>
 </concept>
 <concept>
  <concept_id>10010520.10010553.10010554</concept_id>
  <concept_desc>Computer systems organization~Robotics</concept_desc>
  <concept_significance>100</concept_significance>
 </concept>
 <concept>
  <concept_id>10003033.10003083.10003095</concept_id>
  <concept_desc>Networks~Network reliability</concept_desc>
  <concept_significance>100</concept_significance>
 </concept>
</ccs2012>
\end{CCSXML}

\ccsdesc[500]{Human-centered computing~User studies}
\ccsdesc[300]{Computing methodologies~Intelligent agents}

\keywords{Intelligent User Interfaces, Human-Computer Interaction
}

%\input{0-TODO-list}

%\tableofcontents
%\clearpage

\maketitle
% \linenumbers

\input{doc/01-Introduction}

\input{doc/02-background}
\input{doc/03-Related-Works}
\input{doc/04-Methodology}

\input{doc/05-Results-Find-AI-Inclusivity-Bugs}
\input{doc/06-Results-GenderMag-for-AI}

\input{doc/07-Discussion}
\input{doc/08-Limitations}

\input{doc/09-Conclusion}

\bibliographystyle{ACM-Reference-Format}
\bibliography{000-paper.bib}

\appendix

\end{document}

%% file: z-commands.tex
%%%%%%%% TEXT STYLING COMMANDS %%%%%%%%

\newcommand{\briefCite}[1]{~\citeauthor{#1} (\citeyear{#1})}

\newcommand{\mysection}[1]{\section{#1}}
\newcommand{\mysubsection}[1]{\subsection{#1}}
\newcommand{\mysubsubsection}[1]{\subsubsection{#1}}

\newcommand{\AAA}[1]{\textcolor{blue}{[AAA: #1]}}
\newcommand{\JED}[1]{\textcolor{orange}{[JED: #1]}}
\newcommand{\MMB}[1]{\textcolor{purple}{[MMB: #1]}}
\newcommand{\JN}[1]{\textcolor{teal}{[JN: #1]}}

\newcommand{\colorboxBackgroundForegroundText}[3]%
{\protect\adjustbox{padding=1pt 1pt, bgcolor=#1}%
{\color{#2}#3}}%

\newcommand{\coloredText}[3]{%
    \colorboxBackgroundForegroundText{#1}{#2}{#3}%
}%

%\newcommand{\redact}[1]{X University}

% list of predefined colors: black darkgray gray lightgray white
%	blue, brown, cyan, green, lime, magenta, olive, orange, pink, purple, red, teal, violet, yellow.

%%%%%%%%%%%% Counters for the commands will need
\newcounter{boldifyCounter}
\newcounter{fixmeSectionCounter}
\newcounter{fixmeTotalCounter}
\newcounter{insightCounter}
\newcounter{insightSubCounter}
\newcounter{resultCounter}
\newcounter{adaptWinCounter}

\makeatletter
\@addtoreset{fixmeSectionCounter}{section}
\@addtoreset{fixmeSectionCounter}{subsection}
\@addtoreset{boldifyCounter}{section}
\@addtoreset{boldifyCounter}{subsection}

\makeatother

%%%%%%%%%%%%% Drafting commands %%%%%%%%%%%%%%%%%%%%%%
% FAM: below is for TeXCount to not count boldification words
%TC:macro \boldify [ignore]

\newcommand{\boldify}[1]{}
\ifdefined\boldifyON
	\renewcommand{\boldify}[1]{\par\noindent%
		%\stepcounter{boldifyCounter}%
		\textbf{
        {**}%
	%	~\arabic{section}.\arabic{boldifyCounter}%
		 #1** 
      }%
	}
\fi

\newcommand{\boldifyConcepts}[2]{}
\ifdefined\boldifyON
	\renewcommand{\boldifyConcepts}[2]{\par\noindent%
		%\stepcounter{boldifyCounter}%
		\textbf{{**}%
	%	~\arabic{section}.\arabic{boldifyCounter}%
		 #1**}%
        {
        \tiny
        \begin{itemize}
            #2
        \end{itemize}
        }
	}
\fi

\newcommand{\fakeResult}[1]{}
\ifdefined\fakeResultON
    \renewcommand{\fakeResult}[1]{\par\noindent%
        \stepcounter{boldifyCounter}%
        \textbf{{\color{blue}**FAKE FAKE FAKE**}%
        ~\arabic{section}.\arabic{subsection}.\arabic{boldifyCounter}%
        : #1\\}
    }
\fi

\newcommand{\FIXED}[1]{{\color{blue}#1}}
\newcommand{\FIXME}[2]{}
\ifdefined\fixmeON
	\renewcommand{\FIXME}[2]{\par\noindent%
		\stepcounter{fixmeSectionCounter}\stepcounter{fixmeTotalCounter}%
		{\color{red}\fbox{\color{black}%
			\parbox{.97\linewidth}{%
                \begin{center}
				\textbf{FIXME \arabic{section}.\arabic{subsection}.%
        		\arabic{fixmeSectionCounter} ({\color{red}%
        		\#\arabic{fixmeTotalCounter}})}
                
                \end{center}

                \textbf{#1: }#2
        		}
        		}%
        }
	}
\fi

\newcommand{\Result}[2]{\par\noindent%
    \stepcounter{resultCounter}
    \color{black}\fbox{\color{black}%
		\parbox{.97\linewidth}{%
             \textbf{Result \#\arabic{resultCounter}:}
             \textit{#1}\\#2
        }
    }
}

\newcommand{\adaptWin}[2]{\par\noindent%
    \stepcounter{adaptWinCounter}
    \color{black}\fbox{\color{black}%
		\parbox{.97\linewidth}{%
             \textbf{Adaptation Win \#\arabic{adaptWinCounter}}
             (\textit{#1})\\#2
        }
    }
}

% ---- version with (rounded) boxes
%\newcommand{\INSIGHT}[1]{ \par\noindent%
%    \stepcounter{insightCounter}
%    {\color{black}
%        \begin{tcolorbox}[colback = white, before = \par\noindent]
%        \textbf{\textit{Insight \#\arabic{insightCounter}}: } #1
%        \end{tcolorbox}{\color{black}%
%       }%     
%    }
%}

\newcommand{\note}[2]{}
\ifdefined\noteON
	\renewcommand{\note}[2]{\par\noindent%
		\stepcounter{fixmeSectionCounter}\stepcounter{fixmeTotalCounter}%
		{\color{blue}\fbox{\color{black}%
			\parbox{.97\linewidth}{%
    %             \begin{center}
				% \textbf{FIXME \arabic{section}.\arabic{subsection}.%
    %     		\arabic{fixmeSectionCounter} ({\color{red}%
    %     		\#\arabic{fixmeTotalCounter}})}
                
    %             \end{center}
                {
                \tiny
                \textbf{#1: }#2
        		} 
        		}%
          }%
        }
	}
\fi

\newenvironment{insight}[3]
    %This is for \begin{insight}
    {\par%
    \setlength{\hangindent}{0.025\columnwidth}%
    \setlength{\parindent}{0.025\columnwidth}%
    \textit{Insight-G#1-#2: } #3\\%
}%

%%%%%%%% DRAFT HEADER//FOOTER STYLING COMMANDS %%%%%%%%
\ifdefined\specialFooterON
\usepackage{fancyhdr, datetime}
\pagestyle{fancy}
\fancyhf{}
\pagenumbering{arabic}
\rfoot{\smallbreak{\tiny\today~\currenttime~GMT\quad\quad}Page~\thepage~}
\fi
%%%%%%%% END DRAFT HEADER//FOOTER STYLING COMMANDS %%%%%%%%

% this command is hidden text, intended to be toggled from the main by defining a \draftStatusOn command
\newcommand{\draftStatus}[2]{}% draftStatus takes 2 argument and ignores it when draft statuses are OFF
\ifdefined\draftStatusON
	\renewcommand{\draftStatus}[2]{% ... and it does more when they are on
        \\**#1 says D#2 %
	}%
\fi % end the draft status if. Next line tells the word counter to IGNORE all text in this command (it is hidden!)
%TC:macro \draftStatus 1

\newcommand{\doubleUnderline}[1]{
    \underline{\underline{#1}}
}

\newcommand{\topic}[1]{} %ignore colorization when TOPIC isn't set
\renewcommand {\topic}[1]{%
     {\color{black}{#1}}%
    }
    
\ifdefined\topicON
    \renewcommand{\topic}[1]{%
     {\color{red}{#1}}%
    }
\fi

\newcommand{\fixfixfix}[1]{%
    {\color{red} FIX FIX FIX: #1}
}%

\newcommand{\wrapFigWidth}[1]{%
    6cm%
}%

\newcommand{\textSmallGray}[1]{%
    {\footnotesize\textcolor{gray}{#1}}% 
}%

%% file: z-defined-colors.tex
\definecolor{AbiOrange}{RGB}{251,212,180}
\definecolor{TimBlue}{RGB}{219,229,241}
\definecolor{NegativeAbi}{RGB}{255, 115, 0}
\definecolor{NegativeTim}{RGB}{60, 120, 216}
\definecolor{Eggshell}{RGB}{215,198,186}
\definecolor{WomanOrange}{RGB}{237,125,49}
\definecolor{ManBlue}{RGB}{31,78,121}
\definecolor{AI-Right}{RGB}{180,199,231}
\definecolor{AI-Wrong}{RGB}{255,120,120}
\definecolor{MMH-Original}{RGB}{171,221,241}
\definecolor{MMH-GM}{RGB}{50,205,50}

% MMB: if desired, see this color picker if you want to fiddle to pick a combo that looks nice with our AbiOrange and TimBlue colors: https://www.sessions.edu/color-calculator/
%
\definecolor{InitiallyBlue}{RGB}{33,116,194} %**MMB added this one
\definecolor{DuringInteractionRed}{RGB}{168,62,27}  %**MMB added this one
%\definecolor{WhenWrongYellow}{RGB}{252, 183, 20} **MMB replaced with below
\definecolor{WhenWrongYellow}{RGB}{235,150,60}
%\definecolor{OverTimeGreen}{RGB}{18,125,62} **MMB replaced with below
\definecolor{OverTimeGreen}{RGB}{79,139,69}

\definecolor{gray}{RGB}{166,166,166}

%% file: z-GM-Adapt-Wordings/01-Original-GM-Wording.tex
\newcommand{\questionWordingOriginal}[1]{%
    \begin{tcolorbox}[
            colframe=white,
            title filled = true,
            %colupper = black,
            colback = white,
            boxrule=0.1mm, 
            sharp corners%,
            %title = \textcolor{black}{Original GenderMag Wording},
            %colbacktitle = white
            ]
        {%{\small Original GenderMag Wording}
        %\tcblower
        \footnotesize
        \vspace{-4pt}
        \begin{enumerate}[ label = \textbf{Subgoal: }, leftmargin = 1cm]
            \item \textit{Will [persona] have formed this sub-goal as a step to their overall goal? }
        \end{enumerate}%
        \begin{enumerate}[label = \textbf{Pre-Action: }, leftmargin = 2cm]
            \item \textit{Will [persona] take this action at this step? Yes/no/maybe, and why?}
        \end{enumerate}%
        \begin{enumerate}[label = \textbf{Post-Action: }, leftmargin = 2cm]
            \item \textit{If [persona] does the right thing, will they know that they did the right thing and is making progress toward their goal? Yes/no/maybe, and why?}%
        \end{enumerate}%
        \vspace{-8pt}%
        }
    \end{tcolorbox}%
}%

%% file: z-GM-Adapt-Wordings/02-Post-Action-Fork-Wording.tex
\newcommand{\questionWordingForkPost}[1]{%
    \begin{tcolorbox}[
            colframe=white,
            title filled = true,
            %colupper = black,
            colback = white,
            boxrule=0.1mm, 
            sharp corners%,
            %title = \textcolor{black}{Post-Action Fork GenderMag Wording},
            %colbacktitle = white
            ]
        {
        \vspace{-4px}
        \footnotesize
        % \begin{center}
        %     \underline{Question wording for #1}
        % \end{center}
        \begin{enumerate}[ label = \textcolor{#1}{\textbf{Subgoal: }}, leftmargin = 1cm]
            \item \textcolor{#1}{\textit{Will [persona] have formed this sub-goal as a step to their overall goal? } }
        \end{enumerate}
        \begin{enumerate}[label = \textcolor{#1}{\textbf{Pre-Action: }}, leftmargin = 2cm]
            \item \textcolor{#1}{Will [persona] take this action at this step? Yes/no/maybe, and why?}
        \end{enumerate}
        \begin{enumerate}[ leftmargin = 2cm]
            \item[\coloredText{AI-Right}{black}{\textbf{Post-Action:}}] 
            \noFailureAI{AI-Right}{black}--- 
            \textit{If [persona] \coloredText{white}{black}{believes} that the previous action's  output is correct}, 
            \textcolor{#1}{\textit{will [persona] know that they did the right thing and are making progress toward their goal? Yes/no/maybe, and why?}}
            
            \item[\coloredText{AI-Wrong}{black}{\textbf{Post-Action:}}] 
            \failureAI{AI-Wrong}{black}---
            \textit{If [persona] \coloredText{white}{black}{does not believe} that the previous action's output is correct},
            \textcolor{#1}{\textit{will [persona] know that they did the right thing and are making progress toward their goal? Yes/no/maybe, and why?}}
        \end{enumerate}
        \vspace{-8px}
        }
    \end{tcolorbox}
}

%% file: z-GM-Adapt-Wordings/03-Pre-Action-Fork-Wording.tex
\newcommand{\questionWordingForkPre}[1]{%
    \begin{tcolorbox}[
            colframe=white,
            title filled = true,
            %colupper = black,
            colback = white,
            boxrule=0.1mm, 
            sharp corners%,
            %title = \textcolor{black}{Pre-Action Fork GenderMag Wording},
            %colbacktitle = white
            ]
        \vspace{-4px}
        \footnotesize
        % \begin{center}
        %     \underline{Question wording for #1}
        % \end{center}
        \begin{enumerate}[ label = \textcolor{#1}{\textbf{Subgoal: }}, leftmargin = 1cm]
            \item \textcolor{#1}{\textit{Will [persona] have formed this sub-goal as a step to their overall goal? }}
        \end{enumerate}
        \begin{enumerate}[leftmargin = 2cm]
            \item[\coloredText{AI-Right}{black}{\textbf{Pre-Action}}] 
            \noFailureAI{AI-Right}{black}---
            \textit{If [persona] \coloredText{white}{black}{believes} that the previous action's output is correct},
            \textcolor{#1}{\textit{will [persona] take this action at this step? Yes/no/maybe, and why?}}
            
            \item[\coloredText{AI-Wrong}{black}{\textbf{Pre-Action}}]  
            \failureAI{AI-Wrong}{black}---
            \textit{If [persona] \coloredText{white}{black}{does not believe} that the previous action's output is correct},
            \textcolor{#1}{\textit{will [persona] take this action at this step? Yes/no/maybe, and why?}}
        \end{enumerate}
        \begin{enumerate}[label = \textcolor{#1}{\textbf{Post-Action: }}, leftmargin = 2cm]
            \item  (Picking \textit{one} pre-action) ---\textcolor{#1}{\textit{Will [persona] know that they did the right thing and are making progress toward their goal? Why?}}
        \end{enumerate}
        \vspace{-8px}
    \end{tcolorbox}
}

%% file: z-GM-Adapt-Wordings/04-Unimplemented-Adaptation-Wordings.tex
\newcommand{\questionWordingUnderstandableCheckForkPost}[1]{%
    \begin{tcolorbox}[
            colframe=white,
            title filled = true,
            colback = white,
            boxrule=0.1mm, 
            sharp corners%,
            ]
        \footnotesize
        \vspace{-4px}
        %Begin Subgoal Wording.
        \begin{enumerate}[ label = \textcolor{#1}{\textbf{Subgoal: }}, leftmargin = 0.62cm]
            \item \textcolor{#1}{\textit{Will [persona] have formed this sub-goal as a step to their overall goal? }}
        %End Subgoal Wording
        \end{enumerate}
        %Begin Pre-Action Wording.
        \begin{enumerate}[ label = \textcolor{#1}{\textbf{Pre-Action: }}, leftmargin = 1.5cm]
            \item \textcolor{#1}{\textit{Will [persona] take this action at this step? Yes/no/maybe, and why?}}
        %End Pre-Action Wording
        \end{enumerate}
        %Begin Understandability Wording.
        \begin{enumerate}[ label = \textbf{Understand: }, leftmargin = 1.5cm]
            \item \textit{Does [persona] understand the AI's output at this step? Yes/no, and why?}
        %End Understandability Wording
         \item [] \underline{\textbf{If yes}}
         \begin{enumerate}[ leftmargin = 0.7cm]
           
            \item[\coloredText{AI-Right}{black}{\textbf{Post-Action}}]  
            \noFailureAI{AI-Right}{black}---
            \textit{Assuming that [persona] understands the AI's outputs}, \textit{if [persona] believes that the previous action's  output is correct, 
            \textcolor{#1}{will [persona] know that they did the right thing and are making progress toward their goal? Yes/no/maybe, and why?}}

            \item[\coloredText{AI-Wrong}{black}{\textbf{Post-Action}}] 
            \failureAI{AI-Wrong}{black}---
            \textit{Assuming that [persona] understands the AI's outputs}, \textit{if [persona] does not believe that the previous action's output is correct,
            \textcolor{#1}{will [persona] know that they did the right thing and are making progress toward their goal? Yes/no/maybe, and why?}}
            
        \end{enumerate}
        \item[] \underline{\textbf{If no}}
        \begin{enumerate}[leftmargin = 0.7cm]
            \item[\coloredText{AI-Right}{black}{\textbf{Post-Action}}]  
            \noFailureAI{AI-Right}{black}---
            \textit{Assuming that [persona] \underline{does not} understand the AI's outputs}, \textit{if [persona] believes that the previous action's  output is correct, 
            \textcolor{#1}{will [persona] know that they did the right thing and are making progress toward their goal? Yes/no/maybe, and why?}}

            \item[\coloredText{AI-Wrong}{black}{\textbf{Post-Action}}] 
            \failureAI{AI-Wrong}{black}---
            \textit{Assuming that [persona] \underline{does not} understand the AI's outputs}, \textit{if [persona] does not believe that the previous action's output is correct, 
            \textcolor{#1}{will [persona] know that they did the right thing and are making progress toward their goal? Yes/no/maybe, and why?}}
        \end{enumerate}
    \end{enumerate}

        \vspace{-8px}
    \end{tcolorbox}
}

\newcommand{\questionWordingPreActForkNumFails}[1]{%
    \begin{tcolorbox}[
            colframe=white,
            title filled = true,
            %colupper = black,
            colback = white,
            boxrule=0.1mm, 
            sharp corners%,
            %title = \textcolor{black}{Pre-Action Fork GenderMag Wording},
            %colbacktitle = white
            ]
        \vspace{-4px}
        \footnotesize
        \begin{enumerate}[ label = \textcolor{#1}{\textbf{Subgoal: }}, leftmargin = 0.62cm]

            \item \textcolor{#1}{\textit{Will [persona] have formed this sub-goal as a step to their overall goal? }}

        \end{enumerate}
        %Add in a selection for the number of failures to investigate.
        \begin{enumerate}[label = \textbf{Failures: }, 
                    leftmargin = 1.5cm
                ]
                \item How many ways can the AI product fail?
                \item[ ] F $\leftarrow$ Set of Failures 
        \end{enumerate}
        \begin{enumerate}
                [leftmargin = 1.5cm]

            \item[\coloredText{AI-Right}{#1}{\textbf{Pre-Action}}]  
            \textcolor{#1}{\noFailureAI{AI-Right}{#1}---
            \textit{If [persona] believes that the previous action's output is correct}, \textit{will [persona] take this action at this step? Yes/no/maybe, and why?}}
        \end{enumerate}
        \begin{enumerate}[label = \textbf{For f $\in$ F:}, leftmargin = 1.5cm ]
        \item 
        \end{enumerate}
        \begin{enumerate}[leftmargin =2.2cm]
            \item[\coloredText{AI-Wrong}{black}{\textbf{Pre-Action}}]
            \failureAI{AI-Wrong}{black}---\textit{If [persona] does not believe that the previous action's output is correct because of [failure f]},
            \textcolor{#1}{\textit{will [persona] take this action at this step? Yes/no/maybe, and why?}}
        \end{enumerate}
        \begin{enumerate}[label = \textcolor{#1}{\textbf{Post-Action: }}, leftmargin = 1.5cm]
            \item  \textcolor{#1}{(Picking \textit{one} pre-action) ---\textit{Will [persona] know that they did the right thing and are making progress toward their goal? Why?}}
        \end{enumerate}
        \vspace{-8px}
    \end{tcolorbox}
}

\newcommand{\questionWordingOptionalForkPre}[1]{%
    \begin{tcolorbox}[
            colframe=white,
            title filled = true,
            %colupper = black,
            colback = white,
            boxrule=0.1mm, 
            sharp corners%,
            %title = \textcolor{black}{Pre-Action (Optional) Fork GenderMag Wording},
            %colbacktitle = white
            ]
        \footnotesize
        \vspace{-4px}
        %Begin Subgoal Wording.
        \begin{enumerate}[ label = \textcolor{#1}{\textbf{Subgoal: }}, leftmargin = 1cm]
            \item \textcolor{#1}{\textit{Will [persona] have formed this sub-goal as a step to their overall goal? }}
        %End Subgoal Wording
        \end{enumerate}
        %Begin Optional Fork Wording
        \begin{enumerate}[ label = \textbf{Fork? }, leftmargin = 1.5cm]
            \item \textit{For this action, do you think there could be a meaningful difference between the way your user responds if the AI behaves as intended versus when it does not?}
            
        %End Optional Fork Wording
        \end{enumerate}
            %Begin If/Else clause
            \begin{enumerate}[ label = \textbf{}, leftmargin = 1cm]
                \item \textbf{If ``Yes'':}
                %Begin pre-action fork questions
                \begin{enumerate}[leftmargin = 2.5cm]
                    \item[\coloredText{AI-Right}{black}{\textbf{Pre-Action}}]
                    \noFailureAI{AI-Right}{black}---\textit{If [persona] believes that the previous action's output is correct, 
                    \textcolor{#1}{will [persona] take this action at this step? Yes/no/maybe, and why?}}
                    
                    \item[\coloredText{AI-Wrong}{black}{\textbf{Pre-Action}}]
                    \failureAI{AI-Wrong}{black}---\textit{If [persona] does not believe that the previous action's output is correct, 
                    \textcolor{#1}{will [persona] take this action at this step? Yes/no/maybe, and why?}}
                %End pre-action fork questions
                \end{enumerate}
                \item \hspace{0.5cm}\textbf{Else:}
                %Begin Original Pre-Action Wording
                \begin{enumerate}[label = \textbf{Pre-Action: }, leftmargin = 2.5cm]
                \item \textit{Will [persona] take this action at this step? Yes/no/maybe, and why?}
                %End If/Else clause  
                \end{enumerate}    
    \end{enumerate}
    \begin{enumerate}[label = \textcolor{#1}{\textbf{Post-Action: }}, leftmargin = 2cm]
            \item \textcolor{#1}{\textit{If [persona] does the right thing, will they know that they did the right thing and is making progress toward their goal? Yes/no/maybe, and why?}}
        \end{enumerate}

        \vspace{-8px}
    \end{tcolorbox}
}

%% file: z-shorthand-commands.tex
%-------------------------------------------
%These are the explanataion type commands for Team Game. This should reduce remembering the vocab from the TO-DO list.

\newcommand{\BTW}[1]{%
    Scores Best-to-Worst explanation%
}%

\newcommand{\TT}[1]{%
    Scores Through-Time explanation%
}%

\newcommand{\OTB}[1]{%
    Scores On-the-Board explanation%    
}%

%-------------------------------------------
%These are the bug type commands for Step 2's AI inclusivity bugs, including writing AI inclusivity bug (how many times do I write Ai? AI Inclusivity Bug? AI inclusivity bug?. This should reduce remembering how we format things in the text.

\newcommand{\bugInterpret}[1]{%
    \textit{Interpret AI?}%
}%

\newcommand{\bugWhyShouldI}[1]{%
    \textit{AI: why should I?}%
}%

\newcommand{\bugMoreInfo}[1]{%
    \textit{AI: more info!}%
}%

\newcommand{\bugActionable}[1]{%
    \textit{AI: actionable?}%
}%

\newcommand{\bugInputOutput}[1]{%
    \textit{AI input$\leftrightarrow$output?}%
}%

\newcommand{\bugChanges}[1]{%
    \textit{AI changes?}%
}%

\newcommand{\aiInclBug}[1]{%
    AI inclusivity bug%
}%

\newcommand{\aiInclBugType}[1]{%
    AI inclusivity bug type%
}%

\newcommand{\aiInclBugInst}[1]{%
    AI inclusivity bug instance%
}%

%-------------------------------------------
%These are the bug type commands for Team Game's AI players. This should reduce remembering how we format things in the text.

\newcommand{\xPlayer}[1]{%
    {\color{#1} X-player AI}%
}%

\newcommand{\oPlayer}[1]{%
    {\color{#1} O-player AI}%
}%

%-------------------------------------------
%This is Team Weather's LTE shorthand command. This should reduce remembering how we format things in the text.

\newcommand{\LTE}[1]{%
    $LTE_{#1}$%
}

%-------------------------------------------
%This is for the versions of GenderMag, so we aren't just switching version/adaptation names. This should reduce remembering how we format things in the text.

\newcommand{\OriginalGM}[1]{%
    Original GenderMag%
}%

\newcommand{\postForkGM}[1]{%
    Post-Action Fork GenderMag%
}%

\newcommand{\preForkGM}[1]{%
    Pre-Action Fork GenderMag%
}%

\newcommand{\subgoalForkGM}[1]{%
    ``Subgoal Fork GenderMag walkthrough''%
}%

%-------------------------------------------
%This is to save me typing out the AI failure things, but I'm not really using it. Consider cutting.

\newcommand{\noFailureAI}[2]{%
    \coloredText{#1}{#2}{Believes the AI%
    }%
}%

\newcommand{\failureAI}[2]{%
    \coloredText{#1}{#2}{Doubts the AI%
    }%
}%

%-------------------------------------------
%Got tired of typing out $\rightarrow$ and worrying about what to what.

\newcommand{\XtoY}[2]{%
    #1 $\rightarrow$ #2%
}

%% file: z-quotation-commands.tex
%-----------------------------------------------
%Use this one if you want to talk about how teams found bugs (step 2)

\newcommand{\quotateInsetFindBugs}[6]{%
    \begin{quote}%
        \leftskip-.25in%\parindent% this sets the length of the left margin to the length of an indent
        \rightskip-.25in%\parindent% same deal on the right
        \small
        Team #1: ``%
        \textit{#2}''\\%
        \hspace{1cm}{$\bullet$ Bug Type(s): \textit{#3}}\\%
        \hspace{1cm}{$\bullet$ Problem-Solving Style Value(s): \textit{#4}}%
    \end{quote}%
}%

% FAM: attempt, works but not with multiple teams
% \newcommand{\quotateInsetFindBugs}[6]{%
%     \par\vspace{.2in}% Add some space before the quote
%     \noindent % Remove paragraph indentation
%     \hspace{.25in}% Add initial indent
%     \small% Make text smaller
%     Team #1: ``%
%     \hangindent=.5in% Set hanging indent to 2em (adjust as needed)
%     \hangafter=1% Start hanging indent after first line
%     \textit{#2}''%
    
%     \vspace{0.3em}% Small space before bullet points
%     \noindent\hspace{.25in}$\bullet$ Bug Type(s): \textit{#3} %
    
%     \noindent\hspace{.25in}$\bullet$ Problem-Solving Style Value(s): \textit{#4}
%     \par\vspace{.1in}% Add some space after the quote
% }

%-----------------------------------------------
%Use this one if you want to talk about how teams found bugs using a specific version of GenderMag adaptation (step 2), use ONLY in the adapt section or after.

\def \indentationAmount{0.35cm}

\newcommand{\quotateInsetFindBugsWithVersion}[9]{%
\ifthenelse{\equal{#1}{single}}{%
{\renewenvironment{quote}{%
  \list{}{%
    \leftmargin \indentationAmount
    \rightmargin 0cm
  }%
  \item\relax % Start the list item
}{%
  \endlist % End the list environment
}
\begin{quote}%
\small
\setlength{\hangindent}{5cm}
\hangafter=1
Team #2: ``\textit{#3}''%
\end{quote}%

{$\bullet$ Bug Type(s) Found: \textit{#4}}%

{$\bullet$ Problem-Solving Style Value(s): \textit{#5}}%

{$\bullet$ Walkthrough Version: \textit{#6}}
% \hspace{1cm}{$\bullet$ Which Fork: \textit{#6}}
}
}{%
{\renewenvironment{quote}{%
  \list{}{%
    \leftmargin \indentationAmount % Adjusts left margin
    \rightmargin 0cm % Adjusts right margin
  }%
  \item\relax % Start the list item
}{%
  \endlist % End the list environment
}
\begin{quote}%
    \leftskip-.25in%\parindent% this sets the length of the left margin to the length of an indent
    \rightskip-.25in%\parindent% same deal on the right
    \small
    Team #2: ``\textit{#3}''%
    \end{quote}\\%
    
    \hspace{\indentationAmount}{$\bullet$ Bug Type(s) Found: \textit{#4}}\\%
    
    \hspace{-1cm}{$\bullet$ Problem-Solving Style Value(s): \textit{#5}}\\%
    
    \hspace{\indentationAmount}{$\bullet$ Walkthrough Version: \textit{#6}}
    % \hspace{1cm}{$\bullet$ Which Fork: \textit{#6}}
}
}%
}%

%-----------------------------------------------
%Use this one if you want to talk about how teams fixed bugs (step 2)

\newcommand{\quotateInsetFixBugs}[7]{%
    \begin{quote}%
        \leftskip-.25in%\parindent% this sets the length of the left margin to the length of an indent
        \rightskip-.25in%\parindent% same deal on the right
        \small
        Team #1: ``%
        \textit{#2}''\\%
        \hspace{1cm}{$\bullet$ Bug Type(s) Fixed: \textit{#3}}%
        %\hspace{1cm}{$\bullet$ Problem-Solving Style Value(s): \textit{#4}}%
        %\hspace{1cm}{$\bullet$ Fix(es) Applied: \textit{#5}}%
    \end{quote}%
}%

%-----------------------------------------------
%Use this one if you want to talk about how teams fixed bugs (step 2)

\newcommand{\quotateInsetFixBugsWithVersion}[8]{%
    \begin{quote}%
        \leftskip-.25in%\parindent% this sets the length of the left margin to the length of an indent
        \rightskip-.25in%\parindent% same deal on the right
        \small
        Team #1: ``%
        \textit{#2}''\\%
        \hspace{1cm}{$\bullet$ Bug Type(s) Fixed: \textit{#3}}\\%
        \hspace{1cm}{$\bullet$ Problem-Solving Style Value(s): \textit{#4}}\\%
        \hspace{1cm}{$\bullet$ Problem-Solving Style Value(s): \textit{#5}}%
        %\hspace{1cm}{$\bullet$ Fix(es) Applied: \textit{#5}}%
    \end{quote}%
}%

%-----------------------------------------------
%Use this one if you want to talk about how teams discussed adapting GenderMag (step 2), NOT when they find/fix a bug (see the above 2 commands)

\newcommand{\quotateInsetAdaptGM}[5]{%
    \begin{quote}%
        \leftskip-.25in%\parindent% this sets the length of the left margin to the length of an indent
        \rightskip-.25in%\parindent% same deal on the right
        \small
        Team #1: ``\textit{#2}''\\
        \hspace{1cm}{$\bullet$ Walkthrough Version: \textit{#3}}
    \end{quote}%
}%

%-----------------------------------------------
%Use this one if you want to quote a team but not for any of the purposes above (step 2)

\newcommand{\quotateInsetCommentary}[4]{%
    \begin{quote}%
        \leftskip-.25in%\parindent% this sets the length of the left margin to the length of an indent
        \rightskip-.25in%\parindent% same deal on the right
        \small
        Team #1: ``\textit{#2}''
    \end{quote}%
}%

%% file: z-statistical-commands.tex
\newcommand{\tTestResult}[5]{
    (#1-sided t-test, t(#2) = #3, \textit{p} = #4, \textit{d} = #5)%
}%

\newcommand{\oddsRatio}[3]{%
    (Odds Ratio: #1, CI: [#2, #3])%
}%

%% file: doc/01-Introduction.tex
\section{Introduction
\draftStatus{MMB 10/4/25}{2.5}
}
\label{sec:intro}

\boldify{Look at this first figure, which aimed to serve agriculturalists working on recommending frost-mitigation events.}
\input{figure/TeamWeather-Figures/Intro-Team-Weather-Attn-Grabber}

Suppose an Artificial Intelligence (AI) product team creates a dashboard to help farmers and other agricultural stakeholders apply AI to help find when temperatures sink so low that 50\% of their crop will die (Figure~\ref{fig:Intro-Team-Weather-Attn-Grabber}).
%These 50\% thresholds vary by crop, and each crop's threshold can vary over time, depending on factors like dewpoint, humidity, and air temperature.
%The AI takes historical information for these factors and predicts the 50\% temperature threshold for a crop (yellow line). 
% When that prediction is intersected by the forecasted low temperature (blue line), then the AI shows these risks with a red diamond (two in Figure~\ref{fig:Intro-Team-Weather-Attn-Grabber}). 
%MMB 6/29/25: I think all the above struck-out words are in the way, and not needed at this point in the story.
%
Farmers can use the dashboard to decide whether or not to deploy frost-mitigation techniques to keep their crops above this temperature.
%Would this product be inclusive for diverse agriculturalists, who have diverse demographics~\cite{castillo2023farmlabor} and ways of solving problems?

\boldify{However, farmers are diverse---they might not always want to use the AI. Who would this product work for? Who's left behind?}

However, the AI's user-facing information in Figure~\ref{fig:Intro-Team-Weather-Attn-Grabber} may not be useful to all its intended users.
Farmers and other agricultural stakeholders have diverse backgrounds, economic resources, gender identities, ages, education levels, and agricultural experience levels.
Suppose the farmer does not have much engineering/math background or has limited finances to deploy expensive frost-mitigation techniques~\cite{saxena2023grape}.
This raises the question of whether a wide range of diverse farmers can decide, using this tool, what to actually \textit{do} with their fields.
If not, who would it fail to serve, and how should the tool change to include them?

\boldify{These are inclusive design questions!}
These are human-centered AI questions (HCAI).
The HCI area of inclusive design aims to answer questions like these, so as to create user experiences that are ``usable and understandable by as many people as possible, considering users' diverse needs, backgrounds, and experiences''~\cite{IDF2024inclusivedesign}.
Usability bugs that disproportionately exclude some groups from receiving a product's intended benefits are sometimes termed ``inclusivity bugs''~\cite{miller2021inclusivity}.

\boldify{If these AI products are not inclusive, they may run afoul of \textit{\aiInclBug{}s}, which we define as:}

In the field of AI, some inclusivity bugs are ``under-the-hood,'' such as algorithmic or training data biases.
This paper instead considers ``over-the-hood'' inclusivity bugs in user-facing AI products. 
Recently, research has begun to report the presence of over-the-hood inclusivity bugs in AI products (e.g.,~\cite{anderson2024measuring, hamid-2024-tiis}), but has not yet investigated AI inclusivity bugs that are \textit{unique to user-facing AI products}, which we term ``\textit{\aiInclBug{}s}.''
We distinguish \aiInclBug{}s from other usability bugs by the following defining criteria:
(1)~\aiInclBug{}s exist only in user-facing AI information, making them \textit{AI} usability bugs.
(2)~\aiInclBug{}s disproportionately disadvantage certain groups of AI product users, making them AI \textit{inclusivity} bugs.

%\boldify{Do \aiInclBug{}s exist? What do they look like? Are they common? Can AI product teams find them, and if so, how? Once teams find them, how do they fix them?}
%
%The notion of \aiInclBug{}s raises the following questions:
%Do \aiInclBug{}s exist?
%What do they look like?
%How common are they?
%Can an existing inclusive design methods detect them, or is something new needed?
%Can AI product teams find and/or fix \aiInclBug{}s?
%How?

\boldify{To figure these questions out, we conducted a field study with three Ai product teams, to investigate the following RQs.}

In this paper, we report on a field study with three AI product teams to investigate \aiInclBug{}s: Team Game, Team Weather, and Team Farm.
Team Game was working on explaining an AI-powered game that involved sequential decision making.
Team Weather was working on AI-powered agriculture, whose prototype (Figure~\ref{fig:Intro-Team-Weather-Attn-Grabber}) was predicting whether and when to deploy frost-mitigation approaches.
Finally, Team Farm was working on AI-powered irrigation scheduling.
Through our field study with these AI product teams, we investigated the following research questions:

\begin{enumerate}[label = {\textbf{RQ\arabic*}:}, leftmargin = 1.3cm ]
    \item What \textit{types of user-facing \aiInclBug{}s} do AI product teams find? What do they look like? How common are they?
    \item How do AI product teams \textit{fix} the \aiInclBugInst{}s they find? 
    \item Is an existing inclusive design method (GenderMag in this paper), which was not designed particularly for AI, ``enough'' for AI product teams to be effective at finding user-facing \aiInclBug{}s, or is something AI-specific needed?
    \item If something AI-specific is needed, how does the inclusive design method need to change? How effective are these changes? 
\end{enumerate}

%
%\boldify{Answering these research questions provides the following contributions:}
%\textcolor{red}{MMB 10/4: I commented out the contributions because they give away the answers and I can't figure out how to make them strong without doing so.  So I moved them to conclusion instead.}
%
%In answering these research questions, this paper makes the following contributions:
%
%\begin{itemize}
%    \item Reveals six \aiInclBugType{}s, their frequency, and what they look like. All of these \aiInclBugType{}s arose multiple times in multiple applications, suggesting that all three may be common patterns/pitfalls to AI product inclusivity to diverse users.   
%    \item Shows 3 AI product teams' results of finding and fixing \aiInclBug{}s in their own AI products.
%    \item Introduces GenderMag-for-AI, which addresses a missing link of inclusive design methods for AI products---evaluating how to support diverse users when they do \text{not} believe what the AI product has just claimed/decided.
%\end{itemize}

%% file: figure/TeamWeather-Figures/Intro-Team-Weather-Attn-Grabber.tex
% \begin{figure}[h]
\begin{wrapfigure}{R}{4.5cm}
    \centering
    \includegraphics[width=0.6\linewidth]{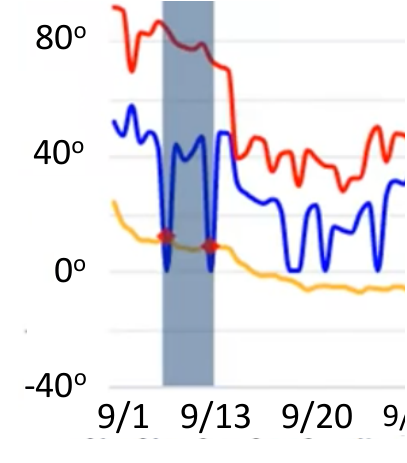}
    \caption{
    An AI-powered agricultural tool for predicting when crops are at risk of dying from cold temperatures (y-axis) over time (x-axis).
    The risky events are where the AI-predictions for this crop (yellow line) intersect with the forecast's low temperature (blue line).
    }
%    (red diamonds). 
    \label{fig:Intro-Team-Weather-Attn-Grabber}
\end{wrapfigure}
% \end{figure}

%% file: doc/02-background.tex
\section{Background: The Gender Inclusiveness Magnifier (GenderMag) Method
\draftStatus{MMB}{2.4: GOOD ENOUGH}}
\label{sec:Background-GM}

%\subsection{GenderMag: The Problem-Solving Style Types \& Values
%}
\label{subsec:GM-Styles-Values}

\boldify{The inclusive design method the teams used was GM, and here's what that is.}

The  inclusive design method the AI product teams used was GenderMag (Gender Inclusiveness Magnifier)~\cite{burnett2016gendermag}.
The GenderMag method is an inclusive design and evaluation method to assist evaluation teams in improving technology products' inclusivity to diverse users.
Multiple empirical studies have shown GenderMag's efficacy at identifying inclusivity bugs and pointing toward fixes~\cite{burnett2016gendermag, guizani2022debug, murphy2024gendermag, padala2020gender, shekhar2018cognitive, vorvoreanu2019gender}.

At GenderMag's core are five \textit{problem-solving styles}, shown in Table~\ref{tab:Background-Persona-Table}.
Each style has a range of problem-solving style \textit{values}, capturing diverse problem-solving approaches.
These five problem-solving style types have repeatedly been shown in research to have strong ties to both problem-solving and gender~\cite{anderson2024measuring,burnett2016gendermag, stumpf2020gender, vorvoreanu2019gender}.

\input{tables/Background-Persona-Table}

\boldify{Each problem-solving style type is a "type" with a range of values \& endpoints. GM has personas to bring some of the values to life}

Table~\ref{tab:Background-Persona-Table} shows these five ranges of values, with distinguished endpoints (columns~2 \& 4).
The set of values in each column are grouped into one of three personas.
%---fictional representations of a group of people who have something in common~\cite{adlin2010essential} (in our case, the same problem-solving style values).
The five problem-solving style values on the left are assigned to the ``Abi (Abigail/Abishek)'' persona. 
The five values on the right are assigned to ``Tim (Timara/Timothy).''
A mix of values are assigned to ``Pat (Patricia/Patrick).''

The principle behind GenderMag is that, when technology \textit{simultaneously} supports the ``endpoint'' personas Abi and Tim, every problem-solving style value within the ranges those endpoints define is also supported.
If those who apply GenderMag identify that something in the technology does not support one of these endpoints, GenderMag defines that occurrence as an \textit{inclusivity bug}~\cite{guizani2022debug}.
These inclusivity bugs are \textit{problem-solving} inclusivity bugs, since they disproportionately impact people with that problem-solving style value. 
The inclusivity bugs identified by the five problem-solving style types in Table~\ref{tab:Background-Persona-Table} are also gender inclusivity bugs.
They are gender inclusivity bugs because these five problem-solving style types capture (statistical) gender differences in how people problem-solve~\cite{anderson2024measuring,burnett2016gendermag,guizani2022debug,stumpf2020gender,vorvoreanu2019gender} (e.g., Figure~\ref{fig:Background-genderFacetGraph}).
%Stumpf et al.~\cite{stumpf2020gender} discussed the  literature behind these five problem-solving style types, which we briefly summarize as:

\input{figure/Background-genderFacetGraph}

\label{subsec:GM-Method-Overview}

\boldify{People use these personas in a specialized cognitive walkthrough, and here's an overview of that walkthrough.}

Software professionals use these personas in a specialized cognitive walkthrough.
%Like other cognitive walkthroughs, GenderMag asks questions at each step, but it also adds the GenderMag personas and focuses each question on the five GenderMag problem-solving style types~\cite{burnett2016gendermag}.
%Teams first select one of the personas and evaluate their product through that persona's perspectives.
%Teams can customize the persona in limited ways, like changing their demographics or matching their backgrounds to suit the evaluated product; 
%however, the persona's problem-solving style values are fixed.
%
As with other cognitive walkthroughs~\cite{mahatody2010CW}, teams apply the GenderMag walkthrough by first choosing a use-case/scenario as an overall goal, and then answering questions about subgoals and actions that the team think the persona ``should take'' to accomplish the overall goal (Figure~\ref{fig:Original_GenderMag_Flowchart}).
\input{figure/Walkthrough-Adaptation-Figures/Original_GenderMag_Flowchart}

%% file: tables/Background-Persona-Table.tex
\begin{table}[h]
\centering

\includegraphics[width = \columnwidth]{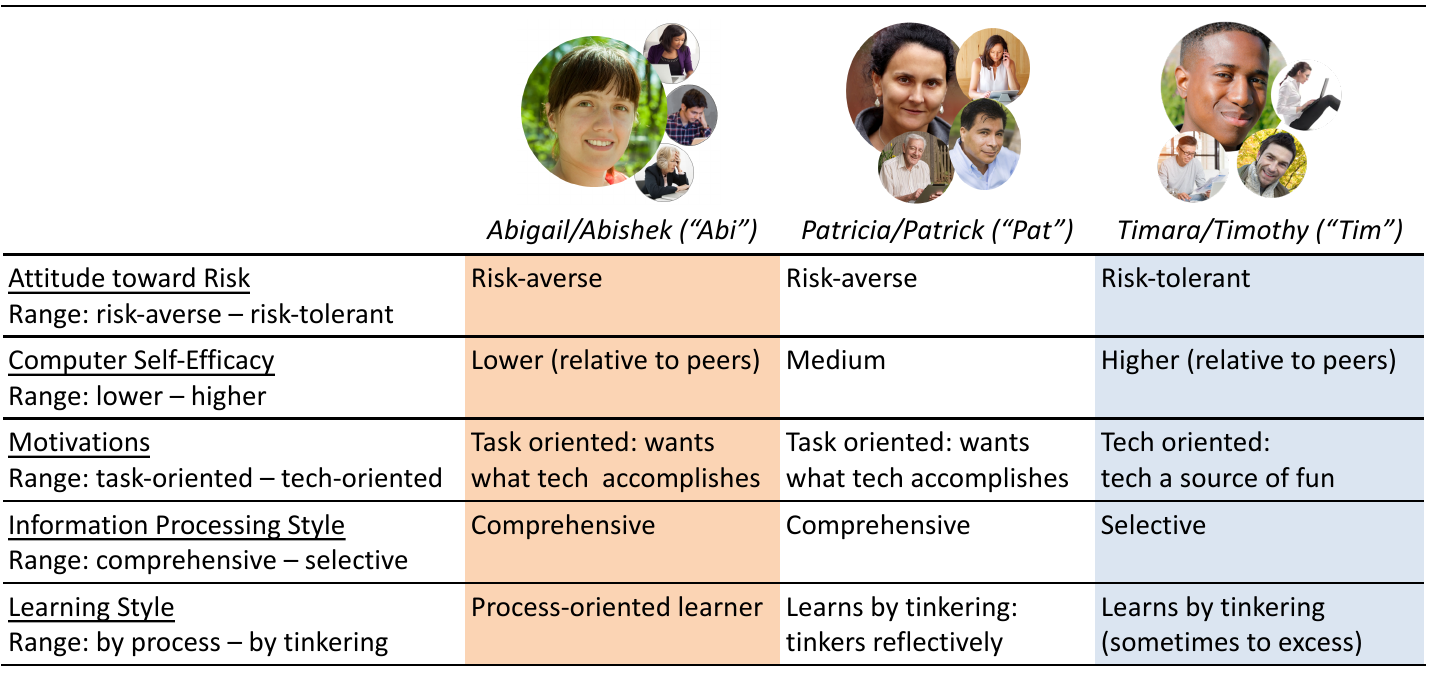}

%alt={there are 5 rows.
%        Row 1 says "Attitude...".  The values are "Risk-averse" (for the Abi persona), ..."
%        Row says ../.}, 

% \end{tabular}
    \caption{The five GenderMag problem-solving style types (rows), and the range of values for each GenderMag persona (columns).
%    Abi represents the \textit{underserved} population, those more likely to be inadvertently left out when using technology, whereas Tim represents the \textit{mainstream} population~\cite{adlin2010essential}. %MMB: Unsubstantiated claim. Also, Adlin cite not correct for this.
    These problem-solving styles have empirically statistically clustered by people's genders (e.g., \cite{anderson2024measuring,burnett2016gendermag,guizani2022debug,stumpf2020gender,vorvoreanu2019gender}).}
    \label{tab:Background-Persona-Table}

\end{table}

%% file: figure/Background-genderFacetGraph.tex
\begin{figure}[h!]
    \centering
    \includegraphics[width = 0.4\linewidth]{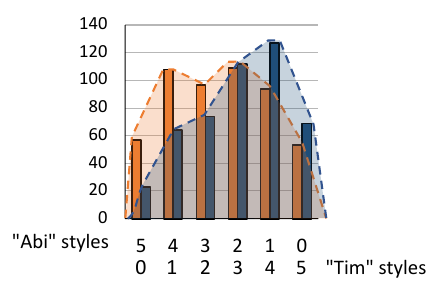}
    \caption{Number of "Abi" problem-solving style values reported by the women (left bar in each pair, bright orange) and men (right bars, dark blue) in Anderson et al.~\cite{anderson2024measuring}. 
    X-axis: number of Abi problem-solving style values (Table~\ref{tab:Background-Persona-Table}). 
    Y-axis: number of participants having this number of Abi problem-solving style values. 
    Statistically, men skewed significantly more toward "Tim" problem-solving style values than women did.}
    \label{fig:Background-genderFacetGraph}
\end{figure}

%% file: figure/Walkthrough-Adaptation-Figures/Original_GenderMag_Flowchart.tex
\begin{figure}[h]
    \centering
    \begin{tabular}{c}
    % \toprule
        \includegraphics[width=0.99\linewidth]{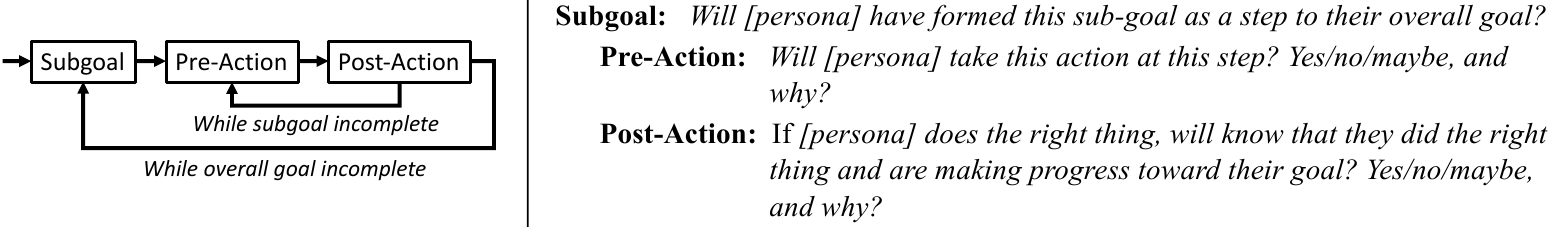}  \\
        % \midrule
        % \questionWordingOriginal{} \\
    % \bottomrule
    \end{tabular}
    \caption{
    \textbf{(Left):} The GenderMag walkthrough graph. Each node represents a step in the walkthrough and is repeated in a loop until the overall goal is complete.
    \textbf{(Right):} The GenderMag walkthrough questions at each step.
    }
    \label{fig:Original_GenderMag_Flowchart}
    % \textcolor{red}{MMB: don't forget to fix this figure like the others (and caption.)}
\end{figure}

%% file: doc/03-Related-Works.tex
\section{Related Works
\draftStatus{MMB}{2.4+ on 10/2/25 (good enough)}
}
\label{subsec:related-works}

\boldify{Some works in inclusivity focus on algorithmic inclusivity, and here's a host of citations for them. However, our paper does NOT FOCUS ON ALGORITHMIC INCLUSIVITY}

In human-AI interaction, inclusivity and related concepts like fairness can be categorized into two broad focuses: 1) an ``under-the-hood'' algorithmic or data focus (i.e., detecting/fixing \textit{algorithms or training data} when they cause harm to some groups over others) and 2) an ``over-the-hood''  focus on \textit{diverse users' experiences with user-facing} aspects of  AI products).
There has been a host of literature for the first (e.g.,~\cite{bird2020fairlearn,chi2021DEIinAIethics-2021, katzman2023taxonomizing,propublica-compas,harrison2020empirical,green2019disparate,buolamwini2018gender,yang2020towards}), but this paper focuses on the second category.

\boldify{There are some guidelines and frameworks, which are analytical, and a few evaluations of those guidelines/frameworks (~\cite{li-MSR-work,anderson2024measuring}), but how to apply these analytical processes is left up to the practitioner. Our work is \textit{not} guidelines but rather a systematic process.}

One direction researchers have taken in this category is taking an analytical approach to assess and improve users' interactions with user-facing AI products.
Guidelines are one example of an analytical approach, which provide general advice on how to improve human-AI interaction~\cite{apple2019guidelines,amershi2019guidelines}.
%Wright et al.~\cite{wright2020comparative} collated a set of guidelines, current as of 2020
Another analytical approach to evaluate human-AI interaction has been to establish frameworks, supporting structures for thinking about and doing human–AI interaction~\cite{long2021approaches}.
One instance of these frameworks is Shneiderman's human-centered AI (HCAI) framework~\cite{shneiderman2020human}.
Their framework's two dimensions assessed 1) how much automation a system has (low $\leftrightarrow$ high) and 2) who/what is in control of the system (computer $\leftrightarrow$ human).
It sought to highlight when designers might encounter negative outcomes for AI-powered systems, like excessive automation/human control. 
There are myriad examples of both guidelines and frameworks for human-AI interaction (e.g.,~\cite{rezwana2023designing,sundar2020rise,wickramasinghe2020trustworthy,xu2023hcai}), but these approaches leave teams to decide how to go about applying them to their AI products.
Our approach is neither a set of guidelines nor a framework, but rather a systematic process for teams to apply to their AI products to improve diversity and inclusion.

\boldify{One part of GenderMag is the use of personas, but most of the AI literature focuses on generating personas using AI, rather than evaluating AI products with personas}

As with our work, some analytical work uses personas to help evaluate tech products.
%The three personas bring GenderMag's five problem-solving style types to life (Section~\ref{subsec:GM-Styles-Values}).
An abundance of persona work involving AI has focused on how AI can help generate personas (e.g.,~\cite{holzinger2022personas,chung2024revolutionizing,shin2024understanding,tan2022generating}), rather than evaluating AI products using personas.
One notable exception is Joshua Puglisi's thesis~\cite{puglisi2023developing}, which constructed four personas to help generate design requirements, evaluated an AI sentiment analysis tool with these personas, and validated the personas' findings with a user study.
Their personas incorporated dimensions of personality (i.e., the Myers-Briggs four personality dimensions~\cite{myers1962myers}), their goals/frustrations, need for cognition, tech saviness, and how social they were.
(The ``need for cognition'' is the dimension most closely aligned with the information processing style problem-solving style presented in our work.)
Our work differs from Puglisi's by focusing on AI inclusivity bugs' disproportionate impact on certain diverse problem-solvers, instead of the impact of personality traits.

\boldify{There's GenderMag, which our work is based on and covered in Section~\ref{subsec:background-GM}. Mentioned there, it uses personas and on CWs (both analytical), and each of these alone have been used to improve/evaluate AI systems (refs). None of these had a focus on diverse people or has been modified/tailored to AI situations}

Other researchers have used analytical evaluation methods such as cognitive walkthroughs while studying AI.
One approach has been to incorporate AI \textit{into} a cognitive walkthrough, such as Bisante et al.~\cite{bisante2024enhancing}, who embedded Open AI's Generative Pre-trained Transformer (GPT) into their tool for cognitive walkthrough evaluations, named CWGPT.
They found that CWGPT independently identified task sequences in the walkthrough, and expert evaluators agreed with 116/128 issues that CWGPT found (93.6\%).

In contrast, in our work \textit{human} teams evaluate AI products, and several other AI researchers have also reported human activities with cognitive walkthroughs.
For example, De Santana et al.~\cite{de2023retrospective} used a cognitive walkthrough as a debriefing tool when users interacted with AI, identifying issues with recency bias, confirmation bias, and trust in the system.
Yildirim et al.~\cite{yildirim2023conversation} reported on 21 student teams who applied cognitive walkthroughs to a conversational agent.
These teams found a wide range of usability problems, identifying how to improve conversational interactions with the AI.
In contrast, our field study reports on \textit{professional AI development teams} using the full GenderMag method, which integrates personas and a specialized cognitive walkthrough to \textit{find and fix AI inclusivity bugs}.

Finally, a few researchers have focused on users' experiences with user-facing AI through a diversity and inclusion lens. 
Some works have uncovered gender differences in perceived fairness of an AI product~\cite{van2021effect}, probability of adopting an AI product~\cite{ofosu2023gender,nouraldeen2023impact}, likability of an AI product~\cite{derrick2014affective}, and awareness of how an AI product operated~\cite{joseph2024artificial}.
Some investigations have considered the five GenderMag problem-solving style types described earlier in Section~\ref{sec:Background-GM}.
For example, Kulesza et al.~\cite{kulesza2012tell} measured the change in their participants' computer self-efficacy, given a ``why''-oriented explanation approach. 
Jiang et al.~\cite{jiang2000persuasive} found that users with higher self-confidence were less likely to accept an AI's proposed solution.
Similar empirical findings have also been found for attitudes toward risk~\cite{schmidt2020calibrating,cohen2020sensitivity}, information processing style and learning style~\cite{nam2023ide,zhang2021information}, and motivations~\cite{shao2021hello,skjuve2024people,li2020understanding}.
Vorvoreanu et al. applied GenderMag to an academic search engine, revealing 10 inclusivity bugs, and fixing six of them~\cite{vorvoreanu2019gender}.
Hamid et al. conducted an empirical study of users interacting with a before-GenderMag vs. after-GenderMag AI-powered game, to determine which was more inclusive to problem-solving users like the Abi and Tim personas~\cite{hamid-2024-tiis}.
They found that the after-GenderMag product was more inclusive, both by persona and by gender. 
Anderson et al.~\cite{anderson2024measuring} considered all five of the problem-solving style types discussed in this paper, finding inclusivity and equity differences for all five types. 

However, none of these works investigates \textit{how} the generators of these user-facing products, \textit{professional AI teams} evaluating their \textit{own} AI products, go about using such a method to find and fix their AI products' user-facing \textit{AI} inclusivity bugs (i.e., inclusivity bugs communicated from the AI to the user).
%Our work also differs by considering whether and how the GenderMag walkthrough should change to address the nuances of AI products.
That is the gap this paper aims to fill.

%------------- Leftovers
%\FIXME{MMB@ALL}{This last sentence is the crucial one---work hard to find out if it's true and to find nearest neighbours.}
%
%\boldify{GenderMag has likewise been successful (refs) and \textit{does} focus on diverse people, and in a few instances used in products that have AI aspects to them with good results (~\cite{vorvoreanu2019gender,murphy2024gendermag}, but nobody has experimented with the method to tailor it to AI. That's the gap this paper fills.}
%
%Although most investigations using GenderMag~\cite{hilderbrand2020engineering,hill2016gendermag,hill2016trials,hill2017gender,murphy2024gendermag} have not been applied to user-facing AI products, a few have begun to do so.
%The three most closely related to our work are Anderson et al.~\cite{anderson2024measuring}, Hamid et al.~\cite{hamid-2024-tiis}, and Vorvoreanu et al~\cite{vorvoreanu2019gender}.
%Anderson et al.'s work was a measurement project---they did not use GenderMag per se, but rather used the GenderMag facets survey to statistically measure how many users in each problem-solving style disliked particular features.
%Hamid et al.'s work was an empirical study of users interacting with a before-GenderMag vs. after-GenderMag AI-powered game, to determine which was more inclusive to problem-solver users like the Abi and Tim personas.
%They found that the after-GenderMag product was more inclusive, both by persona and by gender. 
%to find or fix inclusivity bugs, 

%% file: doc/04-Methodology.tex
\section{Methodology 
\draftStatus{MMB 10/4/25}{top is 2.4: GOOD ENOUGH}}
\label{sec:methodology}

% \FIXME{MMB@anyone}{We can't just drop them into the teams, because readers don't know what we did with them. So need about a parag overviewing the procedures, eg, something like: \\
% "We asked several AI teams whether they were interested in exploring x**, and 3 teams stepped forward. We brought the GenderMag inclusive design and evaluation method to their projects, and observed their use of it, from finding to fixing to outcomes. We begin the specific details of our methodology with each of the three teams' particular context."}

To answer our research questions, we invited several AI product teams to evaluate their AI products' inclusivity, and 3 teams stepped forward:
Team Game, Team Weather, and Team Farm decided to participate.

%-----
\subsection{Team Game
\draftStatus{MMB 10/4/25}{2.4--GOOD ENOUGH, page budget to .75}}
\label{subsec:Team-Game-Participants-Domain}

\boldify{Here's who Team Game were...}

Team Game was a team of computer scientists at a large US university, working on an eXplainable Artificial Intelligence (XAI) approach for sequential decision-making domains.
%A faculty member led the team and the other members were graduate students. % (2 M.S. and 2 PhD).
Four team members identified as men and one as a woman.
Team Game wanted to investigate inclusivity improvements to their explanation design.
%, to enable more diverse users to understand their AI's decision making. 

\boldify{Team Game used Dodge et al.'s explanation interface for MNK, which is...
}

Team Game's domain was sequential decision making for M-N-K games.
M-N-K games are played on $M \times N$ sized boards, where players either 1) win by constructing sequences of length $K$ or 2) draw when no more empty squares remain on the board.
Tic-Tac-Toe is a well-known 3-3-3 instance of M-N-K games. %, where players place either an $X$ or $O$ on the board.
Team Game investigated 9-4-4 games (board size: $9 \times 4$, win sequence length: $4$).

\boldify{The game's input and internal model were...}
\boldify{Team Game had started working with source code from Dodge et al.~\cite{dodge2022people}, but they created a high-fidelity prototype of its front end}

In Team Game's prototype, which was based on Dodge et al.'s source code~\cite{dodge2022people}, each game player was an AI agent with a convolutional neural network trained to play 9-4-4 games (Figure~\ref{fig:Methodology-Team-Game-Interface}).
The gameboard %(Figure~\ref{fig:Team-Game-Gameboard-Callout}) 
(Figure~\ref{fig:Methodology-Team-Game-Interface}, left), was how users saw the games progress.
For each move, the \xPlayer{blue} or \oPlayer{red} placed one of their game pieces in one of the 36 squares.
Each move was labeled with a move number in the top-right of each occupied gameboard square, to help users remember/see which moves had come before which other moves.
%, indexed by $A1-I4$.
When one of the players won, the winning sequence of four squares were highlighted in the winning player's color (the \xPlayer{black}'s blue cells $D2-G2$ in the gameboard).

\input{figure/TeamGame-Figures/Methodology-Team-Game-Interface}

The \xPlayer{blue} decided on its next move by choosing the move with the best score, which it calculated as follows. 
For each move the \xPlayer{black} made, the AI took the current game information as input and calculated three probabilities for each square: the probability it would eventually win the game if it took that square (P(Win)), that it would eventually lose if it took that square (P(Loss)), and that the game would end in a draw (P(Draw)).
The \xPlayer{black}'s ``score'' for each gameboard square was: $Score = P(Win) - P(Loss)$.

\boldify{Team Game had three explanations to explain the \xPlayer{black}'s reasoning to users.}

Team Game had three explanations to explain the \xPlayer{black}'s reasoning to users, updating after each of its moves.
The first was \BTW{} (Figure~\ref{fig:Methodology-Team-Game-Interface}, top right), which shows how the \xPlayer{black} sorted all 36 of its score evaluations in a monotonically decreasing line from the best (left) to worst (right).
The second explanation was \TT{} (Figure~\ref{fig:Methodology-Team-Game-Interface}, middle right), where the \xPlayer{black} provided distributions of all 36 scores through time (moves); which was designed to show users how “confident” the AI was through time.
In the last one was \OTB{} (Figure~\ref{fig:Methodology-Team-Game-Interface}, bottom right), which displayed miniaturized Scores Through-Time explanations for all 36 squares. 
This provided simultaneous temporal and spatial score information.

\subsection{Team Weather
\draftStatus{JN}{2.75}
}
\label{subsec:Team-Weather-Participants-Domain}

\boldify{Here's who Team Weather's participants were, along with their demographic information.}

Team Weather was a team of professional developers and academic researchers from a different large US university than Team Game.
They had six team members, three of whom had a computer science background, and three with natural sciences background (environmental science, atmospheric science, and meteorology).
Four members identified as men, and the remaining two identified as women.
Team Weather's application was already in use across the US state of <anonymized>, and the team wanted to improve it for diverse end users.

% The team lead, who identified as a man, was the systems analyst with a computer science background.
% Of other three remaining men, one had a pure computer science background, one had a hybrid background of agriculture and computer science, and the last was a field meteorologist.
% Of the two women on Team Weather, one was a system technician with a computer science background, and the other was an intern with an environmental science background.

\boldify{Here's their domain of interest (not too detailed)}

Team Weather's AI product's recommendations aimed to help agriculturalists protect their wine grapes from grape injuries sustained from exposure to cold temperatures.
This ability of grapevines to survive cold temperatures is known as their cold hardiness, using their Low-Temperature Exotherm (LTE).
Agriculturalists use these estimated temperature thresholds, which measure the point where they would lose 10\% (\LTE{10}), 50\% (\LTE{50}), and 90\% (\LTE{90}) of their crop yield.
To avoid this, agriculturalists deploy expensive, preemptive frost mitigation methods when the weather forecast estimates a drop below one of these thresholds.

\boldify{...as well as the model that they wanted to use to solve this problem...}

Team Weather's AI was a Recurrent Neural Network (RNN),
%, and full details of its implementation can be found in Saxena et al.~\cite{saxena2023grape}.
The RNN's input was a table of weather information for field locations over time.
Its rows contained air temperature, LTE values, humidity, and dewpoint information, all used to predict when a cold hardiness event might occur.
The model's learning goal was predicting a sequence of LTE estimates for multiple wine grape varieties on a given day.
That LTE would be compared to the low-temperature forecast for that day, helping agriculturalists decide whether and when to deploy frost mitigation methods.

\boldify{...and their interface, such as where a trained AI may present its outputs.}

Figure~\ref{fig:Methodology-Team-Weather-Interface} shows the AI's information in Team Weather's interface.
The AI predicted \LTE{10} values (yellow line), and these values were compared against the forecasted low-temperature (blue line) over time (x-axis).
The AI's recommendations for when agriculturalists should deploy frost-mitigation techniques occurred at the intersection of these two lines, marked by red diamonds.
%to alert the user about potential damage to their crops (described in Saxena et al.~\cite{saxena2023grape}). 

\input{figure/TeamWeather-Figures/Methodology-Team-Weather-Interface}

%-----------------------------------------
\subsection{Team Farm
\draftStatus{JN}{2.75}
}
\label{subsec:Team-Farm-Participants-Domain}

\boldify{Here's who Team Farm were...}

Team Farm was a research team %from the same US university as Team Weather, 
working on AI-powered agricultural applications.
This team had two members, an engineering technician with no computer science background and a data scientist; one identified as a man, the other a woman.
Team Farm wanted to make their AI product more inclusive to agricultural users.

\boldify{...and they were interested in irrigation scheduling, and here's the domain.}

Team Farm's domain was irrigation scheduling for wine grapes.
To support growth of wine grapes, agriculturalists have to give different volumes of water to different grape varieties.
Agriculturalists have monitored grapes through sensor telemetry, which gather information about the grape vines themselves or field block soil.
Agriculturalists develop irrigation schedules from their experiences with historical sensor data on their fields.
However, these may not always accurately reflect the needs of wine grape varieties, which can change across growing seasons.
One example of this comes towards the end of a growing season, where agriculturalists intentionally reduce irrigation frequency (i.e., \textit{deficit} irrigation~\cite{WilliamsDeficitIrrigation}) to increase wine grape quality.

\boldify{Team Farm were developing a dashboard to embed AI-powered irrigation schedules, and they were in the formative stages of gathering sensor data to train an AI.}

Team Farm created their initial dashboard (Figure~\ref{fig:Methodology-Team-Farm-Interface}) in a high-fidelity prototype, since they were just starting to gather sensor data to train an AI.
Team Farm had deployed six sensors across eight fields (Figure~\ref{fig:Methodology-Team-Farm-Interface}'s teardrops).
Three sensors measured soil water content over time (the three lines in the top-right graph).
The other three sensors measured matric potential over time (the three lines in the bottom-right graph).
Their vision for their AI product was to use these two sources of data to predict how much the soil would saturate and how long it would take until the field needed more irrigation.
This information would support agriculturalists' irrigation scheduling by explicitly considering wine grape variants' dynamic needs throughout a growing season.

\input{figure/TeamFarm-Figures/Methodology-Team-Farm-Interface}

\subsection{Procedures
\draftStatus{MMB, 10/4/25}{2.3 -- perhaps good enough?}
}
\label{subsec:03b-Methodology-Procedures-Top}

\boldify{Before we ran any find/fix sessions (detailed below), we introduced each of the three teams to GenderMag, and each of the teams customized their personas.}

We conducted each field study session with one team at a time using the Zoom video conferencing technology.
A team's first session was a pre-evaluation session that introduced the team to GenderMag, the personas and their associated five problem-solving style types, and how GenderMag worked (Section~\ref{sec:Background-GM}).
Each team then picked the persona they wanted to evaluate their AI products with; all three teams picked the ``Abi'' persona (Figure~\ref{fig:Methodology-Blank-Abi}).
Each team then customized Abi's age, location, pronouns, and background/skills using the persona tool on the GenderMag website~\cite{gendermag-website} to make Abi a good fit to their AI product's target audience.
For instance, Team Game made their Abi a 27-year-old mechanical engineer with she/her/hers pronouns who used AI tools to keep organized.
Team Farm made their Abi a 49-year-old farm manager who used he/him/his pronouns.
All three teams' customized personas can be found in the supplemental documents.
% Appendix~\ref{appendix:team_personas}. 

\input{figure/Methodology-Blank-Abi}

\boldify{We met with each team in iterative find-fix sessions, and here's the roles for the find sessions (note: we facilitated a majority of these sessions)}

After a team's initial session, subsequent sessions proceeded as per Figure~\ref{fig:Methodology-Experimental-Overview}.
As the figure shows, teams ran a series of ``find'' sessions, applying a version of GenderMag (at first, this was the ``\OriginalGM{}'') to their AI products.
Each find session had a facilitator to keep the session on track, a driver to navigate the interface, a recorder to take notes, and evaluators to evaluate the system.
In these sessions, 1-2 researchers joined the team as participant-observers, to help with facilitating (making sure each team member weighed in on every evaluation step), recording, and evaluating.
The result of these find sessions was a set of inclusivity bug instances.
These sessions enabled us to answer RQ1 and contributed in part to RQ3.

\input{figure/Methodology-Experimental-Overview}

\boldify{The teams had finds, but how did they fix? They started with triage and then fixed bugs they wanted to. We DID NOT HELP.}

Once each team had a set of inclusivity bug instances, they transitioned to ``fix'' sessions (the right half of Figure~\ref{fig:Methodology-Experimental-Overview}).
Before their first fix session, we gave them a link to the GenderMag design catalog~\cite{GM-design-catalog} as a potential resource.
1-2 researchers joined the fix sessions, but only as observers, without providing any input as to how to fix the bugs.
Thus, the teams alone chose whether and how to fix bug instances, applied their own prioritization processes and design processes, and used whatever technology they wanted to produce the bug fixes.
Team Game directly sketched low-fidelity concepts using Zoom's annotation tool, verbally describing their proposed fixes as well.
Team Farm modified their high-fidelity PowerPoint prototype directly.
Team Weather verbally described their fixes, programming the fixes directly into their AI product.
The teams' fixes enabled us to answer RQ2.

\boldify{Lastly, changes to GenderMag? Here's the procedure for that.}

After a team had used \OriginalGM{} on their AI product for one or two sessions, we asked whether they wanted to change the process; this completed the answer to RQ3.
We then used the teams' change ideas to create new ``Gendermag-for-AI'' variants, which the teams began using instead of the \OriginalGM{} in Figure~\ref{fig:Methodology-Experimental-Overview}, enabling us to answer RQ4.

%-------------
\subsection{Data Analysis
\draftStatus{MMB 10/4/25}{2.4: Good enough?}
}
\label{subsec:meth-data-analysis}

\boldify{To analyse \textit{only} the find data, we first took all responses from the forms (See Section~\ref{subsec:procedures-find}) and put them in a spreadsheet, printing off the ``maybe'' and ``no'' responses and affinity diagrammed.}

To investigate the types of user-facing \aiInclBug{}s the teams found (RQ1), we used Hsieh and Shannon's~\cite{hsieh2005three} conventional content analysis on the teams' ``find'' session data.
In conventional content analysis, categories are extracted directly from the text data.
In our case, the AI product teams had already flagged where these were by answering ``maybe'' or ``no'' on their subgoals, pre-action, and post-action forms they filled out during their evaluation sessions.%
\footnote{We analyzed only the ``maybe'' and ``no'' responses because %hese indicate the presence of a usability/inclusivity bug, unlike
the ``yes'' response indicates no problem.} %end ftnote
Through affinity diagramming, we incrementally added responses, grouping those which contained similar concepts and labeling those categories.
When no more new categories emerged from added data (i.e., data saturation), we stopped adding responses.
This revealed six potential AI inclusivity bug type categories, which became our codeset, shown later in Section~\ref{sec:Results-AI-Bugs}.

We used this codeset to qualitatively code the complete set of evaluation forms. 
To ensure consistency of our use of these codes, two authors each independently coded 20\% of the data, with resulted in 84\% agreement under the Jaccard index\footnote{Jaccard index between two sets: $J(A,B) = \frac{|A \cap B|}{|A \cup B|}$}~\cite{jaccard1908nouvelles}. %end ftnote
Given this level of reliability, one author coded the remaining data.
The results answered RQ1, and also contributed to the answer to RQ4.

\boldify{For the remainder of these data, the data for RQ2--RQ4 were not subjective, and here's why we believe this...}

Answering the remaining research questions did not require qualitative coding, because human judgment was not needed.
For RQ2, the teams had already been explicit about which bugs they decided to fix and how with sketches and screenshots of their fixes. 
For RQ3 and RQ4, the teams explicitly told us changes to the method they would like to see which, combined with the coding results above, completed the answer to RQ4.

%% file: figure/TeamGame-Figures/Methodology-Team-Game-Interface.tex
\begin{figure}[h!]
    \centering
    \includegraphics[width=0.99\linewidth]{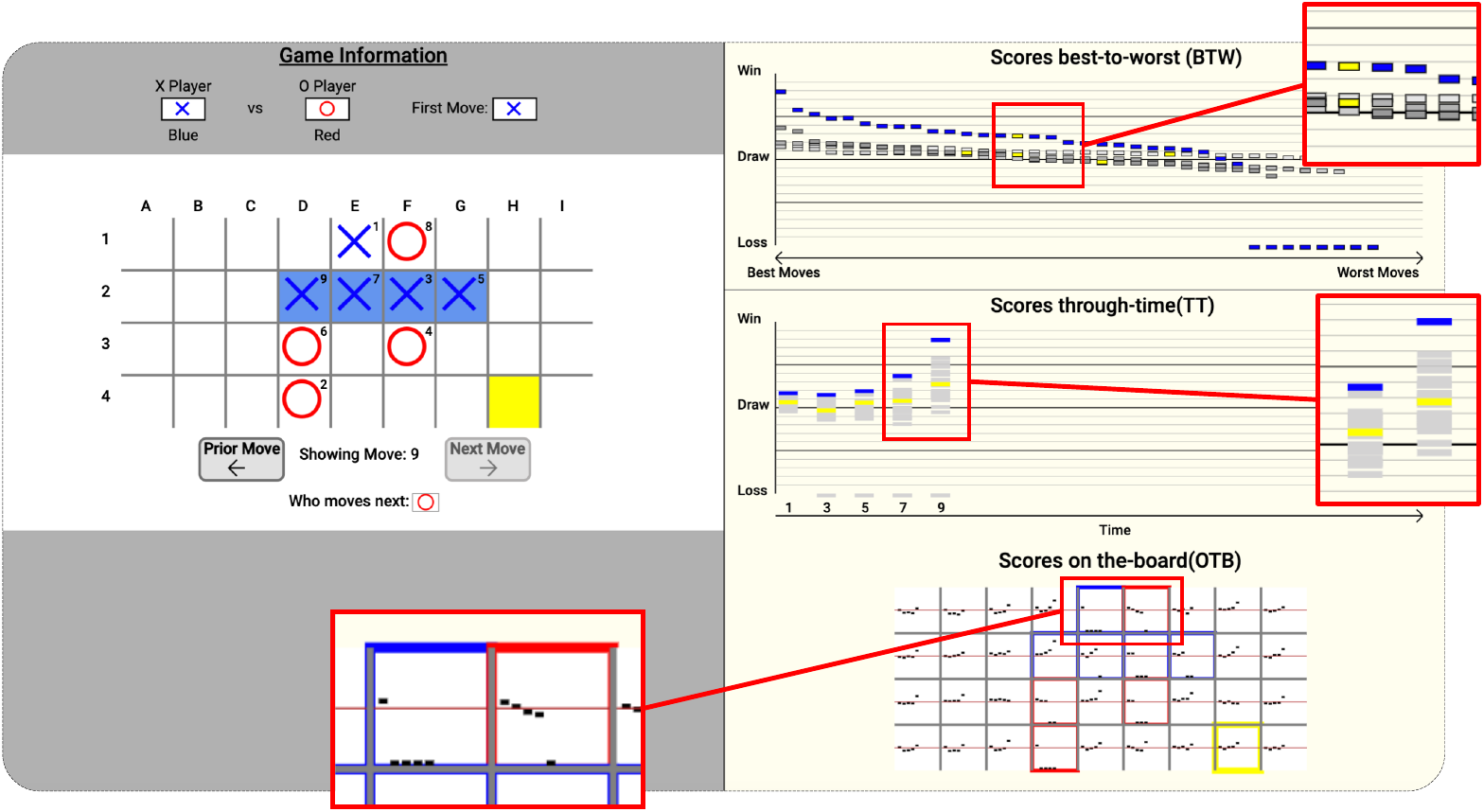}
    \caption{
    Team Game’s eXplainable AI (XAI) interface. 
    On the left is Team Game's gameboard, on the top-right is the Scores Best-to-Worst explanation, on the middle-right is the Scores Through-Time explanation, and on the bottom-right is the Scores On-the-Board explanation. Call-outs with enlarged portions have been superimposed for readability. 
    % A wireframe overlay of Team Game's eXplainable AI (XAI) interface.
    % Users saw both players on the gameboard (left, described in Section~\ref{subsubsec:team-game-gameboard}) and saw   explanations of three types for each decision from the \xPlayer{blue}'s perspective (right, discussed in Sections~\ref{subsubsec:team-game-btw}--\ref{subsubsec:team-game-otb}).}
    }
    \label{fig:Methodology-Team-Game-Interface}
\end{figure}

%% file: figure/TeamWeather-Figures/Methodology-Team-Weather-Interface.tex
\begin{figure}[h]
    \centering
    \includegraphics[width = 0.95\linewidth]{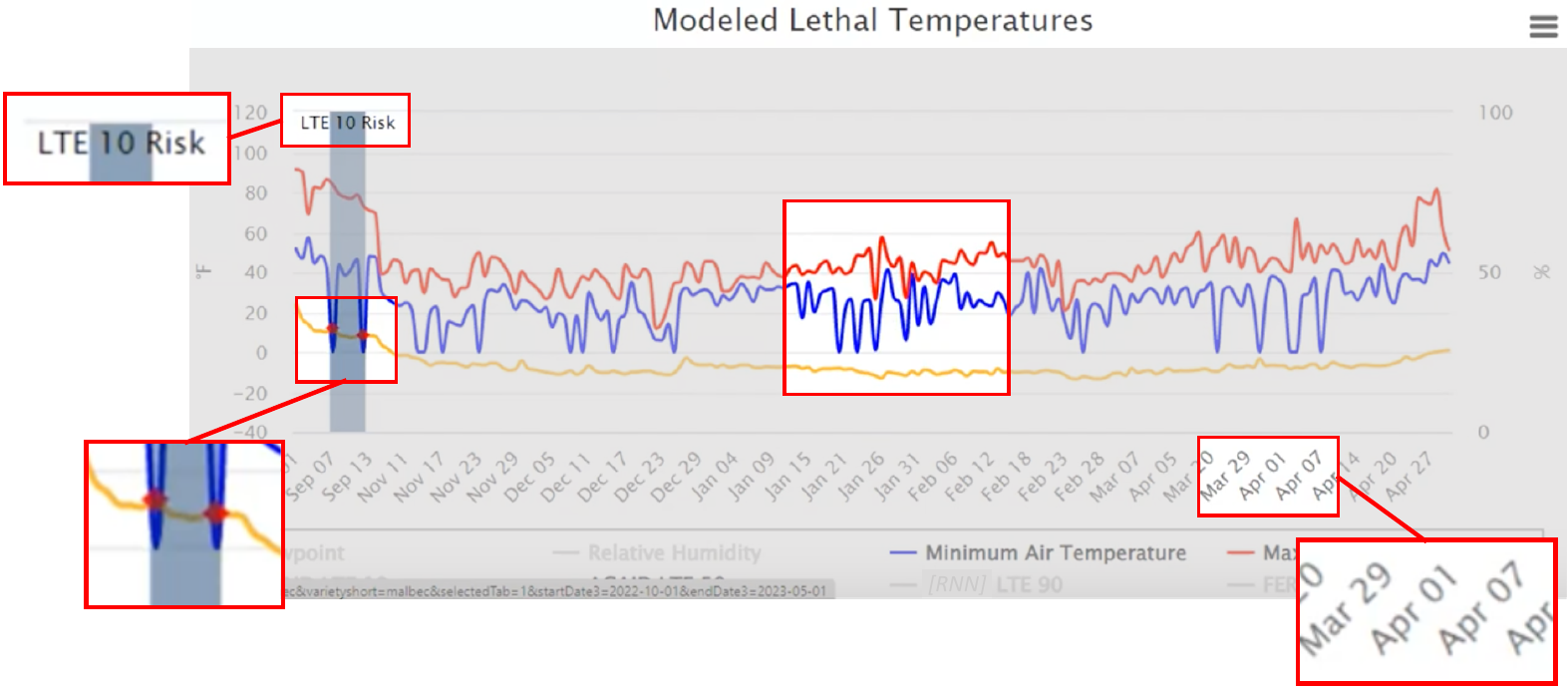}
    \caption{Team Weather AI's \LTE{10} cold-hardiness predictions (yellow line) for different days (x-axis).
    The \LTE{10} predictions were compared to minimum forecasted temperature (blue line), and when \LTE{10} $>$ the minimum, the interface marked these events with a red diamond to help agriculturalists decide whether and how to deploy frost-mitigation methods.
    }
    \label{fig:Methodology-Team-Weather-Interface}
    \FIXME{MMB@anyone}{unreadable figs. See TODO list for how to fix}
    % \FIXME{AAA}{I redacted a lot of things on this interface that would identify either 1) the system name or 2) the school that it's associated with (example: there's the WSU logo and school name in that top-left black square.
    % I'm wondering if I need to make this even more obvious to the reader though.}
\end{figure}

%% file: figure/TeamFarm-Figures/Methodology-Team-Farm-Interface.tex
\begin{figure}[h!]
    \centering
    \includegraphics[width = 0.9\linewidth]{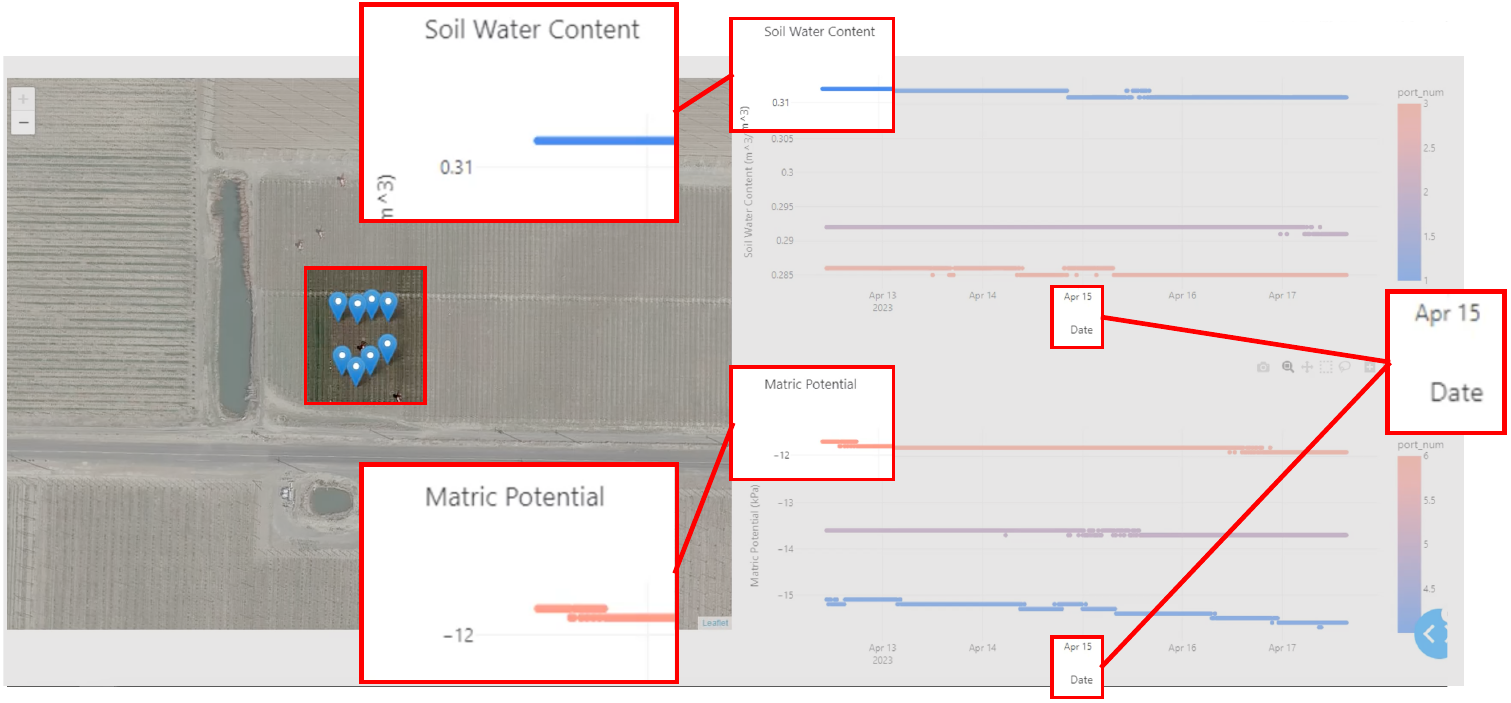}
    \caption{Team Farm's recreated AI-powered irrigation scheduling prototype.
    Their AI would use temporal sensor telemetry of soil water content (top-right) and matric potential (bottom-right) from eight fields (left) to provide decision support for agriculturalists of when (and how much) to irrigate each field.}
    \label{fig:Methodology-Team-Farm-Interface}
    \FIXME{MMB@anyone}{unreadable fig. See TODO list for how to fix}
\end{figure}

%% file: figure/Methodology-Blank-Abi.tex
\begin{figure}[h!]
    \centering
    \includegraphics[width = 0.9\linewidth]{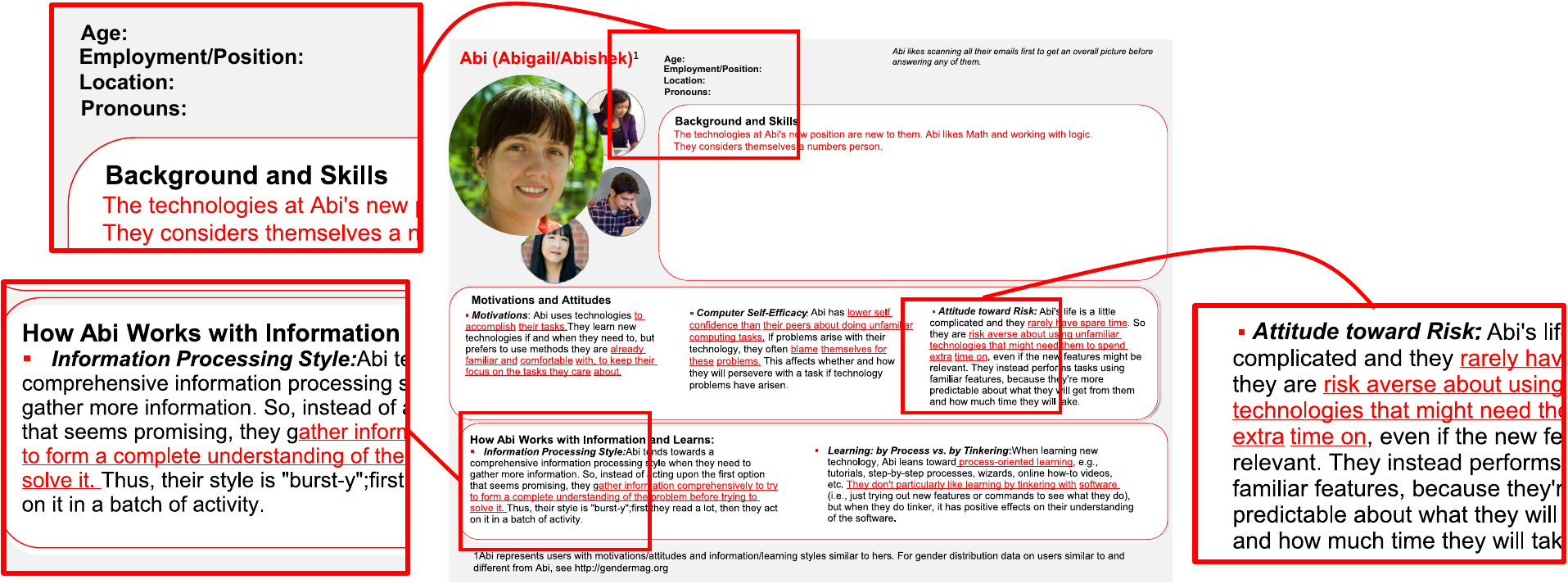}
    \caption{The ``Abigail/Abishek'' (Abi) persona template, which the teams customized to represent their users.
    They could customize the persona's age/employment/location/pronouns, as well as their background and skills (top-left).
    However, they \underline{could not} customize any of the five problem-solving style values (bottom half).}
    \label{fig:Methodology-Blank-Abi}
\end{figure}

%% file: figure/Methodology-Experimental-Overview.tex
\begin{figure}[h]
    \centering
    \includegraphics[width = 0.7\linewidth]{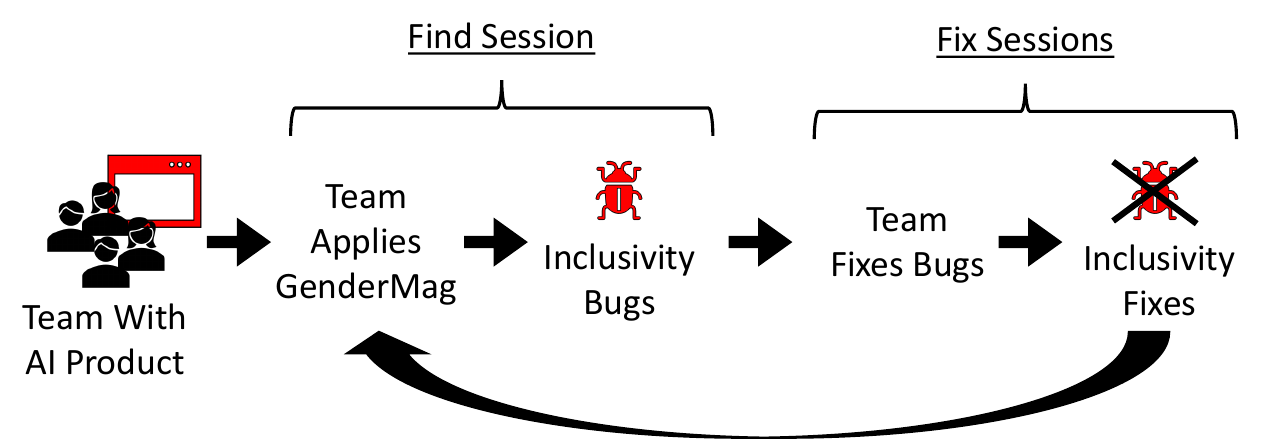}
    \caption{The field study methodology.
    Three teams developing AI products iterated through find/fix sessions.
    Find sessions: teams used a GenderMag walkthrough to evaluate their AI product, producing an inclusivity bug list. 
    Fix sessions: teams fixed inclusivity bugs until they said they were done, prompting another find session.}
    \label{fig:Methodology-Experimental-Overview}
\end{figure}

%% file: doc/05-Results-Find-AI-Inclusivity-Bugs.tex
\section{
Results RQ1 \& RQ2---Six AI Inclusivity Bug Types and the Fixes
\draftStatus{MMB 10/4/25}{top: 2.5--GOOD ENOUGH!}
}
\label{sec:Results-AI-Bugs}

\boldify{Table~\ref{tab:Methodology-Find-Codes} summarizes the inclusivity bug types the teams found -- 83 times!}

We begin with the ``bottom line'' for RQ1.
As Table~\ref{tab:Methodology-Find-Codes} shows, teams' ``find'' sessions revealed six AI inclusivity bug types and 83 bug instances.
Recall from Section~\ref{sec:intro} that an ``AI inclusivity bug'' is a usability bug that also meets two criteria:
(1) The teams explicitly identified the bug in the AI's information, which made the usability bug an \textit{AI} usability bug.
(2) The teams also tied at least one of the persona's \textit{problem-solving style values} (e.g., risk-aversion, comprehensive information processing style, etc.) to the AI usability bug.
An AI usability bug tied with a problem-solving style value means that users with \textit{that} problem-solving style value would be likely to be disproportionately impacted by the bug, which makes an AI usability bug an AI \textit{inclusivity} bug.  

\input{tables/Methodology-Find-Codes}

% \boldify{The inclusivity bugs tended to be associated with certain facet values and we organized them into the following sections.}
% The following subsections present all six AI inclusivity bug types these teams found and the problem-solving style values most frequently associated with them---risk-aversion (Section~\ref{subsubsec:bugs_risk}), comprehensive information processing style (Section~\ref{subsubsec:bugs_info}), and lower self-efficacy (Section~\ref{subsubsec:bugs_se}).
% \textcolor{red}{MMB 9/6/25: this parag doesn't match the boldif. The boldif is ok, but the parag is just a table of contents (which I'm not a fan of including). Solution ideas: (1) cut the parag entirely. OR  (2) Move a variant of the risk portion of it into the top of 5.1.1, but cut the rest of the parag. OR (3) Decide the boldif's point needs to be made here, so rewrite the parag to match the boldif. \\
% FAM: I decided to cut (idea 1), each subsection introduces each facet relation}

%------------
\subsection{Three AI Inclusivity Bug Types \& Risk-Aversion
\draftStatus{MMB 10/5/25}{top: D2.5}
}
\label{subsubsec:bugs_risk}

\boldify{Teams found three AI Inclusivity bug types \bugInterpret{}, \bugInputOutput{}, and \bugWhyShouldI{}) which were frequently associated Abi's risk-aversion.}

Considering risk-averse users was a powerful AI inclusivity bug-finding aid to the teams.
In their evaluation sessions, the teams referred to risk-averse users more often than any other problem-solving style, and for three of the AI inclusivity bug types---\bugInterpret{}, \bugInputOutput{}, and \bugWhyShouldI{}---risk was the top (or tied for top) reason.
%the teams associated the bugs with Abi's risk-aversion in 45.5\% of these bugs' instances.

%--------
\subsubsection{The \bugInterpret{} Inclusivity Bug}
\draftStatus{MMB 10/5/25}{2.4--GOOD ENOUGH}%
\label{subsubsec:bug-interpret}

\boldify{The teams' most frequently identified AI inclusivity bug type was Interpret AI? (27/83)}

``What does this even mean?'' 
\bugInterpret{} was the most frequent AI inclusivity bug type.
In total, the teams identified 27 instances of it.
In these bug instances, the teams decided that the AI's information could be  difficult for their populations to interpret, and they expected this AI inclusivity bug to particularly impact risk-averse users (10/27 instances).
They also expected this bug to affect comprehensive information-processing users, citing it 10 times as well, sometimes in combination with risk aversion. 

\boldify{Risk aversion has arisen in multiple ways in technology settings.}

The attitudes toward risk problem-solving style is nuanced.
It can include not only  well-known technology risks tied with privacy/security, but also risk of producing low-quality work, of wasting too much time, of failing to succeed in harnessing the product's hoped-for benefits, and more~\cite{kim-microsoft2023inclusive, stumpf2020gender}.
In their evaluations, the teams spoke frequently of the latter two aspects.
%
%
%\boldify{For example... The problem was that the BTW explanation was hard to interpret, specifically the change in the explanation.}
%
For example, the risk of wasting time without obtaining benefits figured prominently in Team Game's evaluation of Figure~\ref{fig:Results-4.1-Team-Game-OG-GM-Unclear-Output-Move-2-3}. 
The figure shows the ``before'' state (left) and the ``after'' state (right) for one of Team Game's \bugInterpret{} inclusivity bug instances. 
The after state (right) is where Team Game identified the \bugInterpret{} instance. 

\input{figure/TeamGame-Figures/Results-4.1-Team-Game-OG-GM-Unclear-Output-Move-2-3}

In the before state (left), both the \xPlayer{blue} and \oPlayer{red} had already made the moves shown on the gameboard.
The \BTW{} (bottom left) showed why the \xPlayer{blue} made that move: the \xPlayer{black} had evaluated all 36 squares before it moved, which it explains by drawing a line of blue rectangles from ``best'' move (far left) to ``worst'' (far right).
When the \oPlayer{red} responded with the move shown left, the \xPlayer{blue} re-calculated scores and moved as shown right, which it explains by the updated explanation on the bottom right.
The blue rectangles are the new move's calculations, and the gray rectangles are the previous move's calculations, to show what has changed.
But Team Game decided the \BTW{} was not particularly interpretable, and decided that Abi's perceived risk of time-wasting would outweigh Abi's interest in learning the AI's ``process'':
\quotateInsetFindBugs
%Team Name
{Game}
%Quote Snippet
{...wants to go through the process but does not have the context to forge the relationship. The time investment... to understand what’s going on [vs.] the perceived benefits...}
%Bug type identified
{\bugInterpret{}}
%Problem-solving style type identified
{Risk-Aversion, Process-Oriented Learning.}
%Session (and step) where the comment occurred
{Transcripts 061423-Step2, rowID 91 (session 5)}
%Full quote (from the form), so we don't have to keep going back to excel
{yes-clearly a relationship between the highlighting on the left and the right. No-Learn,At.Risk; Abi wants to go through the process but does not have the context to forge the relationship. The time investment that it would take when to understand what’s going on is based on the perceived benefits at this stage. Maybe the highlighted part of the OTB makes sense but not the BTW and TT.
}

\boldify{hey, this sounds like Blackwell's model of attn investment~\cite{blackwell2002first}, doesn't it?}

Team Game's reasoning is consistent with Blackwell's Model of Attention Investment~\cite{blackwell2002first}.
According to this model, technology users decide whether to spend their attention the same way they decide to spend money: using their own expectations of the cost, of the benefits they can expect, and of the risk (probability) that they will spend the cost but not gain any benefits. 
Team Game's focus on Abi's expection of wasted time (high probability of risk), despite her interest in the hoped-for benefits, fits Blackwell's model well.

\boldify{Once these were found, fixing AI-interpret was pretty stfwd, usually adding or clarifying information given. Team Game did it this way, and others did similar. }

The Blackwell model suggests that key to resolving \bugInterpret{s} is increasing/clarifying the benefits and/or reducing risks/costs in learning how to interpret the AI's output.
And that is what Team Game did, via the risk route: they added a legend (Figure~\ref{fig:Team-Game-AI-Interpret-Legend-Fix}) to reduce the risk of failing to learn what the explanation meant.
In fact, all the teams fixed the \bugInterpret{} bug in ways consistent with the Blackwell model, by increasing the benefits or reducing the attention costs and/or risks (e.g., via adding/improving legends, info boxes, labels) of spending time on uninterpretable AI output the team had spotted.

\input{figure/TeamGame-Figures/TeamGame-Fix-Figures/Team-Game-AI-Interpret-Legend-Fix}

% FAM: maybe summarize the other instances of this bug (was it in explanations) or a summary figure with stamp size pictures (not as good of an idea because the bugs are all similar as it was in the graphs)

%--------
\subsubsection{The \bugInputOutput{} Inclusivity Bug}
\label{subsubsec:bug-input-output}
\draftStatus{MMB 10/5/25}{2.3 (because rather boring)}

\boldify{this bug was particularly interesting as it arose because of the next bug.}
 
\boldify{Team Game found that a \bugInputOutput{} bug instance was responsible. }

``What does this (AI-input) have to do with that (AI-output)?''
In the same moment of play shown above in Figure~\ref{fig:Results-4.1-Team-Game-OG-GM-Unclear-Output-Move-2-3}, Team Game found another AI inclusivity bug, \bugInputOutput{}.
This type of bug is a lack of clarity about whether/how the AI's input(s) relate to its output(s):
% concluding that a second AI inclusivity bug type had been responsible---the \bugInputOutput{} type.
% These three teams identified the \bugInputOutput{} type only nine times, but they associated attitudes toward risk in five instances.
% Tame Game said that the \BTW{} did an insufficient job of relating AI's inputs (the gameboard) and outputs (the \BTW{}):
%
\quotateInsetFindBugs
%Team Name
{Game}
%Quote Snippet
{The relationship between squares on the board and the Best-to-Worst explanation isn’t clear...}
%Bug type identified
{\bugInputOutput{}}
%Problem-solving style type identified
{Risk-Aversion}
%Session (and step) where the comment occurred
{Walkthrough-2-XAI-MNK-061423 2.4b [AFTER]}
%Full quote (from the form), so we don't have to keep going back to excel
{The relationship between squares on the board and the Best-to-Worst explanation isn’t clear. Empirically, in the lab, we can explain things, but those explanations aren’t as obvious in the interface.
}
%This was an \bugInputOutput{} instance because squares on the gameboard were the \xPlayer{black}'s \textit{input}, and the \BTW{} was an \textit{output}.
%These teams identified \bugInputOutput{} instances only nine times, but they associated Abi's risk-aversion in five of them.

\boldify{Most instances of this were seen as specialized instances of Interpret-AI, and indeed, the teams marked both on 7/9 of these. This says the solutions are also a specialized instances of these additions of clarifying info: here CONNECTING inputs w outputs. }

Interestingly, teams seemed to regard most instances of \bugInputOutput{} as specialized instances of \bugInterpret{}, with their verbalizations suggesting both on 7/9 of the \bugInputOutput{} instances. 
Thus, their solutions to \bugInputOutput{} tended to be specialized instances of solutions to \bugInterpret{}.

Specifically, they again added information to increase the user's perceivable benefits or reduce their apparent attention costs and/or risks (e.g., via adding/improving legends, info boxes, labels)---but for \bugInputOutput{}, the new information explicitly connected the AI's input to the AI's output.
%To fix the \bugInputOutput{} bug, teams focused on connecting information to reduce confusion, from using different symbols and visuals to tooltips.
For example, Team Game's solution was to show the connection dynamically.
Whenever users highlighted a score rectangle in the explanation, the corresponding gameboard square became highlighted and vice-versa, and a tooltip  (Figure~\ref{fig:Results-Original-GM-Fix-Team-Game-Tooltip}) would appear to connect the gameboard state (the AI's inputs) to the score rectangles on the \BTW{} (its output).

\input{figure/TeamGame-Figures/Results-Original-GM-Fix-Team-Game-Tooltip}

%--------
\subsubsection{The \bugWhyShouldI{} Inclusivity Bug}
\draftStatus{MMB 10/5/25}{2.4: GOOD ENOUGH?}
\label{subsubsec:bug-why-should}

\boldify{The last bug type that teams frequently associated with risk is the \bugWhyShouldI{} bug type, where Abi thought an action would have no benefit.}

``Why even look at this?''
\bugWhyShouldI{} was the third AI inclusivity bug type associated with risk-aversion particularly frequently (10/19 instances).
These bugs differ from the first two: the first two show at least some user interest in engaging with the AI's information, whereas this bug depicts some risk-averse users not even seeing the point of trying. 
%That is, Abi might view engaging with the AI's information to be a total waste of time, producing no benefit at all.

\input{figure/TeamFarm-Figures/Results-4.1-Team-Farm-OG-GM-Why-Should-I-Graph}

\boldify{Here is an example of what the bug looked like for Abi using Team Farm's AI in Figure~\ref{fig:Results-4.1-Team-Farm-OG-GM-Why-should-i?-Graph}}

Figure~\ref{fig:Results-4.1-Team-Farm-OG-GM-Why-should-I-graph} shows an example of this bug type, which Team Farm found in a pair of irrigation recommendation visualizations based on soil measurements.
%but Team Farm found that Abi would not know why they should spend time trying to understand the critical information
%
%\boldify{Although the information was critical, Abi did not know it was; one concern Team Farm had was that Abi wouldn't want to ``waste'' time.}
%
% abi doesn't know it critical, Abi wouldn't want to waste time on something that didn't seem important
These soil measurements are critical in measuring whether soil can support crop growth, but Team Farm raised concerns that Abi would not know this, and might give up instead of wasting time trying to understand the information:
% FAM: below bug was not fixed, but a similar one is below
\quotateInsetFindBugs
%Team Name
{Farm}
%Quote Snippet
{...[Abi] wouldn't know that looking at this graph for prolonged periods of time is going to help them understand the irrigation issues.}
%Bug type identified
{\bugWhyShouldI{}}
%Problem-solving style type identified
{Risk-Aversion}
%Session (and step) where the comment occurred
{041723-Step2-DemoFarm-Find-1-JN.txt 2a [BEFORE]}
%Full quote (from the form), so we don't have to keep going back to excel
{
They're trying to relate the colors they see, that the colors are related, but they don't know what port number means. You know, they wouldn't know that looking at this graph for prolonged periods of time is going to help them understand on the irrigation issues.
}

\boldify{Team farm fixed a similar bug in a later iteration of their prototype by encouraging curiosity.}
% \boldify{The purpose of changing information for \bugWhyShouldI{} was to promote curiosity, with a reward at the end (surprise-explain-reward)}

In a later version of their graphs (Figure~\ref{fig:Results-Original-GM-Fix-Team-Farm-Tooltip}, left), Team Farm fixed a bug like this by indicating that interaction was available.
% \AAA{Reader has never seen this bug before in a previous section. Unlike the last two, I have to present them the actual quote for this to contextualise for readers what occurred.}
They first added a blue rectangle to indicate that it was clickable (Figure~\ref{fig:Results-Original-GM-Fix-Team-Farm-Tooltip}, right).
For comprehensive information processors like Abi, these clickable instances hint at the possibility of acquiring more information.
Once Abi interacted with this area, Team Farm provided Abi with information about how much and when to irrigate their field:
\quotateInsetCommentary
%Team Name
{Farm}
%Quote Snippet
{We can make [the graph] look more clickable.
We know that Abi doesn't like to take risks...
There's no clear indication that hovering over water content is going to lead to irrigation decisions...}
%Session (and step) where the comment occurred
{070423-Step2-DemoFarm-Fix-5-SA.txt, 58:13:00}
%Full quote (from the form), so we don't have to keep going back to excel
{
But I think it's more than that. Because first of all, as I said, that's why I wanted an icon that Representative not just in computer science, but in anything that you have. So when you go to our website, have you view go, just in a normal sign? I will believe that Abi is familiar. If Abi is not familiar, the icon that you have in the graph is really like you can tell that's not part of the normal flow of the graph, we can make an more look more clickable.
Yes, so that's an indication that actually you can click on that as well, right? Because we know that Abi isn't less, doesn't like to take risks, and doesn't tinker around, it doesn't mean he hasn't used technology before, he has used technology, so he's familiar with a hand, we just need to reduce the tinkering. Yes, I think that will be an indication.
}
%Team Farm's fix was reminiscent of theories like surprise-explain-reward~\cite{wilson2003harnessing}, which has had success in encouraging users to interact with technology and knowing why they interacted.

% In the bugs we categorize as \bugWhyShouldI{}, some users might not see any reason to engage with it.
\boldify{When fixing the why should I bug, teams generally had the strategy of lowering risk/increasing curiosity for Abi as the main problem was hidden information.}

This kind of fix can be regarded as an instantiation of the Surprise-Explain-Reward strategy~\cite{wilson2003harnessing}.
That strategy, inspired in part by Loewenstein's work on curiosity~\cite{loewenstein1994curiosity}, attempts to pique the user's curiosity by surprising them ``just enough'' to inspire them to engage with a particular feature if/when they need to without interrupting them. 
The surprise's job is to deliver the user to a suitable explanation that hints at the benefits that ensue if they further engage, and the ``reward'' is a real-world benefit that the product feature delivers (here, showing exactly when to irrigate).
Team Farm and the other teams may not have known about the Surprise-Explain-Reward strategy, but many of them used it by adding information and interactions to give users like Abi more reason to engage with the AI product feature.

\input{figure/TeamFarm-Figures/Results-Original-GM-Fix-Team-Farm-Tooltip}

% --------------s
\subsection{Two AI Inclusivity Bug Types \& Information Processing Style
\draftStatus{FAM}{2, 10/4/25}
}
\label{subsubsec:bugs_info}

% \AAA{@MMB had a thought to prepare readers for an information processing cycle, like gather-do. I found a book chapter on information processing by Wickens \& Carswell, who point to a cog psych theory for Stimulus $\rightarrow$ Perception $\rightarrow$ Response $\rightarrow$ Response that 11...distinguishes a perceptual stage from one involving the selection and execution of action...''}

%\boldify{Teams identified two more bugs, which they frequently identified by Abi's comprehensive information processing style.}

%Teams frequently associated \bugMoreInfo{} and \bugActionable{} instances with Abi's comprehensive information processing style.

%----------
\subsubsection{The \bugMoreInfo{} Inclusivity Bug}
\draftStatus{FAM}{2.3, 10/4/2025}
\label{subsubsec:more-info}

\boldify{Teams found this AI inclusivity bug  when the information was insufficient for Abi-like information processors}

% \topic{Of the remaining three AI inclusivity bug types (Table~\ref{tab:Methodology-Find-Codes}), the teams associated two of them with Abi's comprehensive information processing style, including \bugMoreInfo{} (13/83).}
``Need more info!''
Teams found nine instances of \bugMoreInfo{}.
Over half of the time (5/9 instances), they decided that these would particularly affect comprehensive information processors like Abi.

\boldify{Figure~\ref{fig:Results-4.1-Team-Farm-OG-GM-AI-More-Info} (left) shows the bug, where the tooltip provided an overview of Plot A's soil status---it was too simple.}

Figure~\ref{fig:Results-4.1-Team-Farm-OG-GM-AI-More-Info} (left) shows one of Team Farm's \bugMoreInfo{} instances.
The AI's assessment (tooltip) of Plot A's soil status was  ``OK,'' so Abi would not have to irrigate this plot of Cabernet Sauvignon grapes.
However, Team Farm questioned whether this summary was oversimplified, insufficient for Abi's comprehensive information processing style:

\quotateInsetFindBugs
%Team Name
{Farm}
%Quote Snippet
{There may not be enough information in the tooltip to give the impression that the `OK' is OK.}
%Bug type identified
{\bugMoreInfo{}}
%Problem-solving style type identified
{Comprehensive Info. Proc.}
%Session (and step) where the comment occurred
{060523-Step2-DemoFarm-Find-2-RG.txt Subgoal #2}
%Full quote (from the form), so we don't have to keep going back to excel
{If Abi sees plot A is okay and they may dive in check the other close plots. Abi also has a comprehensive processing style and by checking the OK status they still may want to know if the status is correct by checking the actual graphs thoroughly. There may not be enough info in the tooltip to give the impression that the OK is OK.
}
This was reminiscent of Kulesza et al.~\cite{kulesza2013too},
who found that if an AI's explanation is too simple, it could lead to increased mental demand and decreased trust in the explanation.

\input{figure/TeamFarm-Figures/Results-4.1-Team-Farm-OG-GM-AI-More-Info}

\boldify{This was a bug as Team Farm wanted Abi to ``...find what's happening with the irrigation in \textit{another} plot,'' but the lack of information prevented Abi from doing this.}

% FAM: Make this paragraph follow boldification
Team Farm regarded this AI inclusivity bug as critical, because it prevented users like Abi from going to Plot B (Figure~\ref{fig:Results-4.1-Team-Farm-OG-GM-AI-More-Info}, right), which was {\color{red}``DRY''} and \textit{needed} Abi to irrigate it.
Since the overview was insufficiently detailed, Team Farm decided that Abi would instead try to validate the irrigation status by investigating the sensor readings from Figure~\ref{fig:Results-4.1-Team-Farm-OG-GM-Why-should-I-graph}:
\quotateInsetFindBugs
%Team Name
{Farm}
%Quote Snippet
{Abi still may want to know if the status is correct by checking the actual graphs thoroughly..}
%Bug type identified
{\bugMoreInfo{}}
%Problem-solving style type identified
{Comprehensive Info. Proc.}
%Session (and step) where the comment occurred
{051523-Step2-DemoFarm-Find-1-JN.txt}
%Full quote (from the form), so we don't have to keep going back to excel
{If Abi sees plot A is okay and they may dive in check the other close plots. Abi also has a comprehensive processing style and by checking the OK status they still may want to know if the status is correct by checking the actual graphs thoroughly. There may not be enough info in the tooltip to give the impression that the OK is OK.
}

% \textcolor{red}{FAM: replacing above example is not a good idea, other viable example is team game but is overused and the other potential examples has uninteresting fixes}
\boldify{Team Farm did not fix this bug instance, but teams in general added more information to address Abi's comprehensive info proc. style.}

While Team Farm did not fix this \bugMoreInfo{} bug, teams fixed four other instances by adding more information in the AI's inputs and/or output where teams (through Abi's problem-solving styles) decided it was not enough.
The teams tied all four of these bug instances to Abi's lower self-efficacy or comprehensive information processing style.

%
%

%----------
\subsubsection{The \bugActionable{} Inclusivity Bug}
\draftStatus{FAM}{2, 10/4/2025}
\label{subsubsec:bug-actionable}

\input{figure/TeamFarm-Figures/Team-Farm-Fix-Figures/Team-Farm-AI-Actionable}

\boldify{The \bugActionable{} bug has to do with why people gather information, to \textit{do} something which was unclear from the AI. Comprehensive information processing was associated the most with this bug.}

``So?  What should I DO?''
Teams found 12 instances where it was not clear how users should act upon an AI's information.
Almost half of the time (5/12 instances), these teams decided this would particularly affect comprehensive information processors like Abi.

% FAM: Moved old bug example (team game actionable? bug at end of a game) to leftovers
\boldify{An example of this is in Team Farm's later iteration on their AI product where it was unclear what Abi should do with the information from the AI.}
When Team Farm later evaluated their solution to the graph in Figure~\ref{fig:Results-Original-GM-Fix-Team-Farm-Tooltip} (right), they decided that their solution was not enough to hint what Abi should do:

\quotateInsetFindBugs
%Team Name
{Farm}
%Quote Snippet
{Will Abi click? No. Will Abi move the cursor? Probably. Exactly on top of the rectangle? Probably not. If I (Abi) don't really understand what that rectangle is for...
}
% bug type
{\bugActionable{}}
% problem solving style(s)
{Comprehensive Info. Proc., Risk-Aversion}
%Session (and step) where the comment occurred
{062723-Step2-DemoFarm-Find-3-RG.txt, 56:15:00}
%Full quote (from the form), so we don't have to keep going back to excel
{I don't think so. [R01: why is that?] Well, you see, the thing is that probably- well I have to be Abi. Will Abi click? No. Will abi move the cursor? Probably. Exactly on top of the rectangle? Probably not. If I don't really understand what that what that rectangle is for, I can just even change the another option underneath the graph, or I can change tabs or just get out of the website at all- or like completely. sorry. so. No. so my answer is yeah, no, no 100\%. 
}

\boldify{Team Farm fixed their bug by making the action they want Abi (the user) to take clearer with a visible icon and label. Other teams made similar fixes to better direct users.}

Figure~\ref{fig:Team-Farm-AI-Actionable} shows how Team Farm tried to make it clearer for Abi what to do about the AI's information.
They replaced the rectangular object with an icon of a faucet with a water droplet to make it more apparent what to do and when. Other teams made similar fixes by giving more hints/directions on what users should do through instructions, changing button/icons, or clarifying language.

\quotateInsetCommentary
%Team Name
{Farm}
%Quote Snippet
{Is Abi going to hover over the rectangle? 
There's no rectangle anymore. 
Now we have an indication point. 
Is Abi going to hover over the indication point? 
Yes...
I wanted an icon that's representative not just in computer science, but in anything that you have.}
%Session (and step) where the comment occurred
{file**, time**}
%Full quote (from the form), so we don't have to keep going back to excel
{Okay. So I think we have solved our problems. Okay. All of them. Well, according to us, yeah. Because I have the different forms here. So I was walking through the forms while we were talking. Okay. I don't think there's something else to solve, because we type all day. Is Abi going to hover over the rectangle? There's no rectangle anymore. Now we have an indication point is Abi going to hover over the indication point? Yes.
But I think it's more than that. Because first of all, as I said, that's why I wanted an icon that Representative not just in computer science, but in anything that you have. So when you go to our website, have you view go, just in a normal sign? I will believe that Abi is familiar. If Abi is not familiar, the icon that you have in the graph is really like you can tell that's not part of the normal flow of the graph, we can make an more look more clickable.
}

%-----------------------------
\subsection{An AI Inclusivity Bug Type \& Self-Efficacy
\draftStatus{FAM}{2.6, 10/4/2025}
}
\label{subsubsec:bugs_se}

% \boldify{Self-efficacy kind of coasted under the radar, but it roared for the last AI inclusivity bug.}
\boldify{The last AI inclusivity bug from Table~\ref{tab:Methodology-Find-Codes} was almost always associated with computer self-efficacy (6/8), here are examples from each team.}

``What's changed?''
These teams associated problem-solving values only 8 times with the \bugChanges{} bug, but \textit{six} of them were Abi's lower self-efficacy.
% FAM (this line is FALSE, SE was used more than 8 times) These teams associated Abi's lower self-efficacy only eight times, but they associated \textit{six} of these instances with the last AI inclusivity bug type---the \bugChanges{} type.
Teams identified such instances when it was not clear how the AI's information changed through time.
Teams decided that this deficit in the visualization made those with lower self-efficacy think they had done something wrong or would blame themselves:

%did an insufficient job of making clear exactly how the information changed through time.
%Two teams identified only seven instances of this AI inclusivity bug type throughout all sessions, and they associated Abi's lower computer self-efficacy in 6/7 instances.

\quotateInsetFindBugs
%Team Name
{Farm}
%Quote Snippet
{Abi would struggle to differentiate between the graphs (previous vs. now) and \underline{would blame themselves}, wondering if they selected the previous plot or not.\normalfont{''}\\
\normalfont{Team Farm: ``}\textit{When Abi clicks on the thing, it’s going to lead them to believe they’ve \underline{done something wrong} because nothing changes even upon clicking.}"\\
\normalfont{Team Game: ``}\textit{The game has progressed. There was no update for the other player. Perhaps she may consider she \underline{may have done something wrong.}}}
%Bug type identified
{\bugChanges{}}
%Problem-solving style type identified
{Lower Self-Eff.}
%Session (and step) where the comment occurred
{**FIXME**}
%Full quote (from the form), so we don't have to keep going back to excel
{
SE - Abi would struggle to differentiate between the graphs (previous vs. now) and would blame themselves, wondering if they selected the previous plot or not.
"Abi will take other actions before hovering over another plot will be to click, because that’s the instructions in the title of the map. Part of the bursty information processing style, where they’re going to try to click before moving on. Because of Abi’s self-efficacy, clicking is going to lead Abi to believe it’s the wrong move and abandon the system (it’s become more trial-and-error, rather than something with a structure).
When Abi clicks on the thing, it’s going to lead them to believe they’ve done something wrong because nothing changes even upon clicking."
The game has progressed. There was no update for the other player. Perhaps she may consider she may have done something wrong.
}

\boldify{These examples not only match the literature but also show the consequences of ignoring them}

The literature provides evidence to support these potential consequences of failing to support users with lower self-efficacy~\cite{bandura1986explanatory}.
For example, research has shown that people with lower self-efficacy, such as Abi, may be less willing to explore features new to them~\cite{grigoreanu2008can}.
Others have found that such people may be more likely to abandon the system when they think barriers are ``too high''~\cite{stumpf2020gender}, something Team Farm also identified:
\quotateInsetFindBugs
%Team Name
{Farm}
%Quote Snippet
{Because of Abi’s self-efficacy, clicking is going to lead Abi to believe it’s the wrong move and abandon the system. It’s become more trial-and-error...}
%Bug type identified
{\bugChanges{}}
%Problem-solving style type identified
{Lower Self-Eff.}
%Session (and step) where the comment occurred
{Jake/Paola-DemoFarm-Walkthrough-1-041723 3.1b [AFTER]}
%Full quote (from the form), so we don't have to keep going back to excel
{"Abi will take other actions before hovering over another plot will be to click, because that’s the instructions in the title of the map. Part of the bursty information processing style, where they’re going to try to click before moving on. Because of Abi’s self-efficacy, clicking is going to lead Abi to believe it’s the wrong move and abandon the system (it’s become more trial-and-error, rather than something with a structure).
When Abi clicks on the thing, it’s going to lead them to believe they’ve done something wrong because nothing changes even upon clicking."
}

\input{figure/TeamGame-Figures/Results-Original-GM-Fix-Team-Game-Tie-Arrows}

\boldify{Teams fixed this bug by changing elements in the interface to grab attention. Here is what Team Game did to reassure Abi.}

When fixing this bug type, teams generally changed or added elements in the interface to grab attention when an action was taken.
Team Game fixed their above bug (no changes for other player) with two additions (1) a legend clarifying how changes appear in the \BTW{} (recall Figure~\ref{fig:Team-Game-AI-Interpret-Legend-Fix}) and (2) adding colored arrows to show how the latest move the \xPlayer{blue} or \oPlayer{red} picked now scored in each explanation (Figure~\ref{fig:Results-Original-GM-Fix-Team-Game-Tie-Arrows}).
The temporary appearance of arrows along with the explanation legend aimed to point out what changed in the interface.

%---------- END
\subsection{Did the fixes help? An external empirical investigation}
\draftStatus{FAM}{2.5, 10/4/2025}

In total, the teams fixed 47 of the 83 AI inclusivity bugs  (Table~\ref{tab:Results_Fixes_by_Bug}).
But did the fixes actually help end users of the AI products?

\input{tables/GM_Fixes/Results_Fixes_by_Bug}

%----- transition to actual test

\boldify{Did fixes work? Hamid et al. answered this for Team Game.}

Team Game undertook an external collaboration with a group of empirical researchers to find out the answer for their AI Product.
Although the result of their external collaboration has been presented elsewhere~\cite{hamid-2024-tiis}, we briefly summarize that work here because of its direct pertinence to Team Game's fixes.
Team Game's external collaboration involved a between-subjects lab experiment, in which 69 participants with no formal AI background worked with Team Game's prototype.
Half worked with the original (pre-GenderMag) prototype and the other half worked with the fixed version~\cite{hamid-2024-tiis}.
The results showed that participants using Team Game's \coloredText{MMH-GM}{white}{Post-GenderMag} version had significantly better mental models (conceptual understanding) of the AI's reasoning than participants who used the \coloredText{MMH-Original}{black}{Original} version (Figure~\ref{fig:Results-FixesWorked-EmpiricalResults}).
A second measure was participants' ability to predict the AI's next move, and for this measure, there was no significant difference between groups.
A third measure was gender equity, and here the difference was again significant.
In fact, the fixes in the \coloredText{MMH-GM}{white}{Post-GenderMag} version improved gender equity of participants' mental model scores by 45\%.

\input{figure/Results-FixesWorked-EmpiricalResults}

%% file: tables/Methodology-Find-Codes.tex
    \begin{table}[h]
        \centering
        \includegraphics[width=\linewidth, trim=0 5.8in 1.5in 1in, clip]{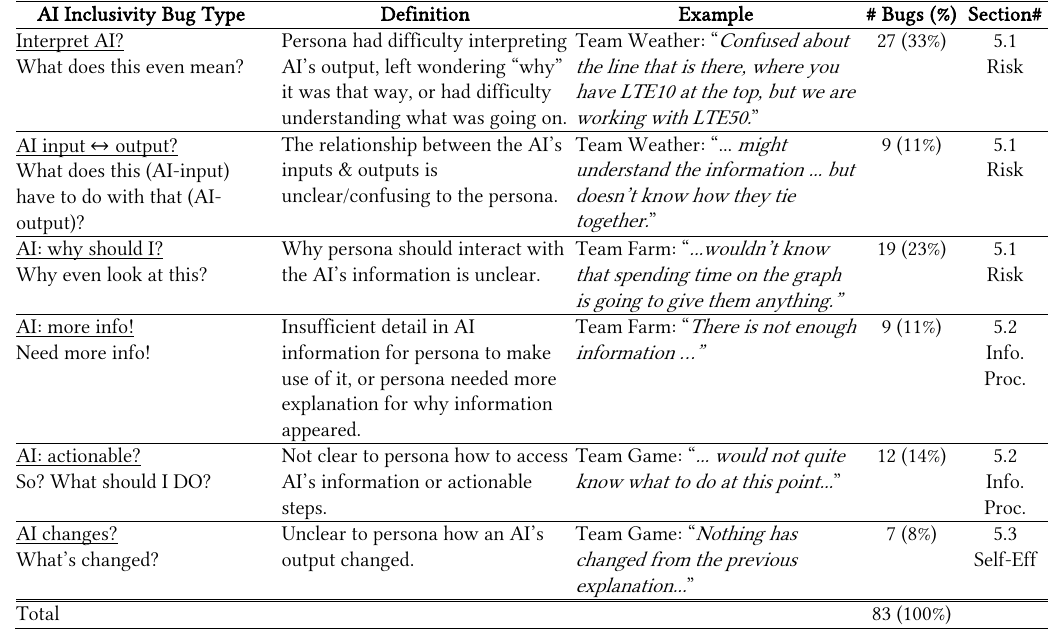} %trim param is amt to crop left, bottom, right, top
 %       \vspace{1mm}
        \caption{The six AI inclusivity bug types (rows). Together, the teams found 83 instances of these bug types.
        }
        \label{tab:Methodology-Find-Codes}
\end{table}

%% file: figure/TeamGame-Figures/Results-4.1-Team-Game-OG-GM-Unclear-Output-Move-2-3.tex
\begin{figure}[h]
    \centering
     \includegraphics[width = 0.95\linewidth]{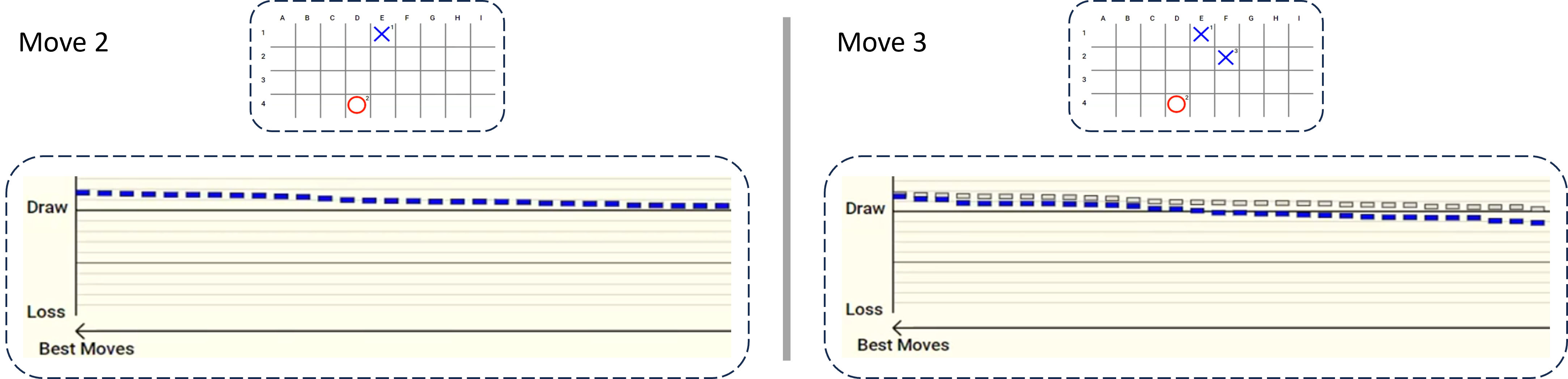}
    \caption{An \bugInterpret{} instance (Team Game).
    \textbf{Left:} Before Bug---the gameboard (top) and \BTW{} (bottom, enlarged for readability).
    \textbf{Right:} Discovered Bug---Team Game noted that the explanation changes (the two lines of rectangles in the explanation) were not straightforward to interpret.}
    \label{fig:Results-4.1-Team-Game-OG-GM-Unclear-Output-Move-2-3}
\end{figure}

%% file: figure/TeamGame-Figures/TeamGame-Fix-Figures/Team-Game-AI-Interpret-Legend-Fix.tex
\begin{figure}[h]
    \centering
    \includegraphics[width=0.49\linewidth]{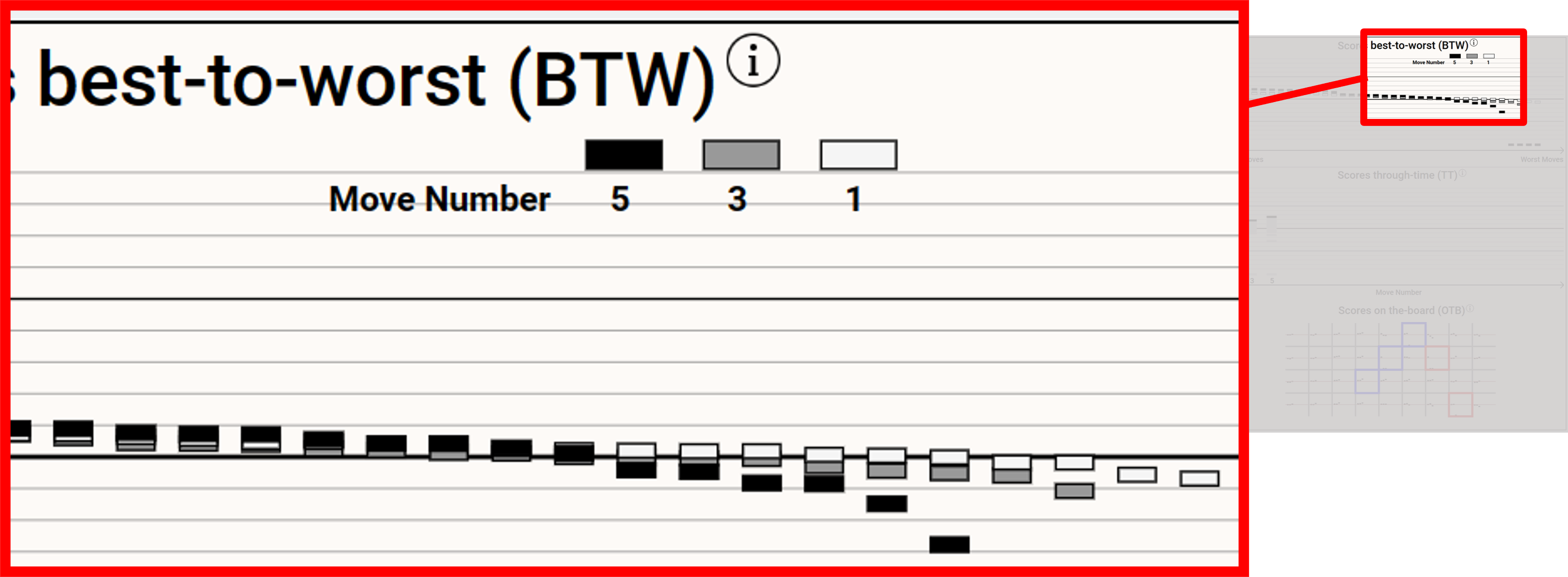}
    % \caption{An \bugInterpret{} fix (Team Game). In Section~\ref{subsubsec:bugs_risk}, Team Game had found that their \BTW{} was not as interpretable as the other explanations.
    % Their fix added a legend, which would show Abi what each color on the explanation meant.}
    
    % FAM: spiking a different result structure, above is the old caption
    \caption{An \bugInterpret{} fix (Team Game). To improve the interpretability of the \BTW{} explanation, Team Game added a legend showing what each color means.}
    \label{fig:Team-Game-AI-Interpret-Legend-Fix}
\end{figure}

%% file: figure/TeamGame-Figures/Results-Original-GM-Fix-Team-Game-Tooltip.tex
% \begin{wrapfigure}{o}{\wrapFigWidth{}}
\begin{figure}[h]
    \centering
        \includegraphics[width= .25\linewidth]{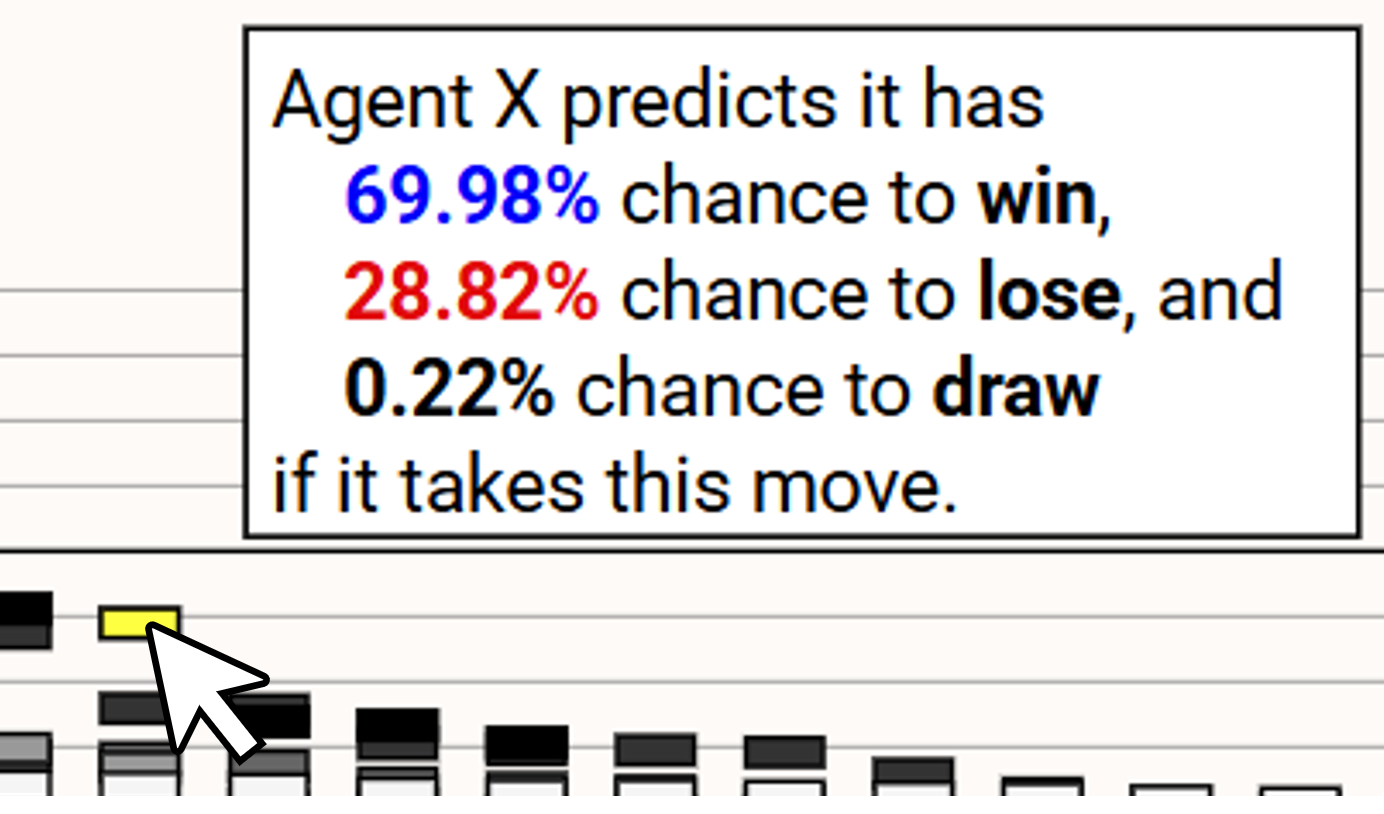}
        %\caption{AI Information (With Tooltip)}
        %\label{fig:Results-Original-GM-Fix-Team-Farm-Tooltip-After}
    \caption{A fix for one of Team Game's \bugInputOutput{} instances.
    The purpose of the fix explicitly mapped how the \xPlayer{black} used its input (moves on the gameboard) to calculate its output (win, lose, draw probabilities).}
    \label{fig:Results-Original-GM-Fix-Team-Game-Tooltip}
\end{figure}
% \end{wrapfigure}

%% file: figure/TeamFarm-Figures/Results-4.1-Team-Farm-OG-GM-Why-Should-I-Graph.tex
\begin{figure}[h]
    \centering
    \includegraphics[width = 0.55\linewidth]{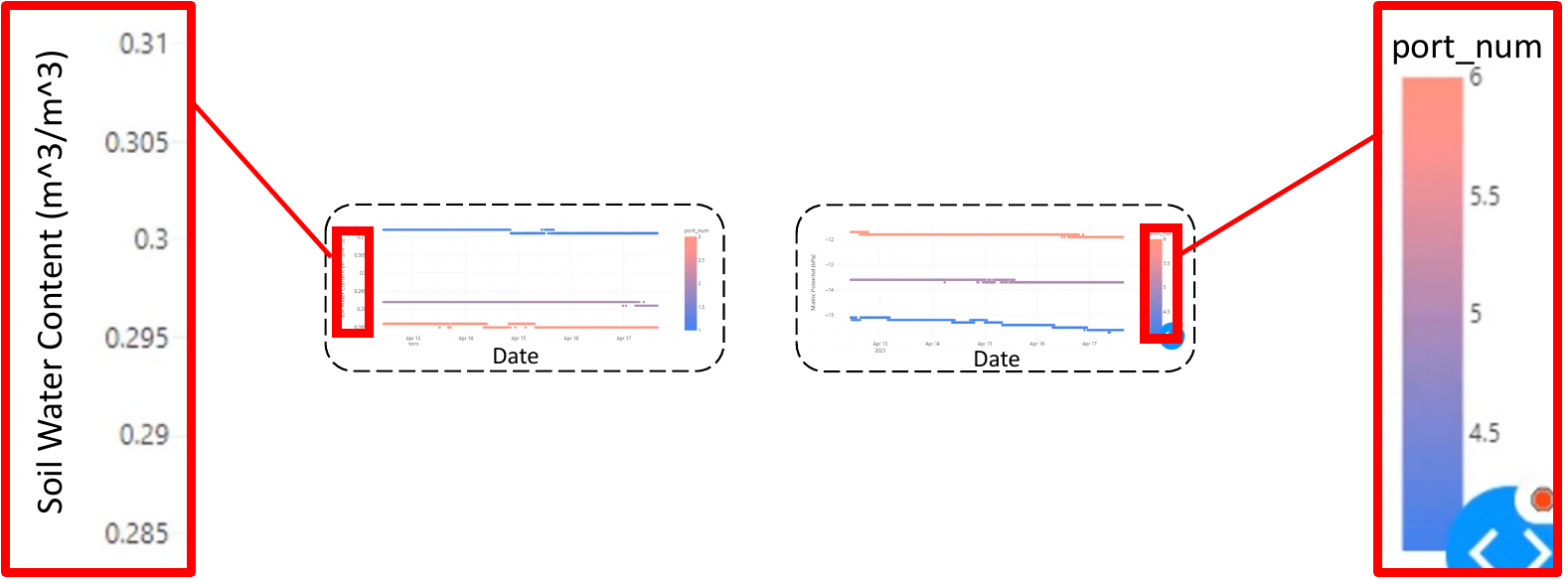}
    \caption{An \bugWhyShouldI{} instance (Team Farm).
    The AI used six real-time sensor readings to make irrigation recommendations, three sensors each for soil water content (left) and soil matric potential.\protect\footnotemark~The colored lines show each sensor's readings, buried at different depths.
    The port\_num (right) identifies each sensor.
    }
    \label{fig:Results-4.1-Team-Farm-OG-GM-Why-should-I-graph}
\end{figure}

\footnotetext{``\textit{Soil matric potential represents the change in energy state of soil water relative to a reference, and it is the component of water potential attributed to the effects of capillary and adsorptive forces acting between liquid, gas, and solid phases}''~\cite{or2005soil}.}

%% file: figure/TeamFarm-Figures/Results-Original-GM-Fix-Team-Farm-Tooltip.tex
\begin{figure}[h]
    \centering
    % \begin{subfigure}[b]{0.4\textwidth}
    %     \centering
    %     \includegraphics[width=\textwidth]{assets/Team-Farm-Assets/Team_Farm_More_Info_Pre_Tooltip.pdf}
    %     %\caption{AI Information (No Tooltip)}
    %     %\label{fig:Results-Original-GM-Fix-Team-Farm-Tooltip-Before}
    % \end{subfigure}
    % \hfill
    % \begin{subfigure}[b]{0.54\textwidth}
    %     \centering
    %     \includegraphics[width=\textwidth]{assets/Team-Farm-Assets/Team-Farm-Fix-Assets/Team_Farm_More_Info_Post_Tooltip.pdf}
    %     %\caption{AI Information (With Tooltip)}
    %     %\label{fig:Results-Original-GM-Fix-Team-Farm-Tooltip-After}
    % \end{subfigure}
    % \includegraphics[width=0.7\linewidth]{assets/Team-Farm-Assets/Team-Farm-Fix-Assets/Team_Farm_More_Info_PreAndPost_Tooltip.png}
    \includegraphics[width=0.7\linewidth]{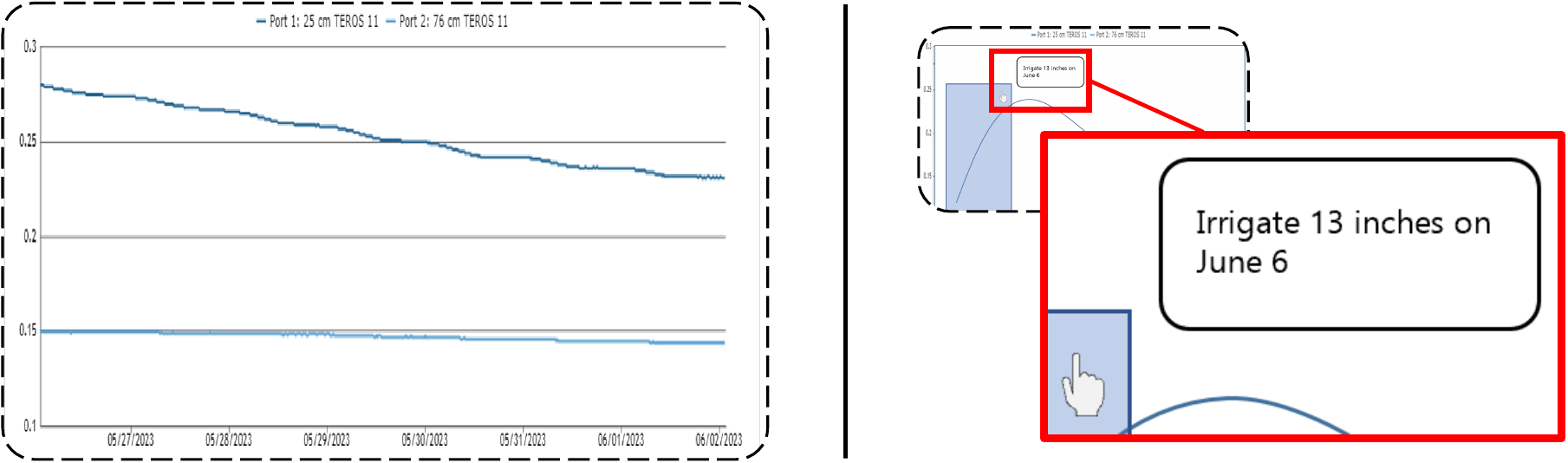}
    \caption{An \bugWhyShouldI{} fix (Team Farm).
    \textbf{(Left)}: Pre-fix, the information did not make clear why Abi should engage with the decreasing sensor readings (risk-aversion).
    \textbf{(Right)}: 
    Team Farm's fix aimed to reduce Abi's time cost to scan the graph (blue rectangle) by making clear why they should---to irrigate the field (the tooltip).
    }
    \label{fig:Results-Original-GM-Fix-Team-Farm-Tooltip}
\end{figure}

%% file: figure/TeamFarm-Figures/Results-4.1-Team-Farm-OG-GM-AI-More-Info.tex
\begin{figure}[h]
    \centering
    \includegraphics[width = .75\linewidth]{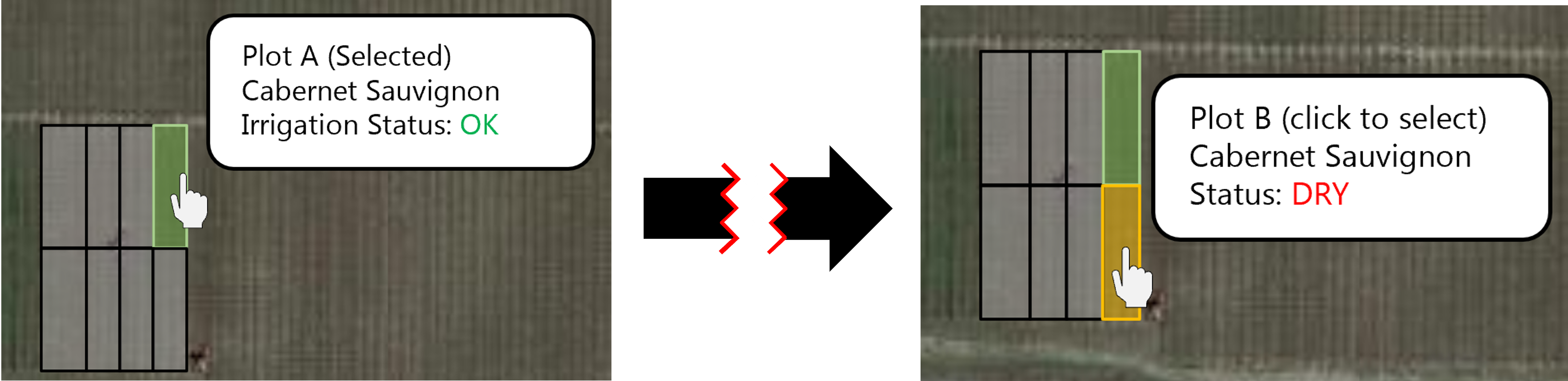}
    \caption{An \bugMoreInfo{} inclusivity instance (Team Farm).
    The team wanted Abi to see an important need for action by transitioning from the ``OK'' Plot A (left) to the {\color{red}DRY} Plot B (right).
    However, the information was insufficient (i.e., just ``OK''), preventing Abi from forming this subgoal (broken arrow).
    %The team noticed that although the AI had classified Plot A as {\color{teal}OK}, ``OK'' was not enough information about Plot A's ``OK-ness'' for Abi's comprehensive information processing style to navigate away to another plot.
    }
    \label{fig:Results-4.1-Team-Farm-OG-GM-AI-More-Info}
\end{figure}

%% file: figure/TeamFarm-Figures/Team-Farm-Fix-Figures/Team-Farm-AI-Actionable.tex
\begin{wrapfigure}{R}{\wrapFigWidth{}}
% \begin{figure}[h]
    \centering
    \includegraphics[width=0.4\linewidth]{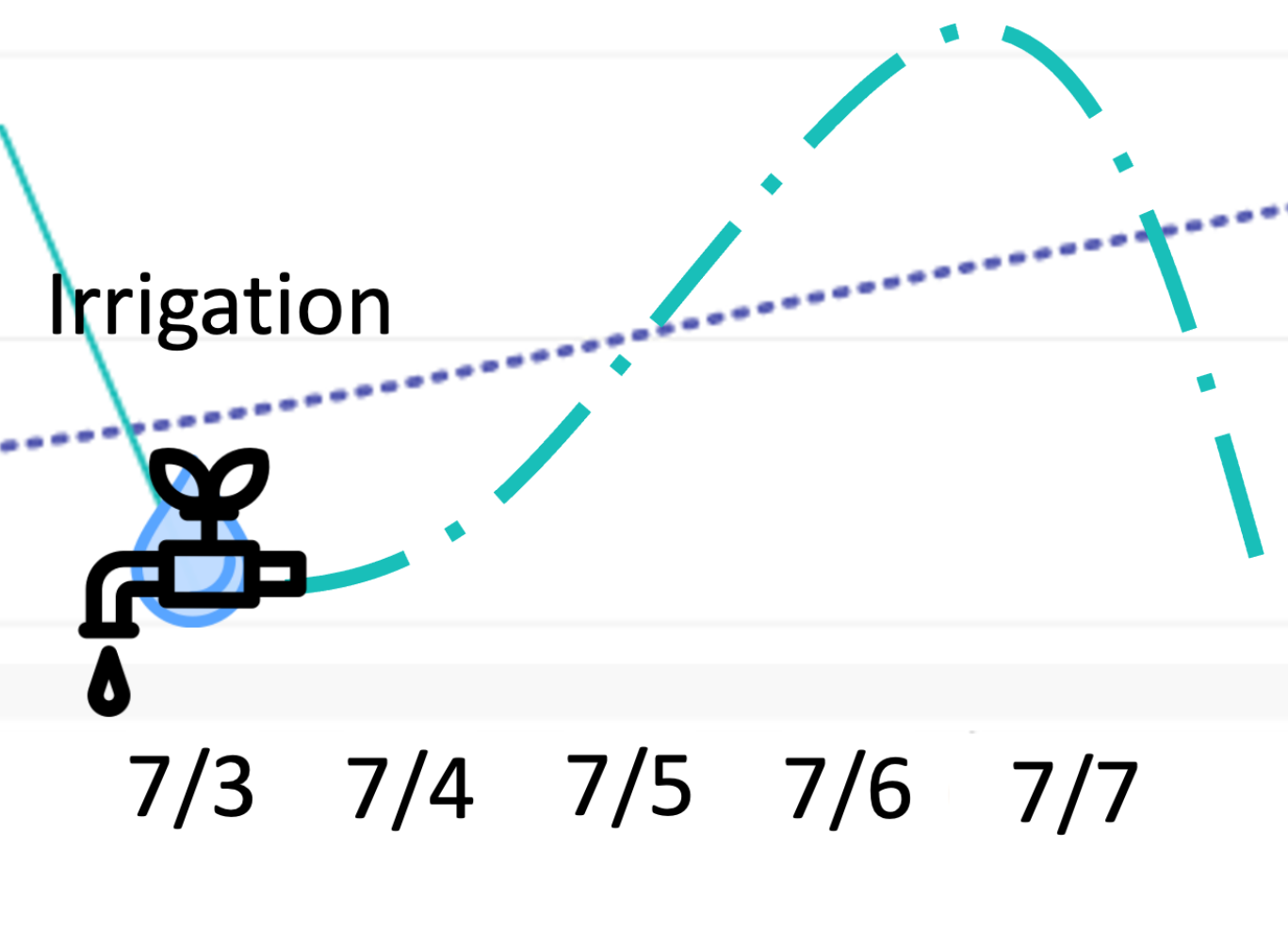}
    \caption{An \bugActionable{} fix (Team Farm).
    Team Farm added information that told the user what to \textit{do} with the AI's information.
    They added an icon to reflect \textit{when} Abi should irrigate (x-axis) and showed the outcomes of irrigating (dashed line, y-axis).}
    \label{fig:Team-Farm-AI-Actionable}
% \end{figure}
\end{wrapfigure}

%% file: figure/TeamGame-Figures/Results-Original-GM-Fix-Team-Game-Tie-Arrows.tex
\begin{figure}[h]
    \centering
    \includegraphics[width = 0.6\linewidth]{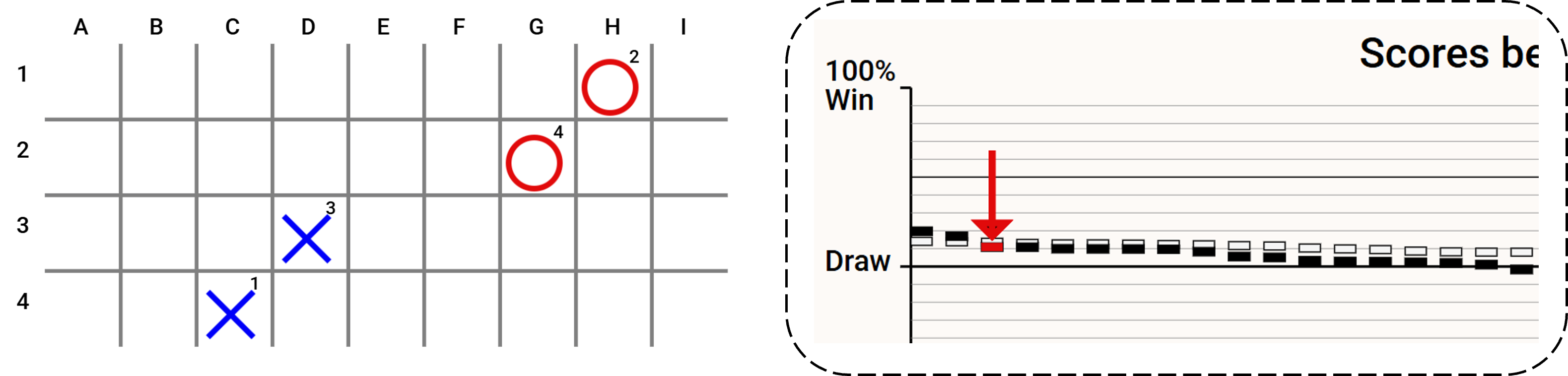}
    \caption{Each time a move is made by the AI players, the score of that move is temporarily highlighted with an arrow in the explanations. On the right, \BTW{} shows how \oPlayer{red} scored on move 4.}
    \label{fig:Results-Original-GM-Fix-Team-Game-Tie-Arrows}
\end{figure}

%% file: tables/GM_Fixes/Results_Fixes_by_Bug.tex
\begin{table}[h]
    \centering
    \begin{tabular}{l|cc|cc|cc}
    \toprule
     \multirow{2}{*}{\textbf{AI Inclusivity Bug}}    & \multicolumn{2}{c|}{\textbf{\# Bugs Found}}    & \multicolumn{2}{c|}{\textbf{\# Bugs Fixed}} & \multicolumn{2}{c}{\multirow{2}{*}{\textbf{\% Bugs Fixed}}} \\
                                            & \multicolumn{2}{c|}{\textbf{(\% Total Found)}} & \multicolumn{2}{c|}{\textbf{(\% Total Fixed)}} &  \\
    \midrule
            {\bugInterpret{}}        & 27 & (33\%)        & 14 & (30\%) & 14/27 & (52\%)  \\
                                                    %& (33\%)    & (30\%) \\
            \hline
            {\bugInputOutput{}}      & 9 & (11\%)      & 7 &  (15\%)  & 7/9 & (78\%) \\
                                                    %& (11\%)    & (15\%) \\
            \hline
            {\bugWhyShouldI{}}       & 19 & (23\%)        & 10 & (21\%)  & 10/19 & (53\%) \\
                                                    %& (23\%)    & (21\%) \\
            \hline
            {\bugMoreInfo{}}       & 9 & (11\%)       & 4 & (9\%) & 4/9 & (44\%)\\
                                                    %& (11\%)    & (9\%) \\
            \hline
            {\bugActionable{}}       & 12 & (14\%)       & 8 & (17\%) & 8/12 & (67\%)\\
                                                    %& (14\%)    & (17\%) \\
            \hline
            {\bugChanges{}}       & 7 & (8\%)      & 4 & (9\%) & 4/7 & (57\%) \\
                                                    %& (8\%)    & (9\%) \\
            \hline
            \hline
            \multicolumn{1}{r|}{Totals:}  & 83 &           & 47 &  & 47/83 & (57\%)\\
    \bottomrule
    \end{tabular}
    \caption{The AI inclusivity bug instances teams found and fixed, by type.
    The teams fixed 47/83 bug instances, fixing each bug type at similar rates.}
    \label{tab:Results_Fixes_by_Bug}
\end{table}

%% file: figure/Results-FixesWorked-EmpiricalResults.tex
\begin{figure}[h]
        \centering
        \includegraphics[width = 0.4\linewidth]{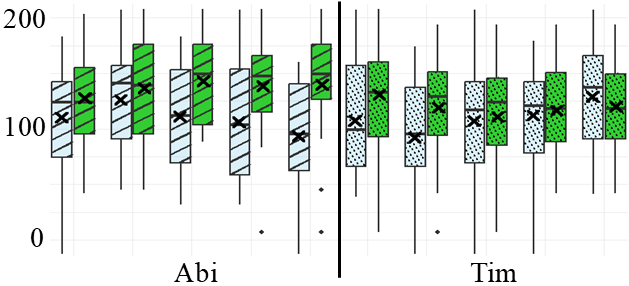}
    \caption{
    From Hamid et al.~\cite{hamid-2024-tiis}, Mental Model Concept Score split by prototype version and participants with an Abi-like or Tim-like problem-solving style.
   \coloredText{MMH-GM}{white}{Post-GenderMag} Abi (striped) and Tim (dotted) participants had higher (better) mental model concepts scores than their \coloredText{MMH-Original}{black}{Original} counterparts. The five problem-solving styles (from left to right): Info, Learn, Motiv, Risk, and SE.
    % From Hamid et al.~\cite{hamid-2024-tiis}.
    % Proportion of \coloredText{MMH-Original}{black}{Original} (n=34) and \coloredText{MMH-GM}{white}{GenderMag-Enhanced} (n=35) participants who mentioned a concept (y-axis) at least once.  
    % Correct concepts (above the dashed line) increased their mental model score, and incorrect concepts decreased it.
    \FIXME{MMB: June 27, 2025}{If THIS paper gets accepted, we need to get reprint permission from TiiS to reprint this figure from his (hopefully) TiiS'25 paper.}
    }
    \label{fig:Results-FixesWorked-EmpiricalResults}
\end{figure}

%% file: doc/06-Results-GenderMag-for-AI.tex
\section{Results RQ3 \& RQ4: Was GenderMag ``Enough''?
\draftStatus{MMB}{2.4 9/27/2025}}
\label{sec:GM-for-AI}

\subsection{RQ3: The \OriginalGM{} and AI \draftStatus{MMB 10/9/25}{top is D3 :-)}}
\label{subsec:RQ3_consider_failure}

\boldify{So they were able to find bugs, yay! But was GenderMag enough for AI? (RQ3) No, they did not consider what could go wrong.}

As the previous section shows, using GenderMag enabled the AI product teams to find inclusivity bugs, and fix them in ways that significantly improved at least one of the products (Team Game's).
%Each team started the field study applying the \OriginalGM{}, an off-the-shelf version of the GenderMag walkthrough, to their AI products.
But only some of these effects came from teams' use of the GenderMag variant we term \OriginalGM{}, and Table~\ref{tab:results_adapt_gm_original_sessions} considers just these.
As the table shows, the teams' use of \OriginalGM{} did enable them to be effective---they found 54 \aiInclBug{s} using it.
However, the table also shows an important omission in the teams' reasoning. 
Although the teams found 54 AI inclusivity bugs with the \OriginalGM{}, fixing 34 of them, their transcripts showed that \textit{none} of them considered the possibility of the persona not believing the AI's outputs.
%\footnote{In software~\cite{kumar2017investigation}, failures are incorrect behavior with respect to expected behavior.
%AI failure outcomes can range from the innocuous (e.g., omitted information, unexpected recommendation) to deadly (e.g., crashed self-driving car).
%We use the umbrella term ``AI failure'' to capture all such outcomes.}

\input{tables/GM_adapt/results_adapt_gm_original_sessions}

\boldify{So, how should we GenderMag an AI's action if Abi thinks the AI did wrong?}

RQ3 asked whether \OriginalGM{} was ``enough'' for the teams' effectiveness, or whether something AI-specific is needed. 
The table suggests that something more AI-specific is needed, because the teams consistently overlooked situations in which the user was unconvinced by the AI.
% Since AI products are non-deterministic, they always have a possibility of failing.
% One area where AI products are more likely to fail is when they encounter Out of Distribution (OOD) situations, where the scenario is significantly different from training data, and detecting such situations an active area of research (e.g.,~\cite{yang2024generalized,mohseni2020self,graham2023denoising}).
%
Given this omission, we invited teams to suggest whether and how to adapt the \OriginalGM{} to better fit AI products.
Teams gave five ideas.

The teams had time to try only two of them, which we describe in the upcoming subsections; the rest of the ideas are enumerated in the supplemental documents. 
We refer to all five adaptations as GenderMag-for-AI variants.

% \boldify{We explore the implemented adaptations to the GenderMag walkthrough in Sections~\ref{subsec:RQ4_initial_change_suggest}--\ref{subsec:RQ4_pre_action_fork}, and discuss their performance.}

\subsection{RQ4: Seeds of Change---Team Game's Trees
\draftStatus{FAM}{** 10/4/25}}
\label{subsec:RQ4_initial_change_suggest}

\boldify{Team Game discussed what made evaluating AI products different (i.e., AI can be wrong wrong) and how would they change the original GenderMag.}
After Session 5, we asked Team Game how they would change the \OriginalGM{} for AI products.
They discussed what made AI products different, concluding that the walkthrough should consider when the AI is wrong (further supporting RQ3):
\quotateInsetAdaptGM
%Note, this is for when they're adapting GenderMag. They're NOT finding bugs. They're discussing the process.
%Team Name
{Game}
%Quote Snippet
{With a [traditional] UI, when something goes wrong, it's either the user's fault or the interface's fault, and the job of this walkthrough is essentially to parse that out...
% And generally speaking, it's going to be like nothing was the user's fault. 
With AI systems, there is a third possibility, and that is that the AI is wrong. 
% And so in this case, you could imagine this case where the agent is not highlighting the second best move as the second best move, that that could be an interface problem. 
% That could be a problem exists between chair and keyboard problem. 
% Or that could be a you know, the AI is just throwing bad values out.
... %there's sort of a third eventuality, 
we need to think about a bit, and I don't know that we've thought that through...}
%Arg3 - Walkthrough Version
{\OriginalGM{}}
%Session (and step) where the comment occurred
{061423-Step2-MNK-Find-2-PV.txt 1:49:59
}
%Full quote (from the form), so we don't have to keep going back to excel
{Ordinarily, when you're doing this on kind of a what I'll say, a non learned function UI, when something goes wrong, you know, it's either the user's fault or the interfaces fault. And the job of this walkthrough is essentially to parse that out. And generally speaking, it's going to be like nothing was the user's fault. But when we deal with AI systems there is a third possibility, and that is that the AI is wrong. And so in this case, you could imagine this case where the agent is not highlighting the second best move as the second best move, that that could be an interface problem. That could be a problem exists between chair and keyboard problem. Or that could be a you know, the AI is just throwing bad values out. So there's sort of a third eventuality, we need to think about a bit, and I don't really know that we've thought that through perfectly.
}

\boldify{They proposed adding a tree-like structure to GenderMag's walkthrough.}
\input{figure/Walkthrough-Adaptation-Figures/Team_Game_Branching_Example}

Team Game suggested changing the \OriginalGM{}'s structure to consider the above:
\quotateInsetAdaptGM
%Note, this is for when they're adapting GenderMag. They're NOT finding bugs. They're discussing the process.
%Team Name
{Game}
%Quote Snippet
{...the [GenderMag] document is linear, [but] the goal structure is a tree...}
%Arg3-Walkthrough Version
{\OriginalGM{}}
%Session (and step) where the comment occurred
{061423-Step2-MNK-Find-2-PV.txt 1:49:59
}
%Full quote (from the form), so we don't have to keep going back to excel
{Another thing I'll highlight is it struck me that we had this kind of, we have this kind of the document is linear, the goal structure is a tree. And there's a little bit poor mapping there in that it would be nice if we could finish a sub goal, leave it and return to the previous sub goal. But under a linear structure that's very difficult to do. Some kind of like nesting might be good, but I don't really know the right way to do that.
}

% \boldify{\textit{How} did we implement this? Figure~\ref{fig:Team_Game_Branching_Example} shows the simplest way to incorporate these two suggestions simultaneously, and it's additive to the Original!}
\boldify{Figure X shows how we interpreted Team Game's suggestion, enabling teams to explore different scenario within the GenderMag walkthrough.}

Figure~\ref{fig:Team_Game_Branching_Example} shows a conceptual view of Team Game's suggestions. 
This tree's two forks could provide teams the opportunity to explore different scenarios.
% The left fork would explore what happens when the AI behaves as expected (i.e., no AI failure).
The left fork would explore what happens when the persona believes the AI.
% The right fork explores the opposite, when the AI behaves unexpectedly (i.e., AI failure).
The right fork explores the opposite, when the persona doubts the AI.
We considered 3 possible places to introduce the forking structure---at the subgoal, the pre-action, and/or the post-action steps (recall Figure~\ref{fig:Original_GenderMag_Flowchart}).
The teams explored two of the possibilities, as we describe next.

%-----------------------------
\subsection{RQ4: The \postForkGM{} \draftStatus{FAM}{2.3 **section may be deleted for space** 9/26/25}}
\label{subsec:RQ4_post_action_fork}

\boldify{We started by forking the post-action question, and we were careful with our wording---we didn't want teams saying ``oh of course it's correct. It's obvious for me.''}

We began by implementing Team Game's suggestions at the last step of the \OriginalGM{}, creating the \postForkGM{} with the workflow shown in Figure~\ref{fig:PostActionFork_GenderMag_Flowchart}.
Like the other GenderMag walkthrough questions, we framed the new \postForkGM{} questions in terms of the \textit{persona's} beliefs---here, of whether the AI is right or wrong---not the \textit{developer's} belief.
%We sought to help teams avoid walkthrough pitfalls similar to the  ``I'' methodology pitfall documented in prior GenderMag literature~\cite{oleson2018pedagogical,hill2016gendermag,shekhar2018cognitive}.

\input{figure/Walkthrough-Adaptation-Figures/PostActionFork_GenderMag_Flowchart}

\boldify{Team Farm used this version in Session 6, and it went badly. Team Farm didn't think that this fork was the right question, suggesting another ``understandability'' question to close a gap.}

% This eventually led Team Farm to question how appropriate the post-action question was when they were ready to consider the first fork of the walkthrough.
Team Farm tested the \postForkGM{}, and they found it confusing.
Their primary feedback to this adaptation was that they thought that this was not the right question to ask at this stage, instead, suggesting adding a question between the pre-action and post-action questions:
\quotateInsetAdaptGM
%Note, this is for when they're adapting GenderMag. They're NOT finding bugs. They're discussing the process.
%Team Name
{Farm}
%Quote Snippet
{...I feel like that's the wrong question to ask. Would it be more pertinent to know if Abi understands what they're seeing first and then whether or not they believe it's correct?}
%Arg3 - Walkthrough Version
{\postForkGM{}}
%Session (and step) where the comment occurred
{061423-Step2-MNK-Find-2-PV.txt 1:49:59
}
%Full quote (from the form), so we don't have to keep going back to excel
{Can I say that I feel like that's the wrong question to ask?
Would it be more pertinent to know if Abi understands what they're seeing first and then whether or not they believe it's correct?
}

% \boldify{This ``post-action fork'' version of the GenderMag walkthrough was so confusing that Team Farm reverted to the Original GM walkthrough and finished their goal (no more sessions).}
% Team Farm were so confused by the \postForkGM{} that they elected to revert to applying the \OriginalGM{}.
% Although they still did not consider perceived AI failures, Team Farm found an additional 12 AI inclusivity bugs in this session.
%Team Farm accomplished Abi's goal with the \OriginalGM{}, and they did not wish to go through another round of GenderMag sessions.

\boldify{However, before Team Farm left, we had a chance to discuss their changes.}

The team reverted back to \OriginalGM{}.
Although they still did not consider whether users may doubt the AI, Team Farm found an additional 12 more AI inclusivity bugs and completing their goal.
In discussing possible GenderMag walkthrough changes for AI products, the topic of potential cognitive taxes came up~\cite{hilderbrand2020engineering}.
Team Farm acknowledged this potential tax, reconsidering their initial suggestion in light of these taxes:
\quotateInsetAdaptGM
%Note, this is for when they're adapting GenderMag. They're NOT finding bugs. They're discussing the process.
%Team Name
{Farm}
%Quote Snippet
{...make it less taxing on the  people evaluating...I think it is okay to skip the step of `does Abi understand that information'...I don't really like the question. I get the point. I just don't know how to change it...}
%Arg3 - Walkthrough Version
{\postForkGM{}}
%Session (and step) where the comment occurred
{062723-Step2-DemoFarm-Find-3-RG.txt 48:30
}
%Full quote (from the form), so we don't have to keep going back to excel
{Yeah, I see what you're saying to try to make it less taxing on the on the people evaluating. I think it is important to ask that step somewhere, but I think it is okay to skip the step of does Abi understand that information and, and force that information into their head.
Yes. Uh huh. And especially I think, then you will put in a lot of weight on how good the model or how bad the model is itself, when you shouldn't, because how good or bad the model is, is not related to the user experience. So it's, there are kind of contrast. And so that's why... I don't really like the question. I get the point. I just don't know how to change it either. Jake, any any ideas of how to change it?
}

%Their last thoughts on the \postForkGM{} were negative, but they did not provide additional adaptation suggestions:
% At the end of the session, Team Farm agreed that adding an additional ``understandability'' question might have imposed 
% We revisit Team Farm's ``understandability'' suggestion in Section~\ref{subsec:Adapt_gm_discussion} to discuss its implications.

% \subsubsection{Mini-Discussion: Why Did The \postForkGM{} Fail This Team?}

% \input{tables/GM_adapt/results_adapt_gm_postfork_sessions}

%-----------------------------
\subsection{RQ4: The \preForkGM{}\draftStatus{FAM}{2.2 10/8/25}}
\label{subsec:RQ4_pre_action_fork}

% \textcolor{red}{**FAM: When referring to the phenomena, make the language about assumptions clear and how the fork brings accuracy e.g.,:\\
% ``by making it explicit we find that we're at least as good...''
% ``now consider the AI failures, now they find brand new bugs they never would have found if they hadn't considered that explicit assumption of perceived AI failure''}
% **FIXME** While the post fork did not work, we wanted to check if a different area would work but it wasn't a guarantee due to cognitive tax concerns...**

\boldify{We went back to Team Game's original suggestion and tried a different area for the fork (pre-action).}

Given that the fork in the post-action did not work for Team Farm, we went back to Team Game's initial suggestions by forking at the pre-action step.
This resulted in the \preForkGM{} with the workflow shown in Figure~\ref{fig:PreActionFork_GenderMag_Flowchart}.

\input{figure/Walkthrough-Adaptation-Figures/PreActionFork_GenderMag_Workflow}

\boldify{This change actually uncovered an added bonus to this change---more flexibility in what the persona would do.}

%giving teams the freedom to adapt the action that they would want a user to take, depending on how the AI behaved.

A difference in the \preForkGM{} from the other variants was in how many actions the teams could evaluate (Figure~\ref{fig:Results-Adapt-GM-PreAction-Forms-Blank}).
With the \OriginalGM{} (left), teams selected only a single action to evaluate in the pre-action step.
However, the \preForkGM{} (right) prodded the teams to set \textit{two} actions: one when Abi believed the AI's previous output (\coloredText{AI-Right}{black}{Pre-Action}) and one when Abi did not (\coloredText{AI-Wrong}{black}{Pre-Action}).
Once the teams evaluated both actions, they chose which one to evaluate in the post-action step. 
% With potentially divergent actions in this adaptation, we changed the procedure for the post-action step, making teams pick one of the actions to force Abi to perform.

\input{figure/Walkthrough-Adaptation-Figures/Results-Adapt-GM-PreAction-Forms-Blank}

\boldify{Both Team Weather and Team Game tested the \preForkGM{}, and when they pursued the \noFailureAI{AI-Right}{black} fork, both \textit{still} found \aiInclBugInst{}s (like in original GenderMag).}

Both Team Weather and Team Game tested the \preForkGM{}, and noticed that the  \noFailureAI{AI-Right}{black} fork performed similarly to the \OriginalGM{} pre-action step. 
Because of this, teams' ability to find \aiInclBug{}s as they had with \OriginalGM{} did not change when the persona did not doubt the AI's output.
But now that they also considered situations when the persona did doubt the AI's output, teams also found new, different bugs compared to the \noFailureAI{AI-Right}{black} fork.

For example, Team Game wanted Abi to take the same action as the \noFailureAI{AI-Right}{black} fork, revisiting the \xPlayer{black}'s previous move, but they found \textit{different} \aiInclBugInst{}s than those in the \noFailureAI{white}{black} fork (Figure~\ref{fig:Results-Adapt-GM-FindBugs-NoFailure}). 
By considering both forks, teams found \aiInclBug{}s that they could not have uncovered with the \OriginalGM{}.

\quotateInsetFindBugsWithVersion
%Note, this is for when they're adapting GenderMag. They're NOT finding bugs. They're discussing the process.
{single}
%Arg1-Team Name
{Game}
%Arg2-Quote Snippet
{Something has gone wrong, and Abi will want to validate why something went wrong before losing the context...}
%Arg3-Bug identified
{\bugInterpret{}}
%Arg4-Problem-solving Style used
{Comprehensive Info. Proc.}
%Arg5-GenderMag Walkthrough version
{Pre-Action Fork Version}
%Arg6-Which Fork
{\failureAI{AI-Wrong}{black}}
%Arg6-Session/transcript/timestamp
{session 7 **FIXME**}
%Arg7-Whole quote
{Because Abi’s attitude toward risk, something has gone wrong and Abi will want to validate why something went wrong before losing the context needed to understand things. With computer self-efficacy reasoning, Abi may blame themselves and feel disheartened and give up because they are not sure what to do at this point and may abandon the system.}

\input{figure/Walkthrough-Adaptation-Figures/Results-Adapt-GM-FindBugs-NoFailure}

\boldify{\textcolor{red}{*FAM@FAM fix flow here**} Teams even made different decisions in the forks, finding a bug on one branch and not the other and vice-versa.}
% \boldify{When Team Game pursued the different forks (i.e., \noFailureAI{AI-Right}{black} and \failureAI{AI-Wrong}{black}), they gave different answers.  This meant that they discovered \aiInclBugInst{}s they would not have uncovered before.}

Team Game chose to always keep Abi's action the same when evaluating both forks, but they gave different answers for whether Abi would take the action or not in each fork.
For instance, when evaluating the \noFailureAI{AI-Right}{black}, Team Game decided that Abi would \textit{not} take the intended action, because Abi's comprehensive information processing style was satiated, so they would not know why they should:

\quotateInsetFindBugsWithVersion
%Note, this is for when they're adapting GenderMag. They're NOT finding bugs. They're discussing the process.
{single}
%Arg1-Team Name
{Game}
%Arg2-Quote Snippet
{The reason I say [no] is, if Abi is already happy with the thing, there's no real reason to be clicking on this button... they have no reason to want this information if they're happy.}
%Arg3-Bug identified
{\bugWhyShouldI{}}
%Arg4-Problem-solving Style used
{Comprehensive Info. Proc.}
%Arg5-GenderMag Walkthrough version
{Pre-Action Fork Version}
%Arg6-Which Fork
{\noFailureAI{AI-Right}{black}}
%Arg6-Session/transcript/timestamp
{**FIXME**}
%Arg7-Whole quote
{The reason why I say this is, if Abi is like, already happy with the thing, and there's no real reason to be clicking on this button. Other than kind of, as mentioned, if they were interested in like, investigating rewind controls, or alternatively, kind of a somewhat corollary argument would be that it's clearly the last button on the interface. And so if they've kind of, like, figured this out, then they've comprehended the whole thing. So I could imagine this sort of comprehensiveness taking over, in addition to just sort of curiosity about time controls, but other than those two reasons, like they have no reason to want this information if they're happy.}

However, while evaluating the other fork (\failureAI{AI-Wrong}{black}), Team Game also gave a different answer.
They said that since Abi did \textit{not} believe the AI's previous output, they would now take the action, providing a different problem-solving style value to explain it:

\quotateInsetFindBugsWithVersion
%Note, this is for when they're adapting GenderMag. They're NOT finding bugs. They're discussing the process.
{single}
%Arg1-Team Name
{Game}
%Arg2-Quote Snippet
{...I'm a yes here... when things go south, Abi is going to be task-oriented and create a task for themselves ...to figure out what went differently from expectation.}
%Arg3-Bug identified
{None}
%Arg4-Problem-solving Style used
{Task-oriented motivations}
%Arg5-GenderMag Walkthrough version
{Pre-Action Fork Version}
%Arg6-Session/transcript/timestamp
{**FIXME**}
%Arg7-Whole quote
{I think I'm a yes here. Think that when things go south, you know, Abi is going to be task oriented and create a task for themselves of trying to figure out what went differently from expectation.}

% -------

\boldify{The \preForkGM{} was a success, enabling a developers to consider such scenarios when they have not before.}
%Thus \preForkGM{} was the most successful GenderMag-for-AI variant (Table~\ref{tab:results_adapt_gm_prefork_sessions}) because it enabled developers to consider when users may doubt the AI when they might not have before.
For Team Weather, \preForkGM{} revealed a different kind of bug. 
It revealed that Team Weather's prototype designers had overlooked the possibility that Abi might doubt the AI's output, Abi's doubts brought about an abrupt dead-end---there was no action in the interface for Abi to take in such a scenario:

\quotateInsetAdaptGM
%Note, this is for when they're adapting GenderMag. They're NOT finding bugs. They're discussing the process.
%Team Name
{Weather}
%Quote Snippet
{If Abi is not satisfied with the output, I don't know what I would expect from Abi...That's not-- I haven't looked at it that way before. It's not the point of view that we would, or that I would typically try to put myself into. The assumption is that you believe me.}
%Arg3-The walkthrough version they observed things with
{\preForkGM{}}
%Session (and step) where the comment occurred
{061423-Step2-MNK-Find-2-PV.txt 1:49:59
}
%Full quote (from the form), so we don't have to keep going back to excel
{But no, really, if Abi is not satisfied with the output, I don't know what I would expect from Abi to-- as far as that goes.
That's not-- I haven't looked at it that way before. So if you have some thoughts or guidance on how to really take that that point of view, I think that'd be useful. Yeah.
Absolutely, it's not the point of view that we would, or that I would typically try to put myself into. The assumption is that you believe me.
}

\boldify{To ensure rigor in our qualitative findings for RQ4, we used triangulation to }

Because the teams started trying \preForkGM{} well into the study period, it did not produce as much data as did their uses of \OriginalGM{}. 
Even so, the evidence of \preForkGM{}'s greater effectiveness over \OriginalGM{} for AI products is very consistent.
First, as Table~\ref{tab:results_adapt_gm_prefork_sessions} shows, the \textit{only} times any team considered the persona doubting the AI was when they used the \preForkGM{}.
%
%To ensure rigor in our qualitative findings for RQ4, we used triangulation which views findings from multiple angles to show whether independent sources of evidence cross-confirm the same conclusions ~\cite{yin2009caseStudies}.
Second, as Table~\ref{tab:Triangulation-Table} shows, 
9 sources of evidence over time, over different teams, and over different data sources (bugs found vs. verbalizations) cross- corroborate the teams' effectiveness when using \preForkGM{}. 

%coming from finding bugs in both forks, unique bugs (one bug in a fork and none in the other), and positive verbal reactions from teams---suggesting that \preForkGM{} was a useful improvement over \OriginalGM{} for AI products. 

\input{tables/GM_adapt/results_adapt_gm_prefork_sessions}

\input{tables/Triangulation-Table}

%% file: tables/GM_adapt/results_adapt_gm_original_sessions.tex
\begin{table}[h]
    \centering
    \begin{tabular}{cccccc}
    \toprule
      \textbf{Session} & \multirow{2}{*}{\textbf{Team}} & \textbf{Which} & \textbf{Considered} & \textbf{\# Bugs} & \textbf{\# Bugs}  \\
       \textbf{Number} & & \textbf{GenderMag?} & \textbf{``Doubts the AI''} & \textbf{Found} & \textbf{Fixed}  \\
    \midrule
       1  & Weather & \OriginalGM{} & No & 3 & 3  \\
       2 & Farm & \OriginalGM{} & No & 13 & 8   \\
       3 & Farm & \OriginalGM{} & No & 8 & 5  \\
       4 & Game & \OriginalGM{} & No & 5 & 3 \\
       5 & Game & \OriginalGM{} & No & 25 & 15 \\
       \hline
       \hline
         &      &          & \multicolumn{1}{c}{Totals:}& 54 & 34 \\
       % 6 & Farm & Post-Action Fork & No  & 12 & 6\\
       % 7 & Weather & Pre-Action Fork & Yes & 9 & 5 \\
       % 8 & Game & Pre-Action Fork & Yes & 8 & 2 \\
    \bottomrule
    \end{tabular}
    \caption{Teams' find sessions applying the \OriginalGM{}.
    All teams successfully found and fixed AI inclusivity bug instances, but \textit{none} of the teams considered whether the persona may doubt the AI.}
    \label{tab:results_adapt_gm_original_sessions}
    \FIXME{AAA}{Move session 5 above the line, get rid of the ``reviewed'' because message should be that they ran all these \OriginalGM{}.
    We want 6/7/8 elsewhere.
    It's a triangulation table. Three teams, 5 sessions, nobody considered AI failures.}
\end{table}

%% file: figure/Walkthrough-Adaptation-Figures/Team_Game_Branching_Example.tex
\begin{wrapfigure}{R}{0.32\linewidth}
        \centering
        % \begin{tabular}{c}
        %     % \includegraphics[width = .99\linewidth]{assets/adaptation_assets/Team_Game_Tree_Example.pdf}
        % \end{tabular}
        \includegraphics[width = .95\linewidth]{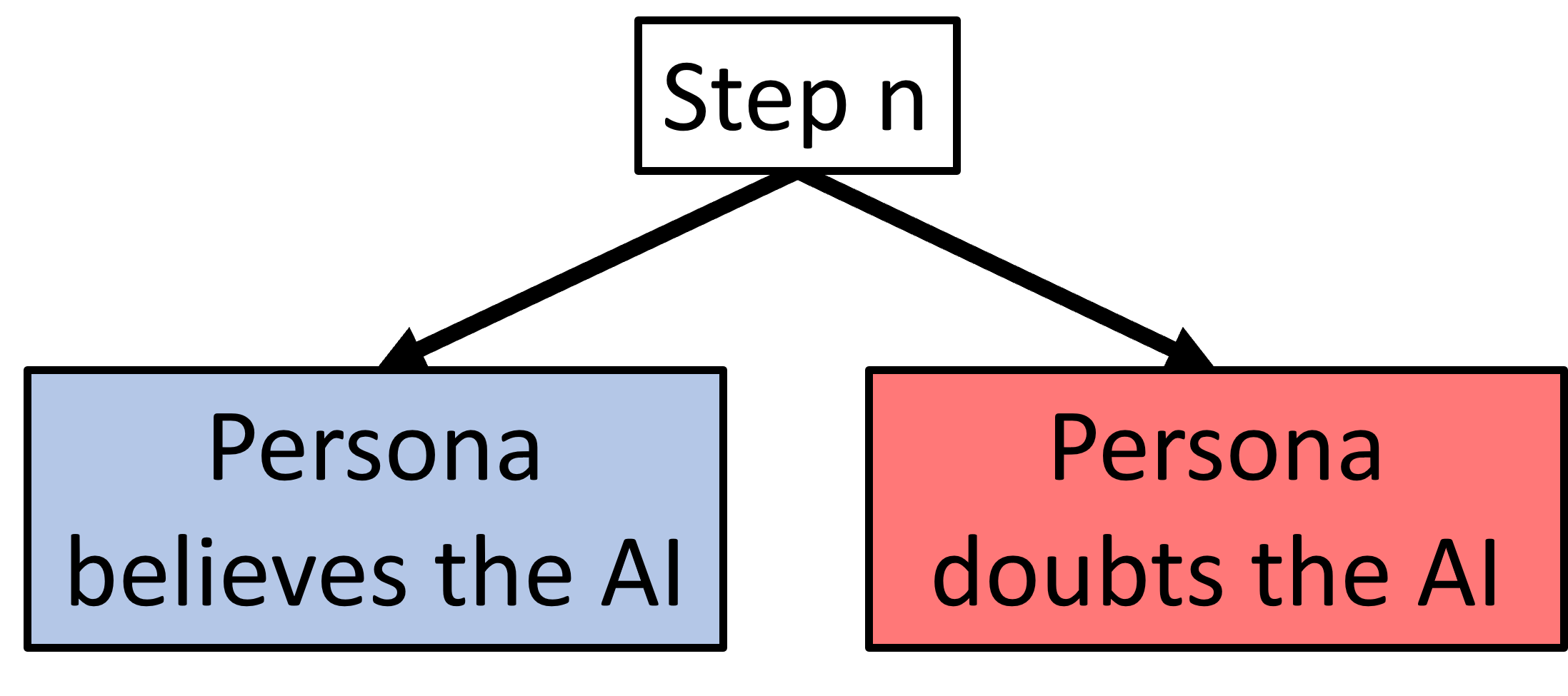}
        \caption{Team Game's suggestion of adding a tree structure.}
        \label{fig:Team_Game_Branching_Example}
\end{wrapfigure}

%% file: figure/Walkthrough-Adaptation-Figures/PostActionFork_GenderMag_Flowchart.tex
\begin{figure}[h]
    \centering
    \begin{tabular}{c}
    % \toprule
        \includegraphics[width=0.99\linewidth]{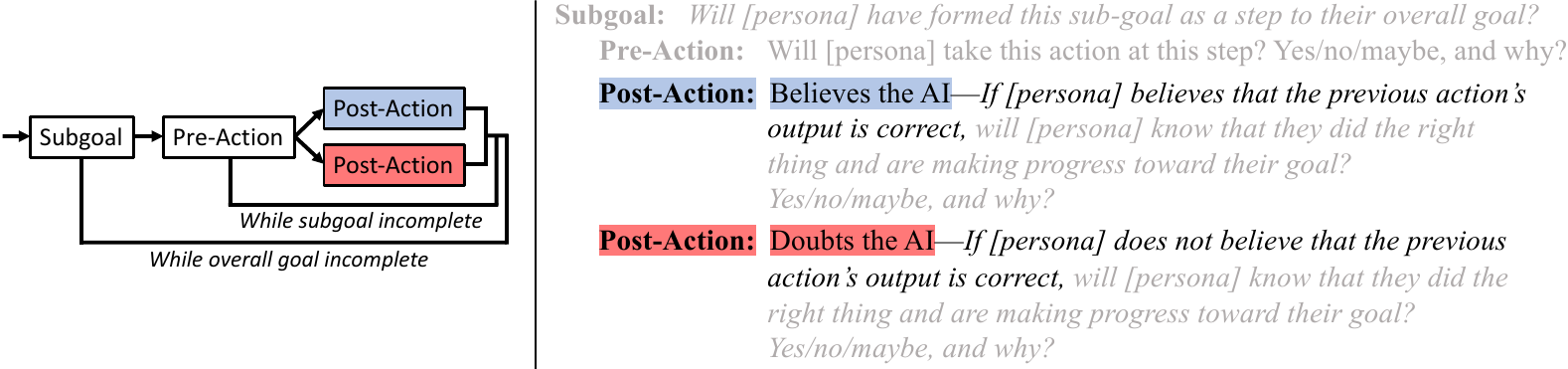}  
    \end{tabular}
    \caption{\textbf{(Left):} The \postForkGM{} graph, forking the post-action step for \noFailureAI{AI-Right}{black} and \failureAI{AI-Wrong}{black}.
    \textbf{(Right):} Question wording for each node in the graph.
    \textcolor{gray}{Gray text}: Unchanged from the Original.}
    \label{fig:PostActionFork_GenderMag_Flowchart}
\end{figure}

%% file: figure/Walkthrough-Adaptation-Figures/PreActionFork_GenderMag_Workflow.tex
\begin{figure}[b]
    \centering
    \begin{tabular}{c}
    % \toprule
        \includegraphics[width=0.99\linewidth]{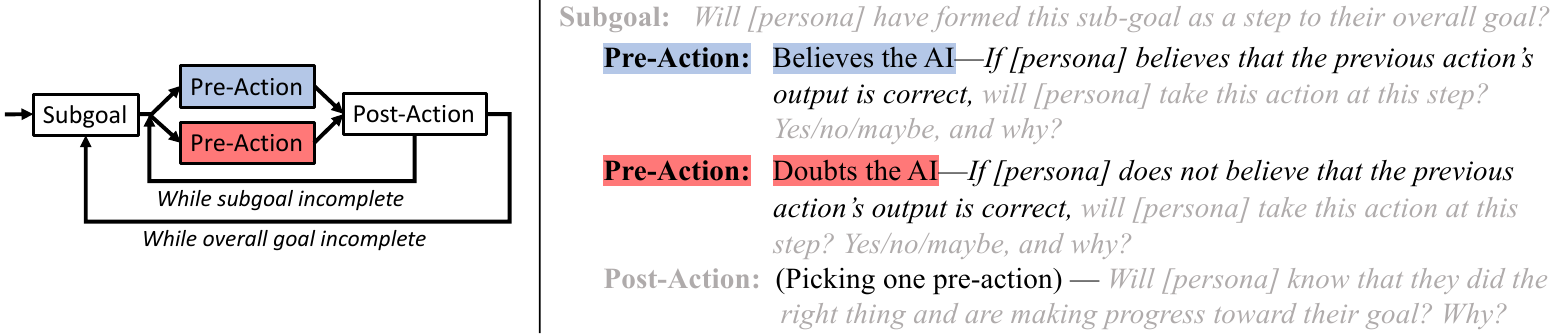}  \\
        % \midrule
        % \questionWordingForkPre{gray} \\
    % \bottomrule
    \end{tabular}
    \caption{\preForkGM{}. \textbf{(Left):} Graphical depiction, forking the pre-action step into 2 circumstances: when the persona believes the AI (\noFailureAI{AI-Right}{black}), and when they do not (\failureAI{AI-Wrong}{black}). \textbf{(Right):}  Question wording for each node. \textcolor{gray}{Gray text}: Unchanged from the Original.
   }
    \label{fig:PreActionFork_GenderMag_Flowchart}
\end{figure}

%% file: figure/Walkthrough-Adaptation-Figures/Results-Adapt-GM-PreAction-Forms-Blank.tex
\begin{figure}[b]
    \centering

    \includegraphics[width=\linewidth]{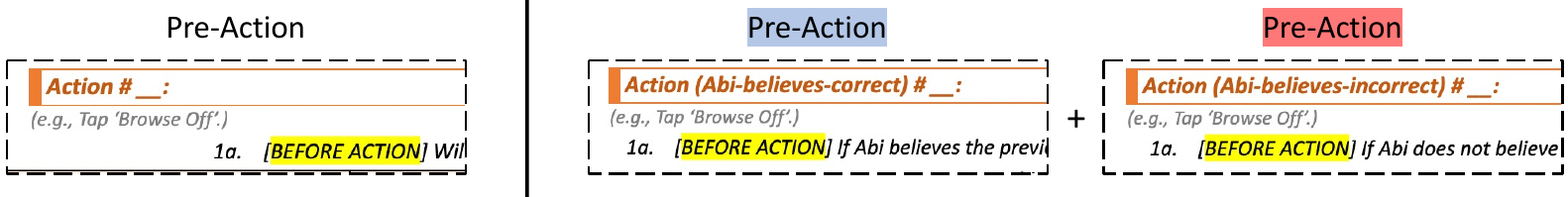}
    
    \caption{The pre-action fork form.
    \textbf{(Left):} For \OriginalGM{}, teams choose \textit{one} action to evaluate.
    \textbf{(Right):} In the \preForkGM{}, teams choose \textit{two} actions to evaluate: one when Abi believes the AI's output (\noFailureAI{AI-Right}{black}) and another when Abi does not (\failureAI{AI-Wrong}{black}).}
    % \textcolor{red}{FIXED! MMB@JN: (1) this fig's source doesn't seem to be in our directory -- pls upload it to the Google folder. (2) Let's go with the line.  But make the line a little longer (extending a few pixels BELOW the bottom of the others) and I think it'll work.}
    % 
    \label{fig:Results-Adapt-GM-PreAction-Forms-Blank}
\end{figure}

%% file: figure/Walkthrough-Adaptation-Figures/Results-Adapt-GM-FindBugs-NoFailure.tex
\begin{figure}[h]
    \centering
    % \begin{subfigure}{0.45\linewidth}
    %     % \includegraphics[width = 0.7\linewidth]{assets/Team-Weather-Assets/Results-PreFork-GM-Find-Weather-InterpretAI.pdf}
    % \end{subfigure}
    % \begin{subfigure}{0.45\linewidth}
    %     \includegraphics[width = \linewidth]{assets/Team-Game-Assets/Team-Game-Interface-Find-3-1830.png}
    % \end{subfigure}
    \includegraphics[width = 0.45\linewidth]{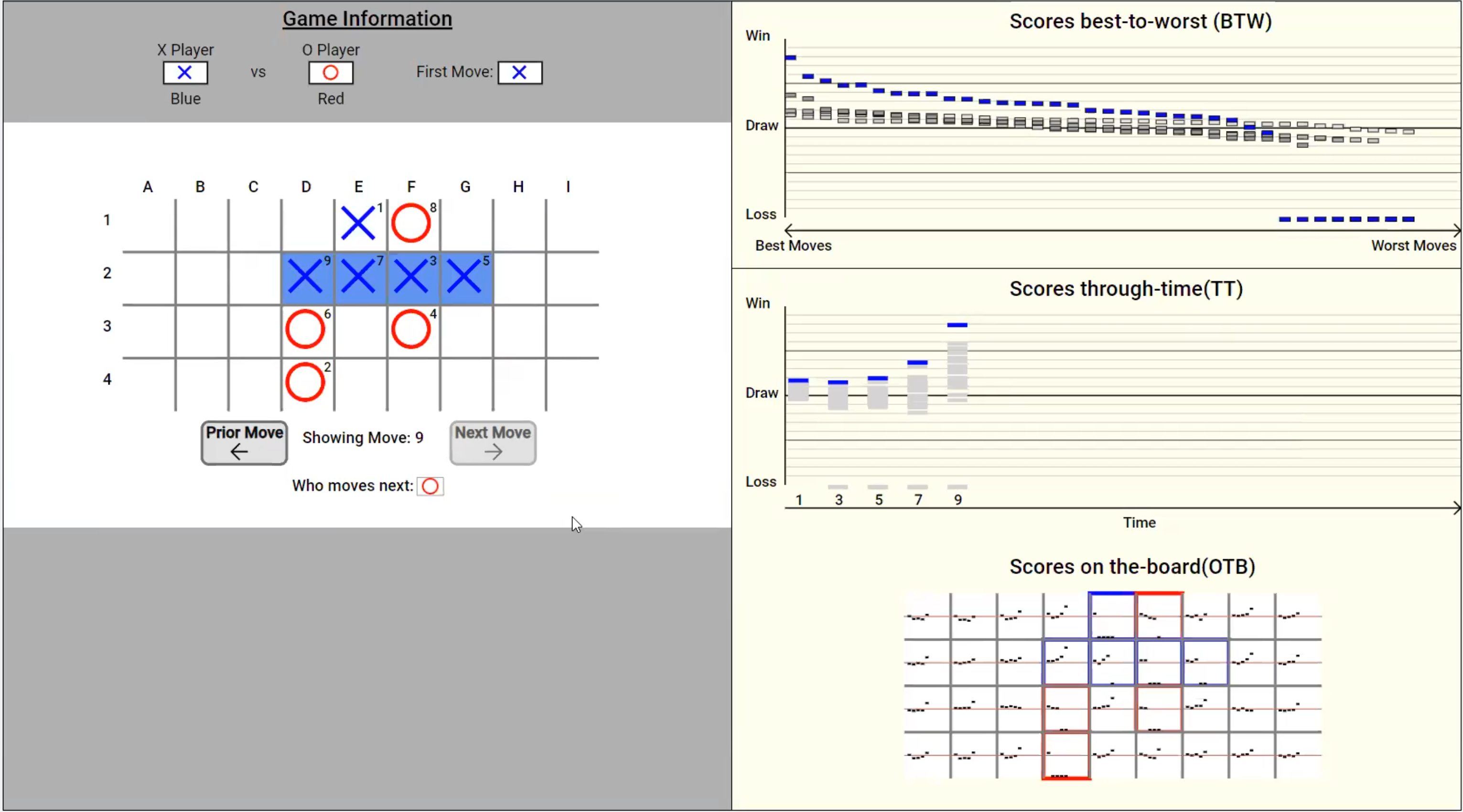}
    % \caption{Two \aiInclBugInst{}s that Team Weather (left) and Team Game (right) found in the \noFailureAI{AI-Right}{black} fork of the \preForkGM{}.
    % This fork made the lack of AI failures explicit (which teams did implicitly with the \OriginalGM{}), but this did not prevent teams from finding issues.
    % }
    \caption{Interface state when  Team Game was evaluating in the \failureAI{AI-Wrong}{black} fork of the \preForkGM{}. The game ended with \xPlayer{black} winning and the intended action was for Abi to go back to previous moves.}
    \label{fig:Results-Adapt-GM-FindBugs-NoFailure}
\end{figure}

%% file: tables/GM_adapt/results_adapt_gm_prefork_sessions.tex
\begin{table}[h]
    \centering
    \begin{tabular}{cccccc}
    \toprule
      \textbf{Session} & \multirow{2}{*}{\textbf{Team}} & \multirow{2}{*}{\textbf{Which GenderMag?}} & \textbf{Considered} & \textbf{\# Bugs} & \textbf{\# Bugs}  \\
       \textbf{Number} & &  & \textbf{\failureAI{AI-Wrong}{black}} & \textbf{Found} & \textbf{Fixed}  \\
    \midrule
        \textSmallGray{1--5}  & \textSmallGray{All 3} & \textSmallGray{\OriginalGM{}} & \textSmallGray{No} & \textSmallGray{54} & \textSmallGray{34}  \\
       % \textSmallGray{1}  & \textSmallGray{Weather} & \textSmallGray{\OriginalGM{}} & \textSmallGray{No} & \textSmallGray{3} & \textSmallGray{3}  \\
       % \textSmallGray{2} & \textSmallGray{Farm} & \textSmallGray{\OriginalGM{}} & \textSmallGray{No} & \textSmallGray{13} & \textSmallGray{8}   \\
       % \textSmallGray{3} & \textSmallGray{Farm} & \textSmallGray{\OriginalGM{}} & \textSmallGray{No} & \textSmallGray{8} & \textSmallGray{5}  \\
       % \textSmallGray{4} & \textSmallGray{Game} & \textSmallGray{\OriginalGM{}} & \textSmallGray{No} & \textSmallGray{5} & \textSmallGray{3} \\
       % \textSmallGray{5} & \textSmallGray{Game} & \textSmallGray{\OriginalGM{}} & \textSmallGray{No} & \textSmallGray{25} & \textSmallGray{15} \\

       6 & Farm & \postForkGM{} & No  & 12 & 6\\
       7 & Weather & \preForkGM{} & Yes & 9 & 5 \\
       8 & Game & \preForkGM{} & Yes & 8 & 2 \\
       \hline
       \hline
         &      &          & \multicolumn{1}{r}{\textit{Totals}:}& 83 & 47 \\
    \bottomrule
    \end{tabular}
    \caption{    
        Teams Weather and Game successfully found and fixed AI inclusivity bug instances with the \preForkGM{}. With this version, teams considered when the persona doubted the AI.
    }
    \label{tab:results_adapt_gm_prefork_sessions}
\end{table}

%% file: tables/Triangulation-Table.tex
\begin{table}[h]
    \centering
    \begin{tabular}{l|c|c|c|c|cc}
        % {\cellcolor{AI-Wrong}\rotatebox[]{0}{Bugs}} & 
        \toprule
        \multirow{2}{*}{} &
        {\textbf{Bug in}} & 
        {\textbf{Bug in}} & 
        {\textbf{Unique bug}} & 
        {\textbf{Verbal}} &
        \multirow{2}{*}{\textbf{Total}} \\
        {} &
        {\textbf{\noFailureAI{AI-Right}{black}}} & 
        {\textbf{\failureAI{AI-Wrong}{black}}} & 
        {\textbf{in fork}} & 
        {\textbf{support}} &
        {}
        \\

    \midrule 
        Team Game &
        \checkmark \checkmark & 
        \checkmark \checkmark & 
        \checkmark \checkmark  & 
        \checkmark &
        7
        
        \\
        \hline
        Team Weather &
        \checkmark  & 
        {} &
        {}  & 
        \checkmark  &
        2
        \\
        \hline
        \hline
        {} & 3 & 2 & 2 & 2 & 9
   %Note, always put a \\ before bottomrule, or the pdf will break. It needs to distinguish between midrule and bottomrule.
        \\
    \bottomrule
    \end{tabular}
    
    \caption{
        Triangulation Table---9 sources of evidence of teams' reactions to the \preForkGM{}. 
    }
    \label{tab:Triangulation-Table}
\end{table}

%% file: doc/07-Discussion.tex
\section{Discussion
\draftStatus{MMB 10/9/25}{2.4?}
}
% \subsection{AI Inclusivity Bugs and Theory
\label{subsec:find_discussion}

\boldify{How 'bout some theory? Let's start with Blackwell's model. }

Several theoretical lenses provide perspectives on the 6 AI inclusivity bug types the teams found.
In this section, we consider how these theoretical lenses may be of practical use to AI product designers.

First and most obvious, the bug instances tie directly to the foundational theories behind  GenderMag's problem-solving style types~\cite{burnett2016gendermag}.
Recall that, whenever a team noticed an AI inclusivity bug, they tied that bug to any problem-solving styles that they expected the bug to particularly impact.
For example, they often tied instances of \bugInterpret{} (``What does this even mean?'') to risk-averse users and to  those who are comprehensive information processors.  
These ties provide the ``why''s behind the AI inclusivity bugs---why users with those problem-solving attributes might experience that particular bug.
For these teams, these why's often pointed the way toward fixes.  
Continuing our example, when they tied an instance of \bugInterpret{} to comprehensive information processing, they tended to fix the bug by making more information available.
Thus, these ties between an AI inclusivity bug and the problem-solving styles behind it sometimes pointed the way toward how to fix the bug.

%Viewing the AI inclusivity bugs through theoretical lenses may bring abstractions that can help AI developers to address future instances of them.
%Toward that end, here we consider the 6 AI inclusivity bug types through two theory lenses.

Another theory that relates to these AI inclusivity bug types is Blackwell's model of attention investment~\cite{blackwell2002first}, already alluded to in Section~\ref{sec:Results-AI-Bugs}.
%
%\boldify{All of these bugs can be viewed as obstructions to receiving the hoped-for benefits in the Blackwell perspective.}
%
Strictly speaking, this model's units are units of attention (similar to time): attention costs or investments (similar to ``time I must spend now'') vs. attention benefits (``time I save later''), modified by the risk (``probability that if I  spend the time I still won't receive the benefits'').
However, for this discussion, we relax this constraint to allow \textit{any} kind of cost/benefit. 

This relaxed model of attention investment enables unifying the 6 AI inclusivity bugs as obstructions to receiving the benefits.
For this discussion, we use Team Weather as a running example. 
Since Team Weather's intended users are agricultural growers, the hoped-for benefit they  gain from using these AI products is to produce higher quality/quantity of crops.
Here, \bugMoreInfo{} could stop some growers from having the information they need to decide whether to spend hard-earned dollars on frost mitigation today to salvage some percentage of their crops, versus avoiding the expenditure while leaving crop survival rate more to chance. 
The other five AI inclusivity bugs---\bugInterpret{}, \bugInputOutput{}, \bugWhyShouldI{}, \bugActionable{}, and \bugChanges{}---can similarly be seen as barriers to some growers receiving the hoped-for benefit of more financially viable farms.

Under this reasoning, the attention investment lens could facilitate an AI product designer using a cost-benefit perspective when ideating potential fixes.
Specifically, it suggests an AI-inclusivity debugging approach driven by: ``how can I fix this in a way that brings to diverse users greater benefit or lower cost and/or a lower probability of failing to receive this benefit?''  

A third theory lens is Norman's Gulf of Execution and Gulf of Evaluation~\cite{norman1986gulfs}.
A Gulf of Execution describes barriers to \textit{doing} something, and a Gulf of Evaluation describes not knowing whether what a user just \textit{did} made an impact (and if so, what).
This theory brings an actionability perspective to the 6 AI inclusivity bugs.

Using this perspective, the \bugMoreInfo{} example above could create an execution gulf, because lack of information could stop some growers from knowing a suitable way to take action on their farm. 
\bugActionable{} directly describes a barrier to taking suitable action. 
\bugWhyShouldI{} suggests a barrier between the goal of a more productive farm and engaging in \textit{any} way with the AI.
\bugInputOutput{} and \bugChanges{} raise gulfs of evaluation, and \bugInterpret{} could produce gulfs of either execution or evaluation.

Using the Norman gulfs could enable an AI product designer to think about fixes from an actionability perspective.
For example, it suggests an inclusivity debugging approach driven by: ``how can I fix this in a way that enables diverse users to take the most appropriate action for their particular farm?''  

%**IFt too? (potential fixes)

\boldify{working on a sunset here}
It remains an open question whether and which of the unifying directions these theories encourage will be useful abstractions beyond the 6 AI inclusivity bug types this paper identified.
We look forward to future researchers' discoveries of new AI inclusivity bugs types that will enable exploration of this question.

%% file: doc/08-Limitations.tex
\section{Limitations
\draftStatus{MMB10/8/25}{2.3: good enough }
}

Every empirical study has limitations~\cite{Wohlin-2012, ko2015practicalempirical}, and ours is no exception.
%
%\boldify{One limitation of our work is that all teams chose Abi and did not evaluate from Tim's perspective}
%
One limitation of our results is that all three teams evaluated their AI products from the perspective of only the Abi persona.
%This means that the 83 instances of the six AI inclusivity bug types all associated Abi's five problem-solving style values (e.g., risk-averse, comprehensive information processing style, etc.).
Ideally, the teams would have also evaluated from the perspective of the Tim persona, so as to work on addressing inclusivity bugs at both endpoints of each problem-solving style.
The teams elected to use only Abi to save time, but this shortcut missed an opportunity to find AI inclusivity bugs across the full range of problem-solving values.

%\boldify{This might also have led to a limitation of the fixes that the teams found}

%A limitation of our results is that only Team Game had an AI expert on their team, although they all worked with AI products.
%This might mean that that the insights for why the teams' personas would not take an action (i.e., an AI inclusivity bug) might have been missing some of the nuances from an AI expert's perspective.
%This limitation might have compounded into our findings in Section~\ref{sec:RQ2-Fix} regarding how teams fixed AI inclusivity bugs.
%AI experts may have found additional nuances to the teams' AI inclusivity bug fixes, which might have resulted in additional discussion or specific guidance.

\boldify{A limitation is that we didn't try all possible combinations of the GM walkthrough. How do we know that the ``everything + kitchen sink'' approach isn't the best?}

A limitation in experimenting with different variants of GenderMag-for-AI was the availability of teams' time to try out all the suggestions to change the \OriginalGM{}.
Between sessions, we only had time to incorporate some of the teams' suggestions.
As previously stated, time was  available to try out only two versions
Thus, although the \preForkGM{} demonstrated the most promise for two teams, there may be a variant that fits AI products better.
Still, the suggested-but-untried variants are shown in the supplemental document, and we invite interested researchers to try out the others.

\boldify{List of bugs not complete.}

Finally, the list of AI inclusivity bugs we identified in our data is unlikely to be a complete list.  
For example, in eXplainable AI (XAI) alone, there are many explanation types that our data do not cover (e.g., counterfactual, feature-importance, saliency maps).
%and levels of granularity (i.e., local and global explanations).
Because this was a field study of particular teams in particular contexts with particular AI products, we expect more AI inclusivity bug types to emerge as other researchers begin to find them in other types of AI products and explanations.

\boldify{Given these limitations, our results are not generalizable, and more studies are needed.}

Given these limitations, our results do not generalize beyond the particular context of our investigation.
Generalization beyond this context is only possible with further investigations in a variety of contexts using a variety of empirical approaches.

%Limitations and threats like these can only be addressed by having additional teams running GenderMag walkthroughs (and other 

%% file: doc/09-Conclusion.tex
\section{Conclusion
\draftStatus{FAM}{?? was 2+, 10/10/2025 9:30 am pst}
}

% \MMB{As of now it's a teeth-gritting D2, but maybe not a strong enough D2 to get you graduated.  I'll mark up soon...}

\boldify{This paper's research questions create four promises}

This paper set out to investigate \aiInclBug{}s, and how to find and fix them.
To investigate this question, we conducted a field study, in which three AI product teams used several variants of the GenderMag inclusive design method to evaluate their own products.
Our field study's results defined the concept of AI inclusivity bugs, revealed 6 types of these bugs with 83 instances, and revealed fixes covering 47 of these instances.

\boldify{Result: six  AI inclusivity bug types (and fixes)}

\aiInclBug{}s are a new concept.
They exist only in an AI's information and \textit{disproportionately impact} some group(s) of AI product users.
Six \aiInclBugType{} categories emerged from the teams' AI products along with their fixes to many of them, answering RQ1 (types of \aiInclBugType{}s) and RQ2 (how teams fixed them):

\begin{itemize}[]
    \item \bugInterpret{} (27 instances): ``What does this even mean?'' Particularly tied with risk-aversion and/or comprehensive information processing. Common fixes added clarifications to features that triggered the bug instances.
    \item \bugInputOutput{} (9 instances): ``What does this (input) have to do with that (output)?'' Particularly tied with risk-aversion. Common fixes were adding explicit ties between the AI's inputs and its outputs. 
    \item \bugWhyShouldI{} (19 instances): ``Why even look at this?'' Particularly tied with risk aversion. Common fixes used instantiations of the Surprise-Explain-Reward strategy.
    \item \bugMoreInfo{} (9 instances): ``Need more info!'' Particularly tied with comprehensive information processing. Common fixes were adding more information.
    \item \bugActionable{} (12 instances): ``So? What should I DO?'' Particularly tied with comprehensive information processing. Common fixes were making possible actions explicit.
    \item \bugChanges{} (7 instances): ``What's changed?'' Particularly tied with self-efficacy relatively low compared to  peers. Common fixes were to be explicit about what changed.
\end{itemize}

%\input{tables/Conclusion-BugTypes}

%Thus, these results confirmed that \aiInclBug{}s exist and showed how these teams both found and fixed them.

% Our second contribution was showing how teams' fixed these six AI inclusivity bug types.
% Once teams found these AI inclusivity bug instances by applying GenderMag, they found solving them to be relatively straightforward.
% In most cases, these fixes were information-oriented, which involved adding, clarifying, mapping, and/or providing follow-up suggestions.

%\boldify{The third contribution was the new GenderMag-for-AI variants.}

%(RQ3) Are current inclusive design methods (like GenderMag) ``enough'' for evaluating AI products?
%and (RQ4) If the answer is no to RQ3, how do inclusive design methods need to change for AI-specific needs?

The AI product teams' work also provided an answer to RQ3, which asks if a current inclusive design method (GenderMag) is ``enough'' for evaluating AI-powered systems.
The teams' work showed that, although the original method did allow them to be effective at finding \aiInclBug{}s, it had a blind spot---it did not prod them to consider users being doubtful of the AI's recommendations/decisions.

Finally, RQ4 asked how the method should change to better accommodate AI-powered systems, which resulted in GenderMag-for-AI variants.
The teams all agreed on the importance of considering both when users believed the AI and when they did not.
Of the methods the teams tried, \preForkGM{} was the most successful, uncovering 
\aiInclBug{}s in situations that the original method overlooked.
As Team Weather put it, before using \preForkGM{}...

%  which gives AI product teams both GenderMag and GenderMag-for-AI to tackle inclusivity issues in their AI products.

% \textcolor{red}{MMB: layout thing: try to make sure the quote below doesn't end up on a new page}

% Notably, one advantage of the \preForkGM{} 
\quotateInsetCommentary
%Note, this is for when they're adapting GenderMag. They're NOT finding bugs. They're discussing the process.
%Team Name
{Weather}
%Quote Snippet
{...if the user doesn't trust it?...it really isn't something that had ever occurred to me at all.}
%Session (and step) where the comment occurred
{**FIXME**
}
%Full quote (from the form), so we don't have to keep going back to excel
{And I think it is really interesting to think about that. How, what if the user doesn't trust it? And, and I'm going to try to put myself in a little thought experiment later today to try to work with that more, because it really isn't something that had ever occurred to me at all.
}

%TC: ignore
% \textcolor{red}{MMB 6/28/25: @AAA, please add acknowledgments here. At least AgAid and my 2 (stop-worked) NSF grants. Also thank the Teams (but don't use their names).}
\section{Acknowledgments}
We are grateful to the AI product teams for their participation and hard work.
This work has been supported in part by  <anonymized> and <anonymized>, and by <anonymized>.
%TC: endignore